\documentclass[aps,prb,reprint,preprintnumbers,superscriptaddress,twocolumn,nofootinbib]{revtex4}
\usepackage{graphicx}
\usepackage{bm,amsmath}
\usepackage{dcolumn}
\usepackage{amsfonts,amssymb}
\usepackage[colorlinks=true, pdfstartview=FitV, linkcolor=red, citecolor=blue, urlcolor=blue]{hyperref}
\usepackage{enumitem}

\usepackage{color}

\graphicspath{{./Figures/}}

\newcommand{\diff}{\mathrm{d}}
\newcommand{\p}{\partial}
\newcommand{\ve}{\varepsilon}
\newcommand{\Diff}{{\mathcal{D}}}

\newcommand{\be}{\begin{equation}}
\newcommand{\ee}{\end{equation}}
\newcommand{\bea}{\begin{eqnarray}}
\newcommand{\eea}{\end{eqnarray}}
\newcommand{\im}{\mathrm{i}}

\newcommand{\calU}{\mathcal{U}}

\newcommand{\tr}{\mathrm{tr}}

\usepackage[normalem]{ulem}

\begin{document}

\title{Anomaly and global inconsistency matching: $\theta$-angles, $SU(3)/U(1)^2$ nonlinear sigma model, $SU(3)$ chains and its generalizations}

\author{Yuya Tanizaki}
\email{yuya.tanizaki@riken.jp}
\affiliation{RIKEN BNL Research Center, Brookhaven National Laboratory, Upton, NY, 11973, USA}

\author{Tin Sulejmanpasic} 
\email{tin.sulejmanpasic@gmail.com}
\affiliation{Philippe Meyer Institute, Physics Department, \'Ecole Normale Sup\'erieure, PSL Research University, 24 rue Lhomond, F-75231 Paris Cedex 05, France}
\affiliation{Institute for Nuclear Physics, University of Mainz, D-55099 Mainz, Germany}

\date{\today}

\begin{abstract}
We discuss the $SU(3)/[U(1)\times U(1)]$ nonlinear sigma model in 1+1D and, more broadly, its linearized counterparts. Such theories can be expressed as $U(1)\times U(1)$ gauge theories and therefore allow for two topological $\theta$-angles. These models provide a field theoretic description of the $SU(3)$ chains. We show that, for particular values of $\theta$-angles, a global symmetry group of such systems has a 't~Hooft anomaly, which manifests itself as an inability to gauge the global symmetry group. By applying anomaly matching, the ground-state properties can be severely constrained. The anomaly matching is an avatar of the Lieb-Schultz-Mattis (LSM) theorem for the spin chain from which the field theory descends, and it forbids a trivially gapped ground state for particular $\theta$-angles. 
We generalize the statement of the LSM theorem and show that 't Hooft anomalies persist even under perturbations which break the spin-symmetry down to the discrete subgroup $\mathbb Z_3\times\mathbb Z_3\subset SU(3)/\mathbb Z_3$. 
In addition the model can further be constrained by applying global inconsistency matching, which indicates the presence of a phase transition between different regions of $\theta$-angles. 
We use these constraints to give possible scenarios of the phase diagram. We also argue that at the special points of the phase diagram the anomalies are matched by the $SU(3)$ Wess-Zumino-Witten model. We generalize the discussion to the $SU(N)/U(1)^{N-1}$ nonlinear sigma models as well as the 't~Hooft anomaly of the $SU(N)$ Wess-Zumino-Witten model, and show that they match. 
Finally the $(2+1)$-dimensional extension is considered briefly, and we show that it has various 't~Hooft anomalies leading to nontrivial consequences. 
\end{abstract}


\maketitle

\section{Introduction}\label{sec:intro}

Spin chain is an important subject of many-body physics, and has been studied extensively both in classical and quantum mechanical contexts. It also gives examples of how striking differences can arise between quantum mechanics and classical analogues. Amongst, the most studied spin chains would be the Heisenberg $SO(3)$ spin chain\footnote{We will also call this system an $SU(2)$ spin chain, to indicate that the quantum version can have states in both integer and half-integer spin representation. }, with the Hamiltonian of the form
\[
H=J\sum_{\langle i,j\rangle}\bm S_i\cdot \bm S_j
\]
where $\bm S_i$ is the spin vector at the lattice site $i$. When $J>0$ the interactions prefer anti-ferromagnetic order.

The quantum variants of such chains were conjectured by Haldane to behave radically different when spin is integer or half-integer~\cite{Haldane:1982rj, Haldane:1983ru}.
In particular, by studying large-dimensional $SU(2)$ representations on each site, Haldane argued that integer and half-integer Heisenberg spin chains fall into different universality classes, the former being gapped while the latter is gappless.
The more modern perspective claims that the gappless nature of half-integer spin chains is understood as a consequence of the Lieb-Schultz-Mattis (LSM) theorem~\cite{Lieb:1961fr, Affleck:1986pq, PhysRevLett.84.1535, Hastings:2003zx}, which is a powerful theorem exploiting the fact that $SO(3)$ spin rotation acts projectively on half-integer spins. More precisely, the LSM theorem proves that either the anti-ferromagnetic chain is gapless or  breaks translational symmetry spontaneously. 
Therefore, the Haldane conjecture may be rephrased that as long as spin-symmetry and lattice translation symmetry are good symmetries, the integer antiferromagnetic spin-chains have trivial ground-states, while half-integer ones are nontrivial\footnote{By a trivial ground state we mean that the system is gapped and ground state is non-degenerate, while the nontrivial ground state is either gapless, breaks some global symmetry or has topological degeneracy}.
The conjecture is confirmed explicitly by exactly solvable systems, like Bethe ansatz on spin-$1/2$ chain~\cite{Bethe:1931hc} and AKLT model for the spin-$1$ chain~\cite{Affleck:1987vf}. 

Generalization of $SU(2)$ chains to $SU(N)$ chains has attracted the interest in various aspects. In fact the LSM theorem is also known for $SU(N)$ chains~\cite{Affleck:1986pq}, showing a nontrivial nature of the ground states depending on the representation. Taking the large representation limit, some spin systems can again be described by nonlinear sigma models which are both asymptotically-free, and have nontrivial ground states. 
For example, the critical nature of the $U(2N)/[U(N)\times U(N)]$ Grassmannian nonlinear sigma model was studied in Refs.~\cite{Affleck:1985wb, Affleck:1988wz}. 
Experimentally, there is a possibility to realize the $SU(N)$ chains via ultracold atoms~\cite{PhysRevLett.91.186402, PhysRevLett.92.170403, 1367-2630-11-10-103033, gorshkov2010two, PhysRevB.86.224409, scazza2014observation, taie20126, pagano2014one, Zhang1467, 0034-4885-77-12-124401, CAPPONI201650}, and theoretical conjectures on $SU(N)$ spin systems can be tested in the future.

Bykov~\cite{Bykov:2011ai, Bykov:2012am} has derived the relativistic sigma models from anti-ferromagnetic $SU(N)$ spin chains with the $p$-box symmetric representation on each site. There, it is shown that the effective theory of a specific spin chain has the flag-manifold target space, $SU(N)/U(1)^{N-1}$. There it was pointed out that such a sigma model allows $N-1$ independent topological $\theta$ terms~\cite{Bykov:2011ai}, $\theta_1,\ldots,\theta_{N-1}$, and they take the specific value~\cite{Bykov:2012am}, $\theta_k=2\pi k/N$. 
Lajk\'o, Wamer, Mila, and Affleck~\cite{Lajko:2017wif} have recently analyzed the phase structure of the $SU(3)$ spin chains with the $p$-box symmetric representation using the $SU(3)/[U(1)\times U(1)]$ nonlinear sigma model. 
They showed the LSM theorem for the $SU(3)/\mathbb{Z}_3$ spin symmetry and the lattice translation for $p\not=0$ mod $3$, and thus the trivial mass gap cannot appear. They also analyzed the lattice strong-coupling limit to gain insight  into the phase diagram, and performed the Monte Carlo simulation to check it using the imaginary $\theta$ angles following Ref.~\cite{PhysRevD.77.056008}. 

In this paper, we shall show that the symmetry itself can constrain the possible phase diagram more strongly. For that purpose, we study the $SU(3)/[U(1)\times U(1)]$ nonlinear sigma model from the viewpoint of the 't~Hooft anomaly matching and global inconsistency matching. 
't~Hooft anomaly is the obstruction to gauging the global symmetry. The consequence of this is that the vacuum cannot be trivially gapped~\cite{tHooft:1979rat, Frishman:1980dq, Coleman:1982yg}  (see also Refs.~\cite{Vishwanath:2012tq, Wen:2013oza, Kapustin:2014lwa, Kapustin:2014zva, Cho:2014jfa, Wang:2014pma, Witten:2015aba,Seiberg:2016rsg,Witten:2016cio, Tachikawa:2016cha, Tachikawa:2016nmo, Gaiotto:2017yup,Wang:2017txt, Tanizaki:2017bam, Komargodski:2017dmc, Komargodski:2017smk, Cho:2017fgz,Shimizu:2017asf, Wang:2017loc, Metlitski:2017fmd,Kikuchi:2017pcp, Gaiotto:2017tne, Tanizaki:2017qhf, Tanizaki:2017mtm, Guo:2017xex, Sulejmanpasic:2018upi, Aitken:2018kky, Kobayashi:2018yuk} for recent developments). 
The 't~Hooft anomaly matching provides the field-theoretic description of the LSM theorem for corresponding lattice quantum systems, and we can reproduce the same constraint on the possible low-energy physics. 
Global inconsistency condition is a more subtle obstruction while gauging the symmetry~\cite{Gaiotto:2017yup, Tanizaki:2017bam, Kikuchi:2017pcp}.

In our nonlinear sigma model, the spin rotation symmetry, $PSU(3)=SU(3)/\mathbb{Z}_3$, is a good symmetry for all the $\theta$ angles, but for special values of the $\theta$ angles there also exists a charge conjugation symmetry $\mathsf{C}$.  At all $\mathsf{C}$-symmetric points, we can gauge the whole $PSU(3)\rtimes \mathsf{C}$ symmetry, so there is no 't~Hooft anomaly for $PSU(3)\rtimes \mathsf{C}$. However, gauging of $\mathsf{C}$ requires a special choice of the discrete $\theta$ parameter of $PSU(3)$ gauge fields, and thus they can be different for different $\mathsf{C}$ invariant $\theta$ angles. When this occurs we say that different regions of the parameter space have a global inconsistency~\cite{Gaiotto:2017yup, Tanizaki:2017bam, Kikuchi:2017pcp}. A consequence of this inconsistency is that either: 1) the two regions are trivially gapped, but one must encounter a phase transition in between or 2) the ground state of one of the two $\mathsf C$-invariant regions is nontrivial.  
By using the matching condition for both 't~Hooft anomaly and global inconsistency, we will find the constraint on the phase diagram that go beyond the LSM theorem. 

We will see that the whole discussion of anomalies and global inconsistencies can be generalized to the $SU(N)/U(1)^{N-1}$ nonlinear sigma models and their linear counterparts.  In particular such models have a $PSU(N)=SU(N)/\mathbb Z_N$ global spin, or flavor, symmetry. They also allow $N-1$ topological $\theta$-angles. At particular values of the $\theta$-angles, they also have $\mathbb Z_N$ global symmetry, which we call the $\mathbb Z_N$ cyclic permutation symmetry\footnote{The name comes from the fact that the relevant models involve $N$ copies of fields which can be mapped cyclically into each other, as we shall see.}. The two symmetries $PSU(N)$ and $\mathbb Z_N$ have a mixed 't Hooft anomaly. Moreover the subgroup $\mathbb Z_N\times\mathbb Z_N\subset PSU(N)$ also has a mixed 't Hooft anomaly with the $\mathbb Z_N$ cyclic permutation symmetry.
When $N$ is even, we also show that there is a 't~Hooft anomaly involving a time-reversal symmetry~\cite{Sulejmanpasic:2018upi}, and the phase diagram can be constrained even when the global spin-symmetry is explicitly broken completely.

It is an interesting question to ask what is the possible conformal field theory if the 't~Hooft anomaly is matched by the existence of gappless excitations. 
In order to explore it, we consider the two-dimensional $SU(N)$ Wess-Zumino-Witten (WZW) model and find the correspondence for symmetries and their 't~Hooft anomaly.  
The anomaly can constrain the possible level number of WZW model. 
The computation of anomaly shall be done by gauging the symmetry of WZW models directly, and we will find the anomaly polynomial described by the $(2+1)$-dimensional symmetry-protected topological (SPT) phase.  
In order to elucidate why these two models have the same 't~Hooft anomaly, we consider a deformation of the WZW model which reduces the $[SU(N)_L\times SU(N)_R]/\mathbb{Z}_N$ global symmetry to the $PSU(N)_V\times \mathbb Z_N$ symmetry. As a result we obtain an $SU(N)/U(1)^{N-1}$ nonlinear sigma model, and hence they must contain the same 't~Hooft anomaly.

The linear sigma model description is also discussed, and it provides a useful consistency check of the phase diagram when all the matter fields are very massive. 
In that limit, the theory becomes a gauge theory of free photons, and we clarify the concrete consequence of the 't~Hooft anomaly and global inconsistency using that example. 
We also propose the circle compactification of the model so that the 't~Hooft anomaly discussed in this paper persists for any size of the compactification radius. Since the model has asymptotic freedom, this provides an opportunity to study the $SU(3)/[U(1)\times U(1)]$ sigma model semiclassically. 

{
We also discuss the $(2+1)$-dimensional version of our model very briefly. While it does not have the $\theta$ terms, it contains a $U(1)\times U(1)$ topological symmetry. 
We show that the model has various 't~Hooft anomalies involving the topological symmetry, indicating that the model cannot be trivially gapped. 
}

The paper is organized as follows. In Sec.~\ref{sec:model}, we explain details about $SU(3)/[U(1)\times U(1)]$ nonlinear sigma model and its symmetries. 
We discuss their 't~Hooft anomaly and global inconsistency in Sec.~\ref{sec:anomaly}, and their implication on the phase diagram is also discussed there. 
Section~\ref{sec:generalization_SU(N)} is devoted to the generalization of our analysis to $SU(N)/U(1)^{N-1}$ nonlinear sigma models, and an anomaly involving time-reversal is found for even $N$. 
We discuss the 't~Hooft anomaly of the $SU(N)$ Wess-Zumino-Witten model in Sec.~\ref{sec:WZWmodel}. 
In Sec.~\ref{sec:linear_model}, we construct the linear sigma model having the same 't~Hooft anomaly and global inconsistency, and perform the analytic computation of the partition function in certain cases. 
In Sec.~\ref{sec:compactification}, we discuss the a small-circle compactification of the nonlinear sigma model, whose phase structure can be adiabatically connected to the large circle limit from the viewpoint of anomaly. 
We discuss the $(2+1)$-dimensional version of the model in Sec.~\ref{sec:2+1D}. 
We make conclusions in Sec.~\ref{sec:conclusion}. 

\section{$SU(3)/[U(1)\times U(1)]$ sigma model and symmetry}\label{sec:model}

An $SU(3)$ spin chain with the $p$-box symmetric representation on each site can be described by a nonlinear sigma model whose target space is the flag manifold $SU(3)/[U(1)\times U(1)]$ with the specific theta terms $\theta=2\pi p/3$ in the large-$p$ limit~\cite{Bykov:2011ai, Bykov:2012am, Lajko:2017wif}. 
We first explain the nonlinear sigma model in Sec.~\ref{sec:details_model}, and discuss its symmetries in Sec.~\ref{sec:symmetry}. To be self-contained, we briefly review its connection with the corresponding lattice spin Hamiltonian in Sec.~\ref{sec:lattice_SU(3)} following Ref.~\cite{Bykov:2011ai, Bykov:2012am, Lajko:2017wif}. 

\subsection{$SU(3)/[U(1)\times U(1)]$ nonlinear sigma model}\label{sec:details_model}

We consider the nonlinear sigma model with the target space $SU(3)/[U(1) \times U(1)]$. The Lagrangian is given by 
\bea
S&=&\sum_{\ell=1}^{3}\int_{M_2} \left[-{1\over 2g}\left|(\diff+\im a_\ell) \bm{\phi}_\ell\right|^2 +{\im \theta_\ell\over 2\pi}\diff a_\ell\right.\nonumber\\
&&\left.+{\lambda\over 2\pi} (\overline{\bm{\phi}}_{\ell+1}\cdot \diff \bm{\phi}_\ell)\wedge (\bm{\phi}_{\ell+1}\cdot \diff \overline{\bm{\phi}}_\ell)\right],
\label{eq:SU(3)/U(1)2NLSM}
\eea
where $\bm{\phi}_\ell=(\phi_{1,\ell},\phi_{2,\ell},\phi_{3,\ell}):M_2\to \mathbb{C}^3$ are three-component complex scalar fields with the constraint, 
\bea
&&\overline{\bm{\phi}}_\ell \cdot \bm{\phi}_{\ell'}=\delta_{\ell \ell'}, \label{eq:orthonormality}\\
&&\ve_{abc}\phi_{a,1}\phi_{b,2}\phi_{c,3}=1, \label{eq:constraint_det}
\eea
and $a_i$ are $U(1)$ gauge fields\footnote{We take the convention that the dynamical gauge fields are denoted by lowercases $a_i$, and that the background ones by uppercases $A,B,C,\ldots$, unless stated explicitly. }. The constraint claims that the $3\times 3$ matrix, $\mathcal{U}=[\bm{\phi}_1,\bm{\phi}_2, \bm{\phi}_3]$, is special unitary, 
\be
\mathcal{U}^\dagger \mathcal{U}=\bm{1}_3,\;\det[\mathcal{U}]=1. 
\ee
As we shall see soon later, the gauge fields obey the constraint $a_1+a_2+a_3=0$ as a consequence of the equation of motion, so the target space is divided by $U(1)\times U(1)$ by gauge invariance and becomes the flag manifold $SU(3)/[U(1)\times U(1)]$. 
The first term is the usual kinetic term of the nonlinear sigma model, and the second one is the topological theta term of the $2$d $U(1)$ gauge theory. 
The last term is the new feature of this nonlinear sigma model, called the $\lambda$-term in Ref.~\cite{Lajko:2017wif}. It is linear both in space and time derivatives, but not topologically quantized to integers unlike the theta terms. It will not be important for our discussion, as it will not be important for the 't Hooft anomaly matching. Furthermore, it is not a universal term of the underlying spin-model and is also perturbatively irrelevant~\cite{Lajko:2017wif}.

The $\lambda$-term may not look gauge-invariant at the first sight, so let us confirm it explicitly. Consider the $U(1)$ gauge transformation, $\bm{\phi}_\ell\mapsto g_\ell \bm{\phi}_\ell$ with $g_\ell:M_2\to  U(1)$, then 
\bea
\overline{\bm{\phi}}_{\ell+1}\cdot\diff \bm{\phi}_\ell
&\mapsto& g_{\ell+1}^{-1}g_\ell \overline{\bm{\phi}}_{\ell+1}\cdot\diff \bm{\phi}_\ell+ \overline{\bm{\phi}}_{\ell+1}\cdot\bm{\phi}_\ell g_{\ell+1}^{-1}\diff g_\ell\nonumber\\
&=& g_{\ell+1}^{-1}g_\ell \overline{\bm{\phi}}_{\ell+1}\cdot\diff \bm{\phi}_\ell, 
\eea
and here we use the orthogonality condition. This proves the $U(1)$ gauge invariance of the $\lambda$-term. 

When $\lambda=0$, the Lagrangian looks like three independent copies of $\mathbb{C}P^2$ nonlinear sigma models, but they are still coupled via the orthonormality constraint. Consequence of the constraint on the topological charges is very important. 
Solving the equation of motion of $a_\ell$, we find that 
\be
a_\ell={\im\over 2}\left(\overline{\bm{\phi}}_\ell \cdot \diff \bm{\phi}_\ell -\diff \overline{\bm{\phi}}_\ell\cdot \bm{\phi}_\ell\right)=\im \overline{\bm{\phi}}_\ell\cdot \diff \bm{\phi}_\ell. 
\ee
As a result of orthonormality, we shall find that 
\be
Q_1+Q_2+Q_3=0, \label{eq:constraint_topological_charges}
\ee
where $Q_\ell={1\over 2\pi}\int \diff a_\ell$ are topological charges $\in\mathbb{Z}$, and thus the Lagrangian contain only two independent $U(1)$ topological charges. Therefore we can always set one of the theta angles equal to zero without loss of generality, and we will set $\theta_2=0$ following Ref.~\cite{Lajko:2017wif} in this and the next sections.

Let us derive this constraint on the topological charge. 
We can solve the constraints (\ref{eq:orthonormality}) and (\ref{eq:constraint_det}) for $\bm{\phi}_3$ uniquely using $\bm\phi_{1,2}$:
\be
\phi_{3a}=\ve_{abc}\overline{\phi}_{b,1}\overline{\phi}_{c,2}. 
\ee
Using this expression, $a_3$ becomes 
\bea
a_3&=&\im \overline{\phi}_{a,3}\diff \phi_{a,3}\nonumber\\
&=&\im \ve_{a b c}\phi_{b,1}\phi_{c,2} \diff (\ve_{a b' c'} \overline{\phi}_{b',1} \overline{\phi}_{c',2})\nonumber\\
&=&\im (-\overline{\bm{\phi}}_1\cdot \diff \bm{\phi}_1-\overline{\bm{\phi}}_2\cdot \diff \bm{\phi}_2). 
\eea
As a result, we find that 
\be
a_1+a_2+a_3=0. \label{eq:constraint_U(1)_field_nonlinear}
\ee
Physical meaning of this constraint is that the sum of $U(1)$ charges of $\bm\phi_\ell$'s must be equal to zero, and this is indeed necessary for the condition (\ref{eq:constraint_det}) having gauge invariance. 
In particular, we obtain the constraint (\ref{eq:constraint_topological_charges}) on the topological charges by taking derivatives. 
Since the Lagrangian is quadratic in $U(1)$ gauge fields $a_\ell$, this constraint obtained by the equation of motion holds at the quantum level. 

\subsection{Global symmetries}\label{sec:symmetry}

Next, we discuss the global symmetry of the model. There are four symmetries of this system:
\begin{itemize}
\item $SU(3)/\mathbb{Z}_3$ flavor symmetry
\item Time reversal $\mathsf{T}$
\item $\mathbb{Z}_3$ permutation symmetry (for special theta's)
\item Charge conjugations $\mathsf{C}$ (for different special theta's)
\end{itemize}
We shall explain these symmetry. 

\textit{Flavor symmetry $SU(3)/\mathbb{Z}_3$:} 
The flavor symmetry acts on $\bm{\phi}_\ell$ as $\bm{\phi}_\ell\mapsto U \bm{\phi}_\ell$ for $U\in SU(3)$. $U$ must be the same for $\phi_{1}$, $\phi_2$, and $\phi_3$, in order to maintain the orthonormality (\ref{eq:orthonormality}) and also the $\lambda$-term, and its determinant must be unity in order to maintain (\ref{eq:constraint_det}).  $SU(3)$ acts faithfully on $\bm{\phi}_i$, but these operators are not $U(1)$ gauge invariant. The center $\mathbb{Z}_3\subset SU(3)$ acts trivially on the local gauge-invariant operators, such as $\overline{\phi}_{a,\ell}\phi_{b,\ell}$, and thus the correct global symmetry is $PSU(3)=SU(3)/\mathbb{Z}_3$. 

\textit{Time reversal $\mathsf{T}$:} 
Time reversal symmetry acts as 
\bea
\mathsf{T}&:& \bm{\phi}_\ell(x,t)\mapsto \overline{\bm{\phi}}_\ell(x,-t),\nonumber\\ && a_{\ell0}(x,t)\mapsto a_{\ell0}(x,-t),\nonumber\\
&&a_{\ell1}(x,t)\mapsto -a_{\ell1}(x,-t). \label{eq:time_reversal}
\eea
The kinetic term is invariant trivially. 
Under this definition of the time reversal, $\int \diff a_\ell$ is invariant under the orientation flip of $M_2$, and thus the theta terms are time-reversal invariant at any theta angles. 
The $\lambda$-term is also invariant as follows: Notice that $\overline{\bm{\phi}}_{\ell+1}\cdot \diff \bm{\phi}_\ell(x,t)\mapsto \bm{\phi}_{\ell+1}\cdot \diff \overline{\bm{\phi}}_\ell(x,-t)$. Since the wedge product anti-commutes, we get one negative sign for the $\lambda$-term, but the linear time derivative gives another negative sign, so the action becomes invariant in total. 

\textit{$\mathbb{Z}_3$ permutation:} 
$\mathbb{Z}_3$ symmetry is the symmetry by the cyclic permutation of the fields 
\be
\bm{\phi}_i\mapsto \bm{\phi}_{\ell+1}\;, \;\;\; a_i\mapsto a_{\ell+1}, \label{eq:Z3_permutation}
\ee
where the label $\ell$ should be identified mod $3$. 
Under this transformation for $\theta_2=0$, the theta term changes as 
\bea
\theta_1 Q_1+\theta_3 Q_3&\mapsto& \theta_1 Q_2+\theta_3 Q_1\nonumber\\
&=&(\theta_3-\theta_1)Q_1-\theta_1 Q_3. 
\eea
In order for the $\mathbb{Z}_3$ permutation to be a symmetry, the theta angles must satisfy 
\be
2\theta_1=\theta_3,\quad \theta_1+\theta_3=0\;\bmod 2\pi. 
\ee
As a result, the $\mathbb{Z}_3$ invariant points are 
\be
(\theta_1,\theta_3)=(0,0), (\pm 2\pi/3,\mp 2\pi/3)\;\bmod 2\pi\mathbb{Z}\times 2\pi\mathbb{Z}. 
\ee

\textit{Charge conjugations $\mathsf{C}$:} 
We again take the convention $\theta_2=0$. Let us define three different charge conjugation operators 
\be
\mathsf{C}_k: \bm{\phi}_\ell \mapsto -\overline{\bm{\phi}}_{-\ell-k}, a_\ell \mapsto -a_{-\ell-k}.  \label{eq:charge_conjugation}
\ee
For example, $\mathsf{C}_2: \bm{\phi}_{1(3)}\mapsto -\overline{\bm{\phi}}_{3(1)}$ and $\bm{\phi}_2\mapsto -\overline{\bm{\phi}}_2$, so $\mathsf{C}_i$ acts on $\bm{\phi}_i$ as a complex conjugation, but other two fields are exchanged in addition to the complex conjugation. The negative sign on $\bm\phi$ fields is necessary for consistency with the constraint (\ref{eq:constraint_det}). Note that the three charge-conjugations differ by a $\mathbb Z_3$ symmetry, so at $\mathbb Z_3$ symmetric point, they really correspond to the same charge conjugation. 

The kinetic and $\lambda$ terms are invariant under this transformation, and the above reordering $\ell\mapsto -\ell-k$ mod $3$ for some $k=1,2,3$ is necessary for invariance of the $\lambda$-term. The theta terms change nontrivially, and they are symmetry only for special theta angles. 

For $\mathsf{C}_2$, $Q_{1(3)}\mapsto -Q_{3(1)}$ and $Q_2\mapsto -Q_2$, and then $\mathsf{C}_2$ is the symmetry only if 
\be
\theta_1 Q_1+\theta_3 Q_3=-\theta_1 Q_3-\theta_3 Q_1\;\bmod 2\pi, \label{eq:charge_conjugation_2}
\ee
for all $Q_{1,3}\in\mathbb{Z}$. This is solved as 
\be
\theta_1=-\theta_3 \;\bmod 2\pi, 
\ee
and $\mathsf{C}_2$-invariant points form parallel lines. 

For $\mathsf{C}_1$, $Q_{2(3)}\mapsto -Q_{3(2)}$ and $Q_1\mapsto -Q_{1}$, and it is a symmetry only if 
\bea
\theta_1 Q_1+\theta_3 Q_3&=&-\theta_1 Q_1 - \theta_3 Q_2\nonumber\\
&=&(\theta_3-\theta_1)Q_1+\theta_3 Q_3 \;\bmod 2\pi,  
\eea
for all $Q_1,Q_3\in \mathbb{Z}$. 
That is, the $\mathsf{C}_1$-invariant points are 
\be
\theta_3=2\theta_1\;\bmod 2\pi, 
\ee 
and they form parallel lines. 
 Similarly, the $\mathsf{C}_3$-invariant points are 
\be
\theta_1=2\theta_3 \;\bmod 2\pi. \label{eq:charge_conjugation_3}
\ee 
These $\mathsf{C}_k$ invariant lines will be shown later in Fig.~\ref{fig:inconsistency}. 
In particular, we should notice that all $\mathsf{C}_k$ are symmetries at the $\mathbb{Z}_3$-invariant points. 
If we define the parity as Euclid $\pi$ rotation of the $\mathsf{C}_k\mathsf{T}$ transformation, then there are also three distinct parity transformations $\mathsf{P}_k$,
\be
\mathsf{P}_k: \bm{\phi}_\ell (x,t)\mapsto -\bm{\phi}_{-\ell-k}(-x,t), 
\ee
and  they are symmetries only for above special theta angles but not for general theta's. By construction, $\mathsf{C}_k\mathsf{P}_k\mathsf{T}$ is always a symmetry, as is required by the $\mathsf{CPT}$ theorem for relativistic field theories. 

\subsection{Lattice $SU(3)$ chains}\label{sec:lattice_SU(3)} 
 D.~Bykov~\cite{Bykov:2011ai, Bykov:2012am}, and also M.~Lajk\'o, K.~Wamer, F.~Mila, and I.~Affleck~\cite{Lajko:2017wif}, have shown that the $SU(3)/[U(1)\times U(1)]$ nonlinear sigma model (\ref{eq:SU(3)/U(1)2NLSM}) is the field theoretic description of a certain $SU(3)$ spin chains in the large representation limit. 
The Hamiltonian in Ref.~\cite{Lajko:2017wif} is given by the antiferromagnetic nearest and next-to-nearest and ferromagnetic next-to-next-to-nearest Heisenberg interaction, 
\bea
H&=&\sum_{j\in\mathbb{Z}} \left[J_1 S^{\alpha}_{\beta}(j) S^{\beta}_{\alpha}(j+1)+J_2 S^{\alpha}_{\beta}(j)S^{\beta}_{\alpha}(j+2)\right.\nonumber\\
&&\left.-J_3 S^{\alpha}_{\beta}(j) S^{\beta}_{\alpha}(j+3)\right], 
\label{eq:lattice_Hamiltonian}
\eea
where $S(j)$ is the $SU(3)$ spin operator of the $p$-box symmetric representation at the site $j$. 
It is interesting to argue that the coupling $J_2,J_3$ of order of $1/p$ are generated by the quantum fluctuation of the nearest neighbor coupling $J_1$, so the results of the nonlinear sigma model are expected to apply for the nearest neighbor Hamiltonian~\cite{Lajko:2017wif}. 
 
Since the symmetric representation can be constructed by the symmetric tensor product of the defining representation, the coherent state of $S(j)$ can be written by $\mathbb{C}P^2(=SU(3)/[SU(2)\times U(1)])$ field\footnote{For more general representations involving antisymmetric products, we need to prepare the bosonic field for each fundamental representation satisfying the orthogonality constraint, and the coherent state for each spin becomes $SU(3)/[U(1)\times U(1)]$ in certain representations (see, e.g., Ref.~\cite{Bykov:2012am}). In the present case, the coherent state of each spin is described by $\mathbb{C}P^2$ because the representation is a totally symmetric tensor, and the flag manifold $SU(3)/[U(1)\times U(1)]$ for nonlinear sigma model comes out from the nature of the antiferromagnetic interaction. The similar thing occurs also for (2+1)-dimensional $SU(3)$ spin magnets~\cite{PhysRevA.93.021606}. }, and we denote the corresponding unit vector field $\Phi(\ell,\tau)\in SU(3)/SU(2)\subset \mathbb{C}^3$. 
To discuss the low-energy physics of this lattice Hamiltonian, it is convenient to consider the three-site unit cell since it contains up to next-to-next-to-nearest neighbor interaction, and decompose the fluctuation into the slow field among unit cells and fast field inside each unit cell. To that end, we decompose the $3\times 3$ complex matrix field for the unit cell into the transverse fluctuation $L$ and slow rotation $\mathcal{U}=[\bm{\phi}_1,\bm{\phi}_2,\bm{\phi}_3]$ given by the $SU(3)$ matrix: 
\bea
&&[\Phi(3j,\tau), \Phi(3j+1,\tau), \Phi(3j+2,\tau)]\nonumber\\
&=&L(j,\tau)\cdot [\bm{\phi}_1(j,\tau), \bm{\phi}_2(j,\tau), \bm{\phi}_3(j,\tau)]. 
\eea
In the large $p$ limit, the fluctuation of $L$ is of order $\mathcal{O}(1/p)$, and can be integrated out. As a result, the $SU(3)/[U(1)\times U(1)]$ nonlinear sigma model is obtained with $1/g=p\sqrt{(J_1J_2+2J_3 J_1+2J_3 J_2)}/(J_1+J_2)$, $\lambda=p{2\pi(2J_2-J_1)}/(3(J_1+J_2))$, and $\theta=\theta_1=-\theta_3={2\pi p/3}$. That is,  the theory lies on the $\mathbb{Z}_3$-invariant point.  
For details, see Ref.~\cite{Lajko:2017wif}. 

This clarifies the origin of the discrete symmetries, $\mathbb{Z}_3$ permutation. The $\mathbb{Z}_3$ permutation originates from the lattice translational symmetry by one lattice unit. Since the low-energy description treats the three consecutive sites as a single unit cell, the translation symmetry act as the $\mathbb{Z}_3$ internal symmetry.

\section{Anomaly, global inconsistency, and phase structure}\label{sec:anomaly}

In this section, we compute the 't~Hooft anomaly and the global inconsistency for the effective theory of $SU(3)$ chains. 
The 't Hooft anomaly is the manifestation of the Lieb-Schultz-Mattis theorem of the lattice model in the continuum low-energy description, and rules out the trivially gapped ground state.
 
Global inconsistency is a more subtle obstruction for gauging the symmetry: When considering the parameter space of the theory, we have enhancement of symmetry at different values of the $\theta$-angles, e.g.~when $(\theta_1,\theta_3)=(0,0), (\pi,0)$ in $SU(3)/[U(1)\times U(1)]$ model. At each point, the symmetry can be gauged, but their gauging is in a sense inconsistent. In such a case, the ground state of of these points in the phase diagram can be trivially gapped, but they cannot \emph{both} be trivially gapped, and hence still give information about the structure of the phase-diagram.

We will now see how the 't~Hooft anomaly and the global inconsistencies arise.


\subsection{Gauging $SU(3)/\mathbb{Z}_3$ flavor symmetry}

In order to detect the mixed 't Hooft anomaly and global inconsistency, we first gauge the $SU(3)/\mathbb{Z}_3$ flavor symmetry. Naively we would promote the covariant derivatives to include a non-abelian gauge field. Indeed we would have to replace
\be
(\diff +\im a_\ell)\bm{\phi}_\ell\rightarrow (\diff+\im a_\ell+\im A)\bm{\phi}_\ell
\ee
where $A$ is the $SU(3)$ gauge field. Seemingly nothing dramatic happens by this promotion. However, we shall see that this is not true when we gauge $SU(3)/\mathbb{Z}_3$, i.e., the $a_\ell$ gauge fields above can no longer be properly quantized gauge fields. 

To see that $a_1,a_2$ are not properly quantized gauge fields, consider a gauge transformation which takes 
\be
\bm{\phi}_\ell \rightarrow U_\ell \bm{\phi}_\ell=e^{i\varphi_\ell}U\bm{\phi}_\ell\;, U\in SU(3)\;, \varphi_3=-\varphi_1-\varphi_2
\ee
Now for this to be a gauge transformation on a compact manifold, e.g. a two-torus $\mathbb T^2$, we must have that the gauge transformation $U_\ell$ is single valued on the torus to ensure the single-valuedness of $\bm\phi_\ell$. 
Since $U\in SU(3)$ should be regarded as the lift of the $PSU(3)$ matrix, it is required to be periodic up to a center $e^{-\im\frac{2\pi}{3}}\bm{1}_3$, which is mapped to unity in the $PSU(3)$ group. 
However $U_\ell$ must be single valued, which requires  $\varphi_1$ and $\varphi_2$ to be  periodic only up to $2\pi/3$, (so that $\varphi_3$ is periodic up to $-4\pi/3=2\pi/3\bmod 2\pi$).  Since the gauge transformation affects the gauge fields $a_\ell\rightarrow a_\ell-\diff \varphi_\ell$, we see that the $a_\ell$ are no longer properly quantized $U(1)$ gauge fields. In particular holonomies $e^{i\int a_\ell}$ are no longer gauge-invariant operators. Rather the gauge field $3a_\ell$ are properly normalized $U(1)$ gauge fields. This in turn implies that the fluxes of $a_\ell$ will be quantized as multiples of $2\pi/3$. 

In fact this deviation of the lack of $2\pi$ quantization is related to a topological invariant of the $PSU(3)=SU(3)/\mathbb Z_3$ gauge bundle, which is a member of the second cohomology $B\in H^2(M_2,\pi_1(PSU(3)))$ (see, e.g., Ref.~\cite{Witten:2000nv}). This topological invariant can also be thought of as the 2-form $\mathbb Z_3$  gauge field~\cite{Kapustin:2014gua, Gaiotto:2014kfa, Aharony:2013hda}, which is necessary to convert the gauge group $SU(3)$ to $PSU(3)$. 

We therefore introduce the background $SU(3)/\mathbb{Z}_3$ gauge fields, which consist of the two ingredients: 
\begin{itemize}
\item $A$-field -- a $SU(3)$ one-form gauge field, 
\item $B$-field -- a $\mathbb{Z}_3$ two-form gauge field. 
\end{itemize}
We realize the $\mathbb{Z}_3$ two-form gauge field as a pair of the $U(1)$ two-form gauge field $B$ and $U(1)$ one-form gauge field $C$ satisfying the constraint $3B=\diff C$. 
To see how the $B$ gauge field arises in the $SU(3)/\mathbb{Z}_3$ gauge theory, we first embed the $SU(3)/\mathbb Z_3$ gauge field into the $U(3)$ gauge field\cite{Kapustin:2014gua, Gaiotto:2014kfa}, 
\be
\widetilde{A}=A+{1\over 3}C\bm{1}_3, 
\ee
where $A$ is traceless, and $C=\tr \widetilde A$.

However, $PSU(3)$ gauge theory and $U(3)$ gauge theory are different in two ways: (1) $U(3)$ gauge field has an extra $U(1)$ photon $C$, and (2) the 't~Hooft flux of $PSU(3)$ bundle is in $H^2(M_2,\mathbb{Z}_3)$, while that of $U(3)$ bundle is in $H^2(M_2,\mathbb{Z})$. 
These differences can be resolved simultaneously~\cite{Kapustin:2014gua} by postulating the $U(1)$ one-form gauge invariance of $B$. 
In fact, what we want to do is to allow the extra $U(1)$ photon to be absorbed by the already existing $a_1,a_2$ photons (recall that $a_3=-a_1-a_2$ due to the constraint). To that end, let us replace the covariant derivatives  $(\diff +\im a_\ell)\bm{\phi}_\ell$  by 
\be\label{eq:cov_der}
(\diff+\im a_\ell + \im \widetilde{A}) \bm{\phi}_\ell\;,
\ee
If we vary $B\mapsto B+\diff \xi$ and
\be
a_\ell \mapsto a_\ell-\xi,\;C\mapsto C+3\xi,\; \widetilde A\mapsto \widetilde A +\xi\bm 1_3\;,
\ee
the above action will be invariant. However notice that the above transformation is not consistent with the constraint $a_1+a_2+a_3=0$. To fix it let us promote this constraint to 
\be
a_1+a_2+a_3+C=0\;.
\ee
which is still manifestly $\mathbb Z_3$ invariant\footnote{We could have also chosen to maintain the constraint $a_1+a_2+a_3=0$, but then we would have had add an extra term $-\im C$ in the definition of the covariant derivative \eqref{eq:cov_der} for, say, $\ell=3$. This would have made the $\mathbb Z_3$-symmetry slightly less manifest, but the discussion remains unchanged.}.

We emphasize that the gauge-variation parameter $\xi$ is a properly normalized $U(1)$ gauge field, so that we call this gauge symmetry a $U(1)$, 1-form gauge symmetry. This effectively gauges the $U(1)$ center of $U(3)$ gauge bundle, and reduces it to the $SU(3)/\mathbb Z_3$ bundle. Locally, the $C$-field can therefore be gauged away, by choosing $\xi=-C/3$, so that there is no photon associated with $C$. But since both $\xi$ and $C$ are properly normalized $U(1)$ gauge fields, the equation $\xi=-C/3$ cannot be satisfied globally. Namely the flux $dC/3$ is gauge invariant mod $2\pi$. Indeed $B=\diff C/3$ is the $\mathbb Z_3$ gauge field of the $PSU(3)$ gauge bundle, which was advertised above.

Further, to maintain this gauge invariance in the $\theta$-terms, we must replace
\begin{align}
&\diff a_{\ell}\rightarrow \diff a_{\ell}+B\;.
\end{align}
Since they are quantized by $2\pi/3$, this forces a $6\pi$ periodicity in $\theta$-angles. Since the $\mathbb Z_3$ exchange symmetry crucially depended on the $2\pi$-periodicity of $\theta$-angles, the $\mathbb Z_3$ symmetry is explicitly broken by the presence of the  $B$-field. This is the source of the 't Hooft anomaly which we will examine more closely in the next section.
We now obtain the fully gauged action,
\bea
&&S_{\mathrm{gauged}}\nonumber\\
&&=\sum_{\ell=1}^{3}\int_{M_2} \left[-{1\over 2g}\left|(\diff+\im a_\ell+\im \widetilde{A}) \bm{\phi}_\ell\right|^2+{\im \theta_\ell\over 2\pi}(\diff a_\ell+B)\right.\nonumber\\
&&\left.+{\lambda\over 2\pi} \{\overline{\bm{\phi}}_{\ell+1}\cdot (\diff+\im\widetilde{A}) \bm{\phi}_\ell\}\wedge \{\bm{\phi}_{\ell+1}\cdot(\diff+\im \widetilde{A}) \overline{\bm{\phi}}_\ell\}\right]. 
\label{eq:PSU3_gauged}
\eea
We should notice that the $\lambda$-term is invariant under $U(1)$ one-form gauge transformations because of the orthogonality constraint. Performing the path integral,
\be
Z[(A,B)]=\int \Diff a\Diff \overline{\bm{\phi}}\Diff \bm{\phi} \exp(S_{\mathrm{gauged}}), 
\ee
we obtain the partition function $Z[(A,B)]$ under the background $SU(3)/\mathbb{Z}_3$ background gauge field. 

\subsection{$SU(3)/\mathbb{Z}_3$-$\mathbb{Z}_3$ anomaly}

Now we turn the mixed 't~Hooft anomaly between the $SU(3)/\mathbb{Z}_3$ flavor symmetry and the $\mathbb{Z}_3$ permutation symmetry. To see it, we show that the partition function under the $SU(3)/\mathbb{Z}_3$ gauge field, $Z[(A,B)]$, is not invariant under the $\mathbb{Z}_3$ permutation at a $\mathbb{Z}_3$-invariant point $(\theta_1,\theta_2,\theta_3)=(2\pi/3,0,-2\pi/3)$.

In the presence of the $SU(3)/\mathbb{Z}_3$ background gauge field, the constraint on the topological charges becomes
\be
\diff a_1+\diff a_2+\diff a_3+\diff C=0, 
\ee
or, equivalently, 
\be
(\diff a_1+B)+(\diff a_2+B)+(\diff a_3+B)=0. \label{eq:constraint_topological_charges_gauged}
\ee

After gauging $SU(3)/\mathbb{Z}_3$ the action is given by \eqref{eq:PSU3_gauged}, with the above $\theta$-terms. The kinetic and $\lambda$ terms are evidently invariant under  $\mathbb{Z}_3$ permutation $\bm{\phi}_\ell \mapsto \bm{\phi}_{\ell+1}$, so we compute the topological term only. At $(\theta_1,\theta_2,\theta_3)=(2\pi/3,0,-2\pi/3)$, 
\bea
S_{\mathrm{top}}&=&\im \int\left({1\over 3}(\diff a_1+B)-{1\over 3}(\diff a_3+B)\right)\nonumber\\
&\mapsto & \im \int \left({1\over 3}(\diff a_2+B)-{1\over 3}(\diff a_1+B)\right)\nonumber\\
&=&S_{\mathrm{top}}-\im \int (\diff a_1+B). 
\eea
Since $\int \diff a_1\in 2\pi\mathbb{Z}$, this term drops off in the path-integral. However the $B$-term $\int B\in {2\pi\over 3}\mathbb{Z}$ contributes a phase, so we have
\be
Z[(A,B)]\mapsto Z[(A,B)]\exp\left(-\im\int B\right)
\label{eq:anomaly_SU(3)_chain_permutation}
\ee
under $\mathbb{Z}_3$ permutation. 
This is the mixed 't~Hooft anomaly between $SU(3)/\mathbb{Z}_3$ and $\mathbb{Z}_3$, implying the generalization of the Haldane conjecture to $SU(3)$ chains. 
There is no local counter term that can eliminate the generation of the $B$-term under the $\mathbb Z_3$ exchange symmetry. Indeed the only counter-terms allowed are 
\be
\im p \int B
\ee 
where $p\in \mathbb Z\bmod 3$, and these are is invariant under the $\mathbb Z_3$ symmetry. 

By anomaly matching argument, the ground state at the $\mathbb{Z}_3$ invariant point, $(\theta_1,\theta_2,\theta_3)=(2\pi/3,0,-2\pi/3)$, cannot be trivially gapped, i.e., the system must have either 
\begin{itemize}
\item spontaneous symmetry breaking (SSB), 
\item topological order, or 
\item conformal behavior. 
\end{itemize}
In $1+1$ dimension, the intrinsic topological order is ruled out by Ref.~\cite{PhysRevB.83.035107}, so the system must either have the spontaneous symmetry breaking or the conformal behavior in the low-energy limit.  

The same statement is obtained by the Lieb-Schultz-Mattis theorem for the lattice $SU(N)$ chain~\cite{Affleck:1986pq, Lajko:2017wif}, and we here provided the field-theoretic counterpart.

\subsection{$SU(3)/\mathbb{Z}_3$-$\mathsf{C}$ global inconsistency}

Here we will discuss a constraint which arises from the $SU(3)/\mathbb Z_3$ symmetry and the charge conjugation $\mathsf C$. As we discussed for generic values of the $\theta$-angles, the charge conjugation is not a symmetry. We always set one of the $\theta$ angles to be zero without the loss of generality, and here we work with $\theta_1$ and $\theta_3$ angles only. 

As we already discussed, there are three distinct ways that we can define the charge conjugation symmetry, and we labeled them by $\mathsf C_k$, $k=1,2,3$, given by \eqref{eq:charge_conjugation}. The three ways differ by the $\mathbb Z_3$ exchange symmetry, and so when $\mathbb Z_3$ is a symmetry (i.e. when $\theta_1=-\theta_3=2\pi/3 s\;, s=0,1,2$) any one of them can be used. Here we will discuss the values of $(\theta_1,\theta_3)$, where $\mathbb Z_3$ permutation symmetry is not necessarily present but there is a sensible $\mathsf C_k$-symmetry for some $k=1,2,3$. 
As we have discussed around Eqs.~(\ref{eq:charge_conjugation_2}-\ref{eq:charge_conjugation_3}), we get that, under $\mathsf C_k$, the $\theta$-angles are mapped as 
\begin{align}
&\mathsf C_1: (\theta_1,\theta_3)\rightarrow (\theta_3-\theta_1,\theta_3)\\
&\mathsf C_2: (\theta_1,\theta_3)\rightarrow (-\theta_3,-\theta_1,)\\
&\mathsf C_3: (\theta_1,\theta_3)\rightarrow (\theta_1,\theta_1-\theta_3,)
\end{align}
Therefore for $\mathsf C_1$ to be a symmetry we must have $\theta_3=2\theta_1\bmod 2\pi$, for $\mathsf C_2$ we must have $\theta_3=-\theta_1\bmod 2\pi$ and for $\mathsf C_3$, $\theta_1=2\theta_3\bmod 2\pi$. 

Since $\mathsf{C}_k$ invariance is trivially true for kinetic and $\lambda$ terms even after gauging $SU(3)/\mathbb{Z}_3$, all we have to discuss is the effect of topological theta terms.  Now let us set $\theta=\theta_1=-\theta_3+\alpha$. If $\alpha=0\bmod 2\pi$ the  $\mathsf{C}_2$ is the symmetry. Upon gauging the $SU(3)/\mathbb Z_3$ symmetry we have that the $\theta$-terms become
\be
S_{\mathrm{top}}={\im \over 2\pi}\int \left\{\theta(\diff a_1+B)+(\alpha-\theta)(\diff a_3 +B)\right\}
\ee
When $\alpha=0$, the $\mathsf C_2$ transformation is clearly a symmetry, if we define that $\mathsf C_2: B\mapsto -B$. However if we now dial $\alpha=2\pi k, k\in \mathbb Z$, we will get that under the transformation
\bea
S_{\mathrm{top}}&=&{\im \over 2\pi}\int \left\{\theta(\diff a_1+B)+(2\pi k-\theta)(\diff a_3 +B)\right\}\nonumber\\
&\mapsto& {\im\over 2\pi}  \int \left\{-\theta(\diff a_3+B)-(2\pi k-\theta)(\diff a_1 +B)\right\}\nonumber\\
&=&S_{\mathrm{top}}-2\im k \int B\quad \mbox{mod}\quad 2\pi.  
\eea
Therefore, the partition function $Z[(A,B)]$ at $(\theta_1,\theta_3)=(\theta,-\theta+2k\pi)$ changes under $\mathsf{C}_2$ as
\be
\mathsf{C}_2: Z[(A,B)]\mapsto Z[(A,B)]\exp\left(-2\im k\int B\right).
\ee 
This, however, is not necessarily a 't~Hooft anomaly, because when gauging $SU(3)/\mathbb{Z}_3$, we have a freedom to add a local gauge-invariant term of the background field. We can define 
\be
Z_n[(A,B)]=Z[(A,B)]\exp\left(\im n \int B\right), 
\ee
where $n$ is called the discrete theta parameter, and $n\in\mathbb{Z}_3$. This gauged partition functions obeys 
\be
\mathsf{C}_2:Z_n[(A,B)]\mapsto Z_n[(A,B)]\exp\left(-2\im (n+k)\int B\right), 
\ee
and thus it becomes $\mathsf{C}_2$ invariant if 
\be
n=2k \;\bmod 3. 
\ee
In this manner, we can always write a local counter term that restores the symmetry on every $\mathsf{C}_2$-invariant line. Therefore, there is no 't Hooft anomaly between $SU(3)/\mathbb{Z}_3$ and $\mathsf{C}_2$. 

However, the local counter term $\int B$ does not allow continuous parameters for its coefficient $n$ in order to satisfy the $U(1)$ one-form gauge invariance. 
In such a case, we can apply the global inconsistency condition: When interpolating adiabatically from $\alpha=2\pi k_1$ to $\alpha=2\pi k_2$, if the local counter terms of $SU(3)/\mathbb{Z}_3$ added for $\mathsf{C}_2$-invariant gauged partition functions at those points are different, then it is called a global inconsistency~\cite{Gaiotto:2017yup,Tanizaki:2017bam, Kikuchi:2017pcp} or a secondary anomaly~\cite{Komargodski:2017dmc}. The conjectured matching condition~\cite{Tanizaki:2017bam, Kikuchi:2017pcp} states that 
\begin{itemize}
\item both are trivially gapped, but they are distinct as the symmetry-protected topological (SPT) phases protected by $SU(3)/\mathbb{Z}_3$, or  
\item one of them has nontrivial ground states as in the case of 't~Hooft anomaly matching. 
\end{itemize}
This consequence obtained by global inconsistency does not have the Lieb-Schultz-Mattis type counterpart.

\begin{figure}[t] 
   \centering
   \includegraphics[width=3.5in]{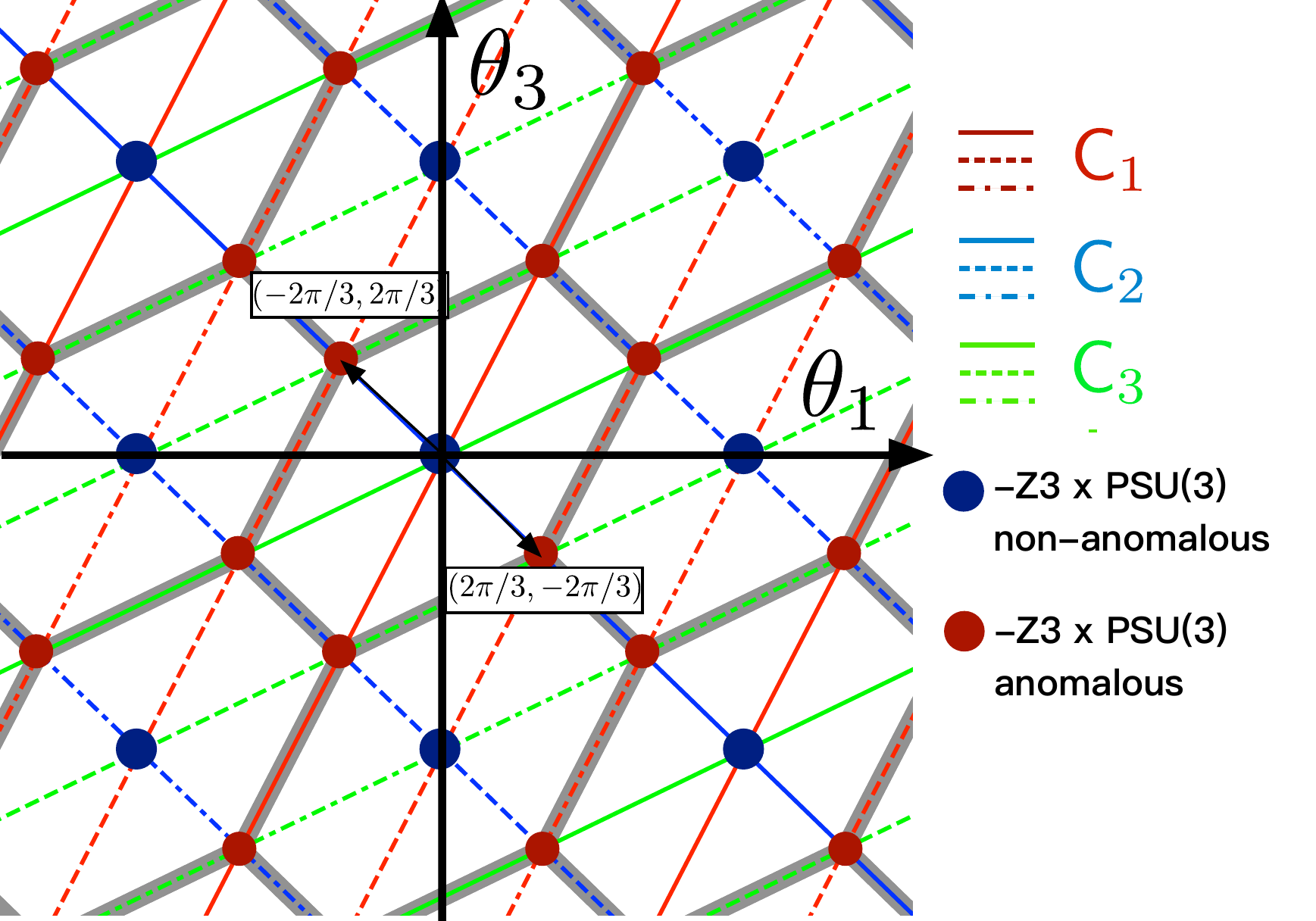} 
   \caption{(color online) The plot of the phase diagram of $SU(3)/[U(1)\times U(1)]$ nonlinear sigma model. 
The $\mathbb{Z}_3$-symmetric points are shown with blobs, and the blue blobs show that there is no 't~Hooft anomaly for $PSU(3)\times \mathbb{Z}_3$ while the red ones show that there is the 't~Hooft anomaly. 
The global inconsistency lines in the $(\theta_1,\theta_3)$ plane for the symmetries $\mathsf C_1,\mathsf C_2,\mathsf C_3$ are sketched as red, blue and green lines. The solid, dashed and dot-dashed lines indicate that different counter-terms are needed to restore the corresponding $\mathsf C$-symmetries when the $SU(3)/\mathbb Z_3$ symmetry is gauged, indicating that there is a global inconsistency between different-type lines (e.g. between solid and dashed). The inconsistency can be saturated either by at least one of these lines having a non-trivial ground state, or that they are separated by a phase transition. }
   \label{fig:inconsistency}
\end{figure}

We can obtain the same conclusion for $\mathsf C_1$ and $\mathsf C_3$ by setting $\theta_3=2\theta+\alpha,\theta_1=\theta$ and $\theta_1=2\theta+\alpha,\theta_3=\theta$, and obtain that there is a global inconsistency between $\alpha=0,2\pi,4\pi \bmod 6\pi$ lines. This situations are sketched in the Fig.~\ref{fig:inconsistency}.

\subsection{Anomaly involving the $\mathbb Z_3\times\mathbb Z_3\subset PSU(3)$ subgroup}

So far we have discussed gauging the full $SU(3)/\mathbb Z_3=PSU(3)$ flavor symmetry and established that there exists an anomaly between it and the $\mathbb Z_3$ cyclic permutation symmetry at the appropriate points in the $(\theta_1,\theta_3)$ phase diagram. Moreover we have also seen that when $(\theta_1,\theta_3)$ are chosen such that the $\mathbb Z_3$ cyclic permutation symmetry is broken, but that a form of charge conjugation is preserved, there exists a global inconsistency between certain regions of the phase diagram, constraining the system significantly more than the LSM theorem.

Here we wish to make a remark that a lot of our discussion applies even to the case of the subgroup $\mathbb Z_3\times\mathbb Z_3\subset PSU(3)$. As we shall see there is a mixed 't~Hooft anomaly between this $\mathbb Z_3\times \mathbb Z_3\subset PSU(3)$ flavor symmetry and the $\mathbb Z_3$ cyclic permutation symmetry. Moreover this anomaly immediately implies the anomaly involving the full $PSU(3)$, so it is more general. We will also use the opportunity to complement the discussion so far, by introducing a slightly different, but equivalent, perspective on the anomaly.

The practical consequence of using the $\mathbb Z_3\times\mathbb Z_3$ subgroup of $PSU(3)$ is that the system may be allowed to break the spin-$PSU(3)$ symmetry all the way down to $\mathbb Z_3\times\mathbb Z_3$, keeping the nontrivial constraints of the anomaly. Further such deformations of the theory will have a richer structure, as spontaneous breaking of the discrete symmetries are not forbidden by the Mermin-Wagner-Coleman theorem. Spontaneous  breaking of the discrete  symmetries also gives rise to domain walls. These too will be constrained by the anomaly as we shall see, and are interesting in their own right.

Before we can argue that there is an anomaly involving the subgroup $\mathbb Z_3\times\mathbb Z_3\subset PSU(3)$, let us first discuss how this group acts on the fields of the theory. To that end consider the lift of the $PSU(3)$ to $SU(3)$. $SU(3)$ group contains two matrices
\begin{align}
&M_C=\begin{pmatrix}
1&0&0\\
0&\mathrm{e}^{\im2\pi/3}&0\\
0&0&\mathrm{e}^{\im4\pi/3}
\end{pmatrix},\quad M_S=\begin{pmatrix}
0&1&0\\
0&0&1\\
1&0&0
\end{pmatrix},
\label{eq:shift_clock_N=3}
\end{align} 
which are dubbed \emph{clock} and \emph{shift} matrices. They satisfy the algebra
\be
M_CM_S=e^{\frac{2\pi \im}{3}}M_S M_C\;.
\ee
i.e. they differ by a center element of $SU(3)$. Further a homomorphism $H:SU(3)\rightarrow SU(3)/\mathbb Z_3$ maps has the center as the kernel, so the two group elements $H(M_C),H(M_S)\in SU(3)/\mathbb Z_3$ commute. Since they also have the property that $M_C^3=M_S^3=\bm{1}_3$, they generate a group $\mathbb Z_3\times\mathbb Z_3\subset SU(3)/\mathbb Z_3$. 

Now we want to promote the global symmetry $\mathbb Z_3\times\mathbb Z_3$ to a gauge symmetry, i.e. we wish to promote the $U(1)^2$ gauge bundle to $\mathbb Z_3^2\times U(1)^2$. Notice however that the $\mathbb Z_3\times\mathbb Z_3$ (just like $SU(3)/\mathbb Z_3$) acts projectively on the fields $\bm \phi_\ell$. 

Before continuing let us first gauge the $PSU(3)$ global symmetry. Before we do, recall that the $PSU(3)$ gauge bundle contains a topological class $H^2(M_2,\mathbb Z_3)$ with $\mathbb Z_3$ coefficients. In fact we have already seen the representative of this topological class. It is the $B$-field used extensively in the discussion so far. This topological class is an obstruction to the lifting of the $PSU(3)$ bundle to $SU(3)$ bundle. To see this let $T_{ij},T_{jk},T_{ki}$ be the transition functions between the three local coordinate charts $U_i,U_j,U_k$ of $M_2$. The cocycle condition on the triple overlap $U_i\cap U_j\cap U_k$ demands
\be
T_{ij}T_{jk}T_{ki}=\mathbb I\;.
\ee
Now let $\tilde T_{ij},\tilde T_{jk}, \tilde T_{ki}$ be the lifts of $T_{ij},T_{jk},T_{ki}$ from $PSU(3)$ to $SU(3)$. The cocycle condition translates into
\be
\tilde T_{ij}\tilde T_{jk}\tilde T_{ki}=z\mathbb I\;, z\in \mathbb Z_3\;.
\ee
In other words the obstructions which cause the cocycle condition of $PSU(3)$ bundle can fail to satisfy the cocycle condition of the $SU(3)$ bundle are classified by the center element $z\in \mathbb Z_3$ of $SU(3)$. 

Now consider the cocycle condition with a nontrivial element $z\in \mathbb Z_3$ and how it affects the  fields $\bm \phi_\ell$ of our theory. The $PSU(3)$ transition functions act as $SU(3)$ matrices on them, so in order to have $\bm \phi_\ell$ to be well defined in the triple intersection, we must compensate the change of phase $z\in \mathbb Z_3$ by an equivalent change of phase in the $U(1)^2$ transition functions. Namely we must have that
\be
\exp\left(\im \varphi_{\ell}^{ij}+\im \varphi_{\ell}^{jk}+\im \varphi_{\ell}^{ki}\right)=\bar z\;.
\ee
where $t_{ij}^\ell=\mathrm{e}^{i\varphi_\ell^{ij}}$ is the transition function for the $U(1)$ gauge bundle acting on the field $\bm\phi_\ell$ (notice that we have a constraint $\sum_{\ell=1,2,3}\varphi_\ell^{ij}=0$). In turn this means that the gauge fields associated with the $U(1)$ gauge bundles are no longer properly quantized, and their fluxes are no longer quantized in multiples of $2\pi$. However their deviation from the quantization is correlated with the value of $B\in H^2(M_2,\mathbb Z_3)$. In other words
\be
\int F_\ell = \int B \bmod 2\pi\;.
\ee
The failure for the abelian fluxes to be properly qunatized is reflected in the loss of the $\mathbb Z_3$ cyclic permutation symmetry, exactly by a value of the $B\in H^2(M_2, \mathbb Z_3)$.

Now we see that nothing will change when we break $PSU(3)$ down to $\mathbb Z_3\times\mathbb Z_3$, defined as above. The $\mathbb Z_3\times\mathbb Z_3$ still acts projectively on the $\bm \phi_\ell$ fields, and the $\mathbb Z_3\times \mathbb Z_3$ bundle is classified by the obstruction to the lifts by $\mathbb Z_3$ central extensions, which we will still call $B$.

Let us see the same thing by another way. We can think of gauging the $\mathbb Z_3\times \mathbb Z_3$ as putting the twisted boundary condition on the 2D manifold. We do this by using the clock and shift matrices $M_C$ and $M_S$ defined above, and twisting the $\phi$-fields with the clock and shift matrices. In other words let us take that
\begin{align}
&\phi_{ \ell,a}(L,t)=\mathrm{e}^{\im\varphi_\ell(t)}{(M_C)_a}^b\phi_{\ell,b}(0,t)\;,\\
&\phi_{\ell,a}(x,\beta)=\mathrm{e}^{\im\tilde \varphi_\ell(x)}{(M_S)_{a}}^b\phi_{\ell,b}(x,0)\\
&a_\ell(L,t)=a_\ell(0,t)-\diff\varphi_\ell(t)\\
&a_\ell(x,\beta)=a_\ell(x,0)-\diff\tilde\varphi_\ell(x)\;.
\end{align}
where $\varphi_\ell$ and $\tilde\varphi_\ell$ at the moment undetermined phases, with the constraint that $\varphi_3=-\varphi_1-\varphi_2$ and $\tilde \varphi_3=-\tilde\varphi_1-\tilde\varphi_2$. By setting $t=\beta$ in the first equation and $x=L$ in the second, we have
\bea
\phi_{\ell,a}(L,\beta)&=&\mathrm{e}^{\im\varphi_\ell(\beta)}{(M_C)_a}^b\phi_{\ell,b}(0,\beta)\nonumber\\
&=&\mathrm{e}^{\im\varphi_\ell(\beta)+\im\tilde \varphi_\ell(0)}{(M_CM_S)_a}^b\phi_{\ell,b}(0,0),\\
\phi_{\ell,a}(L,\beta)&=&\mathrm{e}^{\im\tilde \varphi_\ell(L)}{(M_S)_{a}}^b\phi_{\ell,b}(L,0)\nonumber\\
&=&\mathrm{e}^{\im\tilde\varphi_\ell(L)+\im\varphi_\ell(0)}{(M_SM_C)_a}^b\phi_{\ell,b}(0,0). 
\eea
Since the LHS of the two lines above are equal, we must have that
\be
\varphi_\ell(\beta)+\tilde\varphi_\ell(0)=\tilde\varphi_\ell(L)+\varphi_\ell(0)+\frac{2\pi}{3}\bmod 2\pi\;.
\ee
from which it follows that
\bea
\int \diff a_\ell&=&[\varphi_\ell(\beta)-\varphi_\ell(0)]-[\tilde\varphi_\ell(L)-\tilde\varphi_\ell(0)]\nonumber\\
&=&\frac{2\pi}{3}\bmod 2\pi\;.
\eea

The deviation from the $2\pi$ quantization can be seen as the cup product between the $\mathbb Z_3$ gauge fields for the two generators of $\mathbb Z_3$, which means that we can identify 
\be
B=\frac{3}{2\pi} A^1\wedge A^2\;,
\ee
where $A^1$ and $A^2$ are the $\mathbb Z_3$ gauge fields for the two generators of  $\mathbb Z_3\times \mathbb Z_3$. Indeed if we think of $A^1$ and $A^2$ as embedded in the $U(1)$ gauge group, the above term is gauge invariant under $A^{1,2}\rightarrow A^{1,2}+\diff \varphi^{1,2}$. 

\subsection{The Phase structure}\label{sec:phase_diagram_SU(3)}

\begin{figure*}[t]
\centering
	\includegraphics[width=0.4\textwidth]{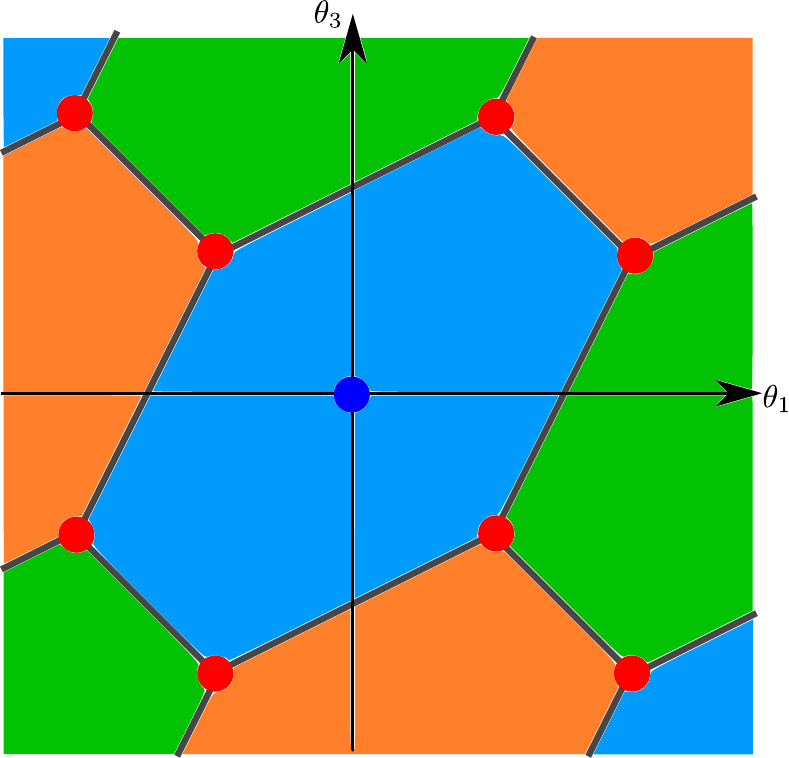}\hspace{1.em}
	\includegraphics[width=0.4\textwidth]{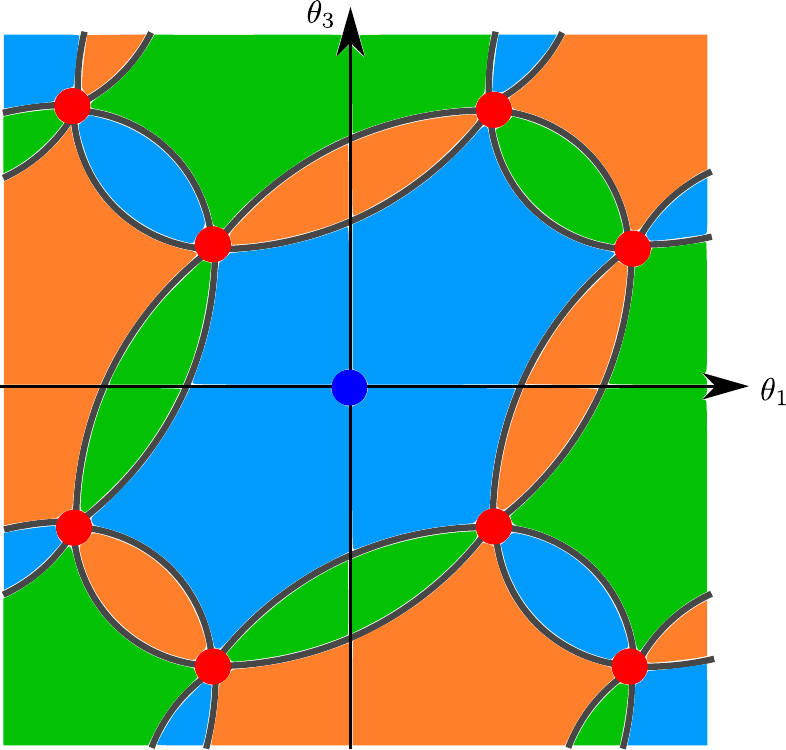}
\caption{ (color online) 
{ Possible scenarios consistent with global inconsistency. The red blobs are $\mathbb{Z}_3$ symmetric points with $PSU(3)\times \mathbb{Z}_3$ 't~Hooft anomaly, and the origin (blue blob) is the $\mathbb{Z}_3$ symmetric point without anomaly. Blank regions painted with different colors (light blue, orange, green) all correspond to trivially gapped phases, but they are different as SPT phases protected by $PSU(3)$ symmetry.  (Left) The global inconsistency is matched by the spontaneous breaking of $\mathsf{C}$ on thick gray lines. (Right) The global inconsistency is matched by the phase transitions lines (gray curves) separating distinct $\mathsf{C}$-symmetric trivial vacua.} 
}
\label{fig:phase_diagrams}
\end{figure*}

In this section, we discuss details how the 't Hooft anomaly and the global inconsistency constrain the phase diagram of the $SU(3)/[U(1)\times U(1)]$ sigma model. 
Figure~\ref{fig:inconsistency} shows one of the possible phase diagrams consistent with the matching condition when the nonvanishing mass gap is assumed everywhere. The red and blue blobs indicate the $\mathbb{Z}_3$-invariant points, red being the points with a 't~Hooft anomaly. 
The thin lines (blue, green, red and solid, dashed dotted) indicate that the system has a charge-conjugation symmetry (i.e. either $\mathsf C_1,\mathsf C_2,\mathsf C_3$-symmetry. 
The thick gray lines show the first-order phase transitions, on which the charge conjugation is spontaneously broken. 
While this is a minimal way to saturate the global inconsistency, it is not the only way. Indeed we will soon discuss a more exotic scenario of the phase diagram (see Fig.~~\ref{fig:phase_diagrams}). Let us first discuss the standard scenario depicted in Fig.~\ref{fig:inconsistency}.

The red $\mathbb{Z}_3$-invariant  points have a mixed 't Hooft anomaly between $SU(3)/\mathbb{Z}_3$ and $\mathbb{Z}_3$ permutation, requiring matching with a nontrivial vacuum. 
The $PSU(3)$ symmetry cannot be spontaneously broken due to the Coleman-Mermin-Wagner theorem~\cite{Coleman:1973ci, mermin1966absence}, the possible choice of the low-energy theory is SSB of $\mathbb{Z}_3$ or the conformal field theory (CFT). 
 The $SU(3)$ spin chain discussed in Ref.~\cite{Bykov:2011ai} has the trimerized phase and the $\mathbb{Z}_3$ symmetry is spontaneously broken. 
Also, the strong-coupling analysis of (\ref{eq:lattice_Hamiltonian}), given in Ref.~\cite{Lajko:2017wif}, shows that the anomaly is matched by SSB of $\mathbb{Z}_3$ permutation (This should be compared with a free photon theory of the linear version of the sigma model, discussed in Sec.~\ref{sec:large_mass}). 
Ref.~\cite{Lajko:2017wif} also performed Monte Carlo simulation at the imaginary $\theta$ angles, and extrapolate the mass gap with the ansatz $(c_1+c_2\theta^2)/(1+c_3\theta^2)$ indicated by Ref.~\cite{PhysRevD.77.056008}.  It claims that the gappless excitation appears for $g<g_c\simeq 2.55$~\cite{Lajko:2017wif}, and then the 't~Hooft anomaly is matched by some CFT if this were really the case.
To get more conclusive remark, it would be quite appealing if the result with the real theta angles is directly obtained via the lattice dualization \cite{Gattringer:2015baa,Bruckmann:2015sua,Bruckmann:2015hua,Sulejmanpasic2017}. 

The nature of the conformal field theory is one of the open questions and still under debate.  Numerical results of exact diagonalization~\cite{PhysRevB.93.155134} suggest that it is $SU(3)_2$ Wess-Zumino-Witten (WZW) model for $p=2$, but the authors also mention that the crossover towards the gapped or $SU(3)_1$ WZW phase may occur as the system size becomes larger.  Ref.~\cite{Lajko:2017wif} also argues that it is $SU(3)_1$ WZW model for all $p\not=0$ mod $3$. 
To constrain the possible CFT from anomaly matching, we have to compute the 't~Hooft anomaly of the $SU(3)$ WZW model. 
We will give a detailed analysis on the anomaly of $SU(N)$ WZW models in Sec.~\ref{sec:WZWmodel}, and we claim that the anomaly matching condition also admits the crossover from the $SU(3)_2$ WZW model toward the $SU(3)_1$ WZW model.

Now consider deviations from the $\mathbb Z_3$ symmetric points. In Ref.~\cite{Lajko:2017wif} it was argued that the deformation along one of the the $\mathsf C$-invariant lines, depicted in Fig.~\ref{fig:inconsistency}, will result in a flow away from a CFT, because, in the absence of the $\mathbb Z_3$ symmetry, $SU(3)_1$ WZW theory has relevant perturbations driving it away from conformality. Using the strong-coupling analysis, they have argued that the lines connecting the red $\mathbb Z_3$ invariant points along $\mathsf C$-invariant lines are phase separating lines which break the corresponding $\mathsf C$-symmetry spontaneously. These phase-separating lines are depicted as thick gray lines on Fig.~\ref{fig:inconsistency}.

We now argue that the global inconsistency between $\mathsf{C}$-invariant lines makes this picture robust. To argue this, we will assume that the system at $\theta_1=\theta_3=0$ has a trivial mass gap, which is consistent with some previous studies~\cite{Bykov:2011ai, Lajko:2017wif, PhysRevB.75.060401, PhysRevB.75.184441, PhysRevB.80.180420} although some others show no indication of the mass gap~\cite{PhysRevB.93.155134}. 
We believe that this is a reasonable assumption since the nonlinear sigma model is asymptotically free and has no imaginary terms in the action at $\theta=0$ (up to irrelevant and non-universal $\lambda$-terms), which are typically trivially gapped in $1+1D$. 

When the trivial mass gap is assumed at blue points of Fig.~\ref{fig:inconsistency}, we must have that as we move from such a point all the way to the thick gray lines, we must either encounter a phase transition on the way or have a non-trivial ground state on the gray lines, matched by breaking $PSU(N)$-symmetry, $\mathsf C$-symmetry or a CFT. 
Mermin-Wagner-Coleman theorem prevents the first, while there is no obvious candidate for the last option, leaving the breaking of the $\mathsf C$-symmetry as an obvious choice. This is consistent with the picture of Ref.~\cite{Lajko:2017wif}. We argue that a similar discussion on the phase diagram using the global inconsistency can be found in previous study of the 4d gauge theories~\cite{Gaiotto:2017yup, Tanizaki:2017bam} and also of the quantum mechanics on a circle~\cite{Kikuchi:2017pcp}.

We should make a  comment that while this is a natural way to saturate the global inconsistency it is not the only way. We could imagine that the thick gray lines of Fig.~\ref{fig:inconsistency} splits into two phase-separating lines as one goes from one nontrivial $\mathbb Z_3$ (red points) to another, causing the vacuum on the $\mathsf C$-invariant line to be trivial. 
{ We illustrate this behavior in Fig.~\ref{fig:phase_diagrams}. The left figure shows the conventional scenario explained above, while the right one gives an exotic one. }
This scenario however seems contrived to us for the model at hand, but it should be possible to achieve by some deformations of the linear sigma model where more tunable parameters are allowed.

When the charge-conjugation $\mathsf{C}$ is spontaneously broken, we can consider the domain wall connecting two vacua. 
Since the partition functions of these vacua are different by $\im \int B$ under the $SU(3)/\mathbb{Z}_3$ background gauge field, the difference must be compensated by a nontrivial domain wall~\cite{Anber:2015kea, Sulejmanpasic:2016uwq, Komargodski:2017smk}. In fact in this case the domain wall is an $SU(3)$-spin triplet. 
In other words, these two vacua are trivially gapped\footnote{Let us clarify terminologies to avoid possible confusions. Phase is called trivially gapped if it has a mass gap with unique ground state on any closed spatial manifolds (i.e., no spontaneous symmetry breaking nor topological order). Trivially gapped states can be nontrivial as symmetry-protected topological phases, which detects the degeneracy of boundary states on open spatial manifolds. } but distinct as SPT phases protected by $SU(3)/\mathbb{Z}_3$, so the fundamental representation of $SU(3)$ is excited on the domain wall without any energy cost. 
In Figs.~\ref{fig:phase_diagrams}, we paint different colors for distinct SPT phases protected by $PSU(3)$ symmetry. Phase transition lines required by global inconsistency must exist to describe these different SPT phases.

Finally we recall that the anomalies and global inconsistencies remain even when the global $PSU(3)$ symmetry group is reduced down to $\mathbb Z_3\times \mathbb Z_3$. The theory, however, will not show conformal behavior at anomalous $\mathbb Z_3$ cyclic permutation symmetry invariant points, but will instead be saturated by a breaking either the $\mathbb Z_3\times \mathbb Z_3$ (a N\'eel phase) or the $\mathbb Z_3$ cyclic permutation symmetry (the VBS phase). What makes this scenario interesting is that the system will support domain walls, all of which will have anomaly inflow and therefore carry nontrivial (i.e. degenerate) particle excitations (note that domain walls are particles in 1+1D).

\section{Generalization to $SU(N)/U(1)^{N-1}$ nonlinear sigma model}\label{sec:generalization_SU(N)}

In this section, we will show that the whole analysis on anomalies and global inconsistencies in the previous section \ref{sec:anomaly} can be extended to the $SU(N)/U(1)^{N-1}$ nonlinear sigma model. We take the same form of the Lagrangian 
\bea
S&=&\sum_{\ell=1}^{N}\int_{M_2} \left[-{1\over 2g}\left|(\diff+\im a_\ell) \bm{\phi}_\ell\right|^2 +{\im \theta_\ell\over 2\pi}\diff a_\ell\right.\nonumber\\
&&\left.+{\lambda\over 2\pi} (\overline{\bm{\phi}}_{\ell+1}\cdot \diff \bm{\phi}_\ell)\wedge (\bm{\phi}_{\ell+1}\cdot \diff \overline{\bm{\phi}}_\ell)\right],
\eea
where $\bm{\phi}_\ell:M_2\to \mathbb{C}^N$ satisfies the  constraint, 
\bea
&&\overline{\bm{\phi}}_\ell\cdot \bm{\phi}_{\ell'}=\delta_{\ell \ell'}, \\
&&\ve_{f_1 f_2\ldots f_N}\phi_{f_1 1}\phi_{f_2 2}\cdots \phi_{f_N N}=1. \label{eq:constraint_det_N}
\eea
The equation of motion of $a_\ell$ gives the constraint on the gauge field, 
\be
\sum_{\ell=1}^{N}a_\ell=0, 
\label{eq:constraint_gauge_field_N}
\ee
and this is necessary for gauge invariance of (\ref{eq:constraint_det_N}).  
Since the sum of topological charges vanishes, we can put one of the theta parameters equal to $0$. We take the convention $\theta_N=0$. The kinetic and $\lambda$ terms are invariant under the following symmetries, 
\begin{itemize}
\item $SU(N)/\mathbb{Z}_N$ flavor symmetry. 
\item Time reversal $\mathsf{T}: \bm\phi_\ell(x,t)\mapsto \overline{\bm\phi}_\ell(x,-t)$. 
\item $\mathbb{Z}_N$ permutation symmetry, $\bm\phi_\ell\mapsto \mathrm{e}^{2\pi\im (\ell+1)/N}\bm\phi_{\ell+1}$ and $a_\ell\mapsto a_{\ell+1}$. 
\item Charge conjugations, $\mathsf{C}_k$: $\bm{\phi}_i\mapsto (-1)^N\overline{\bm{\phi}}_{-i-k}$ and $a_\ell\mapsto-a_{-\ell-k}$. 
\end{itemize}
The first two symmetries $SU(N)/\mathbb{Z}_N$ and $\mathsf{T}$ are symmetries at any theta angles, while the last two, $\mathbb{Z}_N$ permutation and $\mathsf{C}_k$, are symmetries only for special theta angles. 

For $\mathbb{Z}_N$ permutation and charge conjugations, an appropriate phase factors must be multiplied so that those transformations become consistent with the constraint on the determinant, (\ref{eq:constraint_det_N}). Although it does not affect the following anomaly and global inconsistency argument\footnote{The emergent anomaly for $\mathbb{C}P^1$ model argued in Ref.~\cite{Metlitski:2017fmd} is related to this extra phase factor (see below Eq.(8) of the reference), but we will not discuss it here. } , let us make a brief comment for clarity. We perform the $\mathbb{Z}_N$ permutation to the left hand side of (\ref{eq:constraint_det_N}), then 
\bea
&&\ve_{f_1 f_2\ldots f_N}\phi_{f_1 1}\phi_{f_2 2}\cdots \phi_{f_N N}\nonumber\\
&\mapsto&\ve_{f_1 f_2\ldots f_N}\phi_{f_1 2}\phi_{f_2 3}\cdots \phi_{f_N 1}\mathrm{e}^{{2\pi\im\over N}{N(N+1)\over 2}}\nonumber\\
&=&(-1)^{N-1}\mathrm{e}^{\pi\im (N+1)}\ve_{f'_1 f'_2\ldots f'_N}\phi_{f'_1 1}\phi_{f'_2 2}\cdots \phi_{f'_N N}. \quad
\eea
The negative sign coming out of the epsilon tensor for even $N$ is exactly canceled by the additional phase factor, and the transformation is consistent with the constraint (\ref{eq:constraint_det_N}). 
This suggests that we do not need such factors for odd $N$ as in the case of $N=3$. Indeed, we can eliminate those factors by $U(1)^{N-1}$ gauge transformations for odd $N$ but it is impossible for even $N$. The easiest way to understand it is to perform the $\mathbb{Z}_N$ permutation $N$ times, then $\bm{\phi}_\ell\mapsto (-1)^{N-1}\bm{\phi}_{\ell}$. For even $N$, the $\mathbb{Z}_N$ permutation acts projectively on $\bm\phi$ fields\footnote{For $N=2$, the model is the familiar $\mathbb{C}P^1$ nonlinear sigma model. There, the charge conjugation is defined as $\bm\phi_1\mapsto \im \sigma^y \overline{\bm\phi}_1$ for consistency with the spin $SU(2)$ rotation. Doing this charge conjugation twice, we get $\bm\phi_1\mapsto -\bm\phi_1$. Here, we have argued that the same thing is true for larger even $N$. We should still call the global symmetry as $\mathbb{Z}_N$, because such phases does not appear on gauge-invariant operators. }. 

\subsection{Permutation symmetry, $\mathbb{Z}_n$ subgroup, and $SU(N)/\mathbb{Z}_N$-$\mathbb{Z}_n$ anomaly}

Let us first consider the $\mathbb{Z}_N$ permutation. Further, let $n$ be a divisor of $N$, so we can consider a subgroup $\mathbb{Z}_n\subset \mathbb Z_N$ that maps $a_\ell\mapsto a_{\ell+N/n}$. The change of the topological theta term is given by 
\be
\Delta S_{\mathrm{top}}=\im \sum_{\ell=1}^{N}{\theta_{\ell-N/n}-\theta_\ell\over 2\pi}\int \diff a_\ell. 
\ee
In order for $\Delta S_{\mathrm{top}}=0$ mod $2\pi$ for arbitrary topological charges, we find the condition, 
\be
\theta_{\ell+N/n}=\theta_\ell+\alpha,\quad \mbox{mod} \quad2\pi,
\ee
for some constant $\alpha$ because of (\ref{eq:constraint_gauge_field_N}). Repeating this transformation $n$ times, we obtain that $n\alpha=0$ mod $2\pi$, and thus 
\be
\alpha={2\pi p\over n}
\ee
for some $p=0,1,\ldots, n-1$. In addition, we still have $(n-1)$ free parameters $\theta_1,\ldots,\theta_{N/n-1}$, so $\mathbb{Z}_n$-invariant points form $(N/n-1)$-dimensional planes. 
In particular the $\mathbb{Z}_N$-symmetric points are given by 
\be
\theta_\ell={2\pi p \ell\over N}\;\, \mbox{mod}\, 2\pi,
\ee 
for some $p=0, 1,\ldots, N-1$. 

We can discuss the mixed 't Hooft anomaly between $SU(N)/\mathbb{Z}_N$ and $\mathbb{Z}_n$ permutation, where $n$ is a divisor of $N$. By gauging $SU(N)/\mathbb{Z}_N$, we introduce the $SU(N)$ one-form gauge field $A$ and $\mathbb{Z}_N$ two-form gauge field $B$. The important thing is that the $U(1)$ field strength, $\diff a_\ell$, is no longer gauge invariant under a one-form $U(1)$ gauge symmetry, and it must be replaced by $\diff a_\ell+B$. As a consequence, the constraint on the field-strength becomes 
\be
\sum_{\ell=1}^{N}(\diff a_\ell+B)=0. 
\ee
Now, let us compute the effect of $\mathbb{Z}_n$ permutation under these background gauge fields $(A,B)$. As an example of $\mathbb{Z}_n$-invariant plane, we take 
\be
\theta_{i+j N/n}=\theta_i+{2\pi p\over n}j
\ee
for $i=1,\ldots,N/n$ and $j=0, 1,\ldots, n-1$, $\theta_{N/n}$ is fixed so that $\theta_N=0$. 
Under the $\mathbb{Z}_n$ permutation, the change of the topological action at this point does not vanish mod $2\pi$, but it becomes 
\be
\Delta S_{\mathrm{top}}=-\im{N\over n}p\int B, \quad \mbox{mod}\quad 2\pi. 
\ee
Since $\int B\in {2\pi\over N}\mathbb{Z}$, this gives the nontrivial phase to the change of the partition function $Z[(A,B)]$. This is the $SU(N)/\mathbb{Z}_N$-$\mathbb{Z}_n$ 't Hooft anomaly. 

\subsection{$SU(N)/\mathbb{Z}_N$-$\mathsf{C}$ anomaly and global inconsistency}
Next, we consider the charge conjugation symmetry. $\mathsf{C}_k$ changes the topological action by 
\be
\Delta S_{\mathrm{top}}=-\im \sum_{\ell=1}^{N}{\theta_{-\ell-k}+\theta_\ell\over 2\pi}\int \diff a_i. 
\ee
In order for $\Delta S_{\mathrm{top}}=0$ mod $2\pi\im$, we get 
\be
\theta_{-\ell-k}+\theta_\ell=\beta\quad \mbox{mod}\quad 2\pi,
\ee
for some constant $\beta$. 

Gauging $SU(N)/\mathbb{Z}_N$, the above change of the topological term is replaced by 
\be
\Delta S_{\mathrm{top}}=-\im \sum_{\ell=1}^{N}{\theta_{-\ell-k}+\theta_\ell\over 2\pi}\int (\diff a_\ell+B). 
\ee
The $\mathsf{C}_k$ invariance without background $B$ only requires $\theta_{-\ell-k}+\theta_\ell=\beta$ mod $2\pi$ because $\int \diff a_\ell \in 2\pi \mathbb{Z}$, but this is not true with $B$ since $\int B\in {2\pi\over N}\mathbb{Z}$. This derives the $SU(N)/\mathbb{Z}_N$-$\mathsf{C}_k$ mixed 't~Hooft anomaly or global inconsistency depending on whether the anomaly can be canceled by the local counterterm with the discrete lavel $\im n\int B$. 

\subsection{Anomaliey and inconsistency involving $\mathbb Z_N\times\mathbb Z_N\subset SU(N)/\mathbb Z_N$}

Recall that much of the discussion of the anomalies in the spin systems having a $SU(3)/\mathbb Z_3$ global flavor symmetry remained even if only $\mathbb Z_3\times\mathbb Z_3\subset SU(3)/\mathbb Z_3$ was preserved. Generalization of this argument to linearized $SU(N)/U(1)^{N-1}$ models is straightforward. 

The relevant $\mathbb Z_N\times\mathbb Z_N$ symmetry can be seen being generated by $SU(N)$ matrices\footnote{Note that the pre-factor was chosen so that the determinant is unity.}
\begin{align}
&M_C=\omega^{\frac{N-1}{2}}\begin{pmatrix}
1&0&\dots&\dots&0\\
0&\omega&0&\dots&\vdots\\
\vdots&0&\omega^2& &\vdots\\
\vdots&\dots&\dots&\ddots&\vdots\\
0&\dots&\dots&\dots&\omega^{N-1}
\end{pmatrix}\;,\nonumber\\
&M_S=\omega^{\frac{N-1}{2}}\begin{pmatrix}
0&1&\dots&\dots&0\\
\vdots&0&1&\dots&\vdots\\
\vdots& \vdots&\ddots& &\vdots\\
0&\dots&\dots&\ddots&1\\
1&\dots&\dots&\dots&0
\end{pmatrix},
\end{align}
with $\omega=\mathrm{e}^{2\pi\im/N}$. Then, they commute up to a center element, $M_SM_C=\omega M_CM_S$, and $M_C^N=\bm{1}$ and $M_S^N=(-1)^{N-1}\bm{1}$.
Considering the homomorphism $H:SU(N)\rightarrow PSU(N)$, which sends the center elements of $SU(N)$ to unity in $PSU(N)$, makes it clear that $H(S)$ and $H(C)$ generate $\mathbb Z_N\times\mathbb Z_N\subset PSU(N)$. 

As before, we have that when we gauge this $\mathbb Z_N\times \mathbb Z_N$ symmetry by twisting the index $f$ of $\phi_{\ell,a}$ fields by an $M_S\in SU(N)$ matrix in one direction and $M_C\in SU(N)$ in the other direction, we are forced to have fractional fluxes for $U(1)$ gauge fields $a_\ell$. In fact, for all $\ell=1,\ldots,N$, 
\be
\int \diff a_\ell=\int B\bmod 2\pi\;,
\ee
where $B$ is the $\mathbb Z_N$ two-form gauge field of the $PSU(N)$ gauge bundle.

The change in the action is
\be
\Delta S_{\mathrm{top}}=\im p \int B\;,
\ee
where we set $\theta_\ell=\frac{2\pi p \ell}{N}$, so that the $\mathbb Z_N$ was a symmetry prior to gauging the $\mathbb Z_N\times\mathbb Z_N\subset PSU(N)$. This gives the mixed 't~Hooft anomaly for $\mathbb{Z}_N\times \mathbb{Z}_N\subset  PSU(N)$ flavor symmetry and $\mathbb{Z}_N$ permutation. 
The anomaly and global inconsistency discussed in previous subsections can also be found in the same manner.

\subsection{$\mathsf{C}$-$\mathsf{T}'$ and $\mathbb{Z}_N$-$\mathsf{T}'$ 't~Hooft anomaly for even $N$}

Here we will discuss anomalies involving time-reversal symmetry, which were discussed in \cite{Sulejmanpasic:2018upi} for the $N=2$ case.

The time-reversal symmetry $\mathsf{T}$ acts on $\bm\phi_i$ as 
\be
\mathsf{T}: \bm\phi_\ell (x,t)\mapsto \overline{\bm\phi}_\ell (x,-t). 
\ee
When $N$ is even, we can also define 
\be
\mathsf{T}': \bm\phi_\ell(x,t)\mapsto \mathcal{T}\overline{\bm\phi}_\ell(x,-t), 
\ee
with 
\be
\mathcal{T}=\underbrace{(\im \tau^2)\otimes \cdots \otimes (\im \tau^2)}_{N/2}\in SU(N). 
\ee
The matrix $\mathcal{T}$ satisfies $\mathcal{T}^\dagger=\mathcal{T}^t=-\mathcal{T}$ and $\mathcal{T}^2=-\bm{1}_N$.  
$\mathsf{T}'$ thus satisfies $\mathsf{T}'^2=(-1)^{N_{\bm\phi}}$, where $N_{\bm\phi}$ counts the number of $\bm\phi_i$'s, and generates $\mathbb{Z}_2$ symmetry on gauge-invariant operators. 
Using this time-reversal symmetry, we can put the theory on non-orientable manifolds with the structure $\mathrm{Pin}^{\tilde{c}}=\mathrm{Pin}_-\ltimes_{\mathbb{Z}_2} U(1) $. 
This means that we put the background gauge field for $\mathsf{T}'$ symmetry. 

After gauging $\mathsf{T}'$, the $U(1)$ gauge fields $\diff a_\ell$ should obey 
\be
\int F_\ell =\pi\int w_2(TM_2) \;\bmod 2\pi, 
\ee
where $w_2(TM_2)$ is the second Stiefel-Whitney class of the tangent bundle of $M_2$. That is, $\pi w_2$ plays the role of $(N/2)B$ if we want to make a correspondence to the analysis with $SU(N)/\mathbb{Z}_N$ background gauge fields. (For derivation of this, see Ref.~\cite{Sulejmanpasic:2018upi}.)

Now, we can do the same analysis for $\mathbb{Z}_N$ permutation and $\mathsf{C}_k$ symmetries. Considering special theta angles with these additional symmetries, we discuss the change of the partition function under those transformations with $w_2(TM_2)$. 

As an example, let us take a $\mathbb{Z}_N$-symmetric point $\theta_\ell=2\pi \ell/N$. The $\mathbb{Z}_N$ permutation changes the topological term as 
\be
\Delta S_{\mathrm{top}}=-\im \pi\int w_2(TM_2), 
\ee
and thus the partition function on $\mathbb{R}P^2$, $Z(\mathbb{R}P^2)$, changes the sign under $\mathbb{Z}_N$ permutation. This means that there is the mixed 't~Hooft anomaly between $\mathbb{Z}_N$ and $\mathsf{T}'$ symmetries. 

As another example, let us take a $\mathsf{C}_1$-symmetric point, $\theta_1=\pi$ and $\theta_\ell=0$ for $\ell\ge 2$. Again, the change of the topological term under $\mathsf{C}_1$ is given by 
\be
\Delta S_{\mathrm{top}}=-\im \pi\int w_2(TM_2), 
\ee
and we find the mixed anomaly between $\mathsf{C}_1$ and $\mathsf{T}'$. This mixed anomaly for spin systems without spin rotational symmetries was first found in the study~\cite{Sulejmanpasic:2018upi} of $\mathbb{C}P^1$ nonlinear sigma model at $\theta=\pi$ (see also Refs.~\cite{chen2011, PhysRevB.93.104425}).

\section{$SU(N)$ Wess-Zumino-Witten model}\label{sec:WZWmodel}

As we have mentioned briefly in Sec.~\ref{sec:phase_diagram_SU(3)}, the $SU(3)/[U(1)\times U(1)]$ nonlinear sigma model at the $\mathbb{Z}_3$-symmetric point is believed to be gappless. 
It is therefore an important and interesting problem to ask which conformal field theory appears in the low-energy limit. 
The anomaly matching condition tells us that the conformal field theory must have the same 't~Hooft anomaly. 

In this section, we compute the 't~Hooft anomaly of $(1+1)$-dimensional $SU(N)$ Wess-Zumino-Witten (WZW) model and make connections with the $SU(N)/U(1)^{N-1}$ nonlinear sigma model. The classical action of the level-$k$ $SU(N)$ WZW model is defined as~\cite{Wess:1971yu, Witten:1983tw, Witten:1983ar} 
\be
S=-{|k|\over 8\pi}\int_{M_2} \tr[\diff U\wedge \star \diff U^\dagger]+k \Gamma_{\mathrm{WZ}}[U], 
\ee
where $U$ is the $SU(N)$-valued scalar field on $M_2$, and the Wess-Zumino term $\Gamma_{\mathrm{WZ}}$ is defined by
\be
\Gamma_{\mathrm{WZ}}[U]={\im\over 12\pi}\int_{M_3}\tr[(U^\dagger \diff U)^3]. 
\ee
Here, $M_3$ is the $3$-dimensional manifold with $\p M_3=M_2$, and $U$ is extended to $M_3$. By imposing the condition that $\exp(S)$ is independent of this extension, the level $k$ is quantized to integers, $k\in\mathbb{Z}$.
Since the parity flips the sign of the level $k$, it is often considered only for $k>0$ and the level $-k$ shows the same conformal behavior. 

The model has the $[SU(N)_L\times SU(N)_R]/\mathbb{Z}_N$ global symmetry. $SU(N)_L\times SU(N)_R\ni (g_L,g_R)$ acts on $U$ as $U\mapsto g_L U g_R^\dagger$, but the symmetry group must be divided by $\mathbb{Z}_N$ since the diagonal center $\mathbb{Z}_N\subset SU(N)_L\times SU(N)_R$ does not change $U$. 
The subgroup $(SU(N)/\mathbb{Z}_N)_V \times (\mathbb{Z}_N)_L$ is of our interest. In previous studies~\cite{Gepner:1986wi, Furuya:2015coa, Numasawa:2017crf}, it was shown that the modular invariance of CFT and $(\mathbb{Z}_N)_L$ cannot be gauged (or, orbifolded) simultaneously. Our interest here is the anomaly matching between the UV theory ($SU(N)/U(1)^{N-1}$ nonlinear sigma model) and the IR conformal behavior ($SU(N)$ WZW), and we are interested in the anomaly of the common global symmetries. $g_V\in SU(N)_V$ acts on $U$ as $U\mapsto g_V U g_V^\dagger$, and $\omega^\ell \in (\mathbb{Z}_N)_L$ acts on $U$ as $U\mapsto \omega^\ell U$ with $\omega=\mathrm{e}^{2\pi\im/N}$ and $\ell=0,1,\ldots, N-1$. 
For connection with the main part of this paper, we would like to identify the $PSU(N)$ flavor symmetry as $SU(N)_V/\mathbb{Z}_N$ and the $\mathbb{Z}_N$ permutation symmetry as $(\mathbb{Z}_N)_L$.

\subsection{'t~Hooft anomaly of WZW model}

In order to gauge $(SU(N)/\mathbb{Z}_N)_V$, we introduce the $SU(N)$ one-form gauge field $A_V$ and the $\mathbb{Z}_N$ two-form gauge field $B$. 
The $\mathbb{Z}_N$ two-form gauge field is realized as the pair of the $U(1)$ two-form gauge field $B$ and the $U(1)$ one-form gauge field $C$ satisfying the constraint, 
\be
NB=\diff C
\ee
and we construct the $U(N)$ gauge field $\widetilde{A}_V$ by 
\be
\widetilde{A}_V=A_V+{1\over N}C\bm{1}_N. 
\ee
The naive minimal coupling procedure is to replace $U^\dagger \diff U$ by 
\be
U^\dagger D_V U = U^\dagger (\diff U + \im \widetilde{A}_V U -\im U \widetilde{A}_V). 
\ee
It is important to notice that the constraint and the covariant derivatives are invariant under the $U(1)$ one-form gauge transformation, 
\be
\widetilde{A}_V\mapsto \widetilde{A}_V+\xi,\; C\mapsto C+N\xi, \; B\mapsto B+\diff \xi. 
\ee
If we do this minimal coupling procedure for $\Gamma_{\mathrm{WZ}}$, however, it becomes dependent on the choice of $M_3$. We will reconsider this to compute the anomaly soon later, and we will find that $SU(N)_V/\mathbb{Z}_N$ itself does not have the anomaly but has a mixed anomaly with $(\mathbb{Z}_N)_L$. 

In order to gauge $(\mathbb{Z}_N)_L$, we introduce the $\mathbb{Z}_N$ one-form gauge field $A_L$. As we have done above, it is convenient to realize it by the $U(1)$ one-form gauge field satisfying the constraint
\be
N A_L=\diff \varphi, 
\ee
where $\varphi$ is the $2\pi$-periodic scalar field. 
To do it, we first regard $U\in SU(N)$ as $U\in U(N)$ with the constraint 
\be
\det(U)=1. 
\ee
In gauging $\mathbb{Z}_N$, we replace the constraint by 
\be
\mathrm{e}^{\im\varphi}\det(U)=1, \label{eq:constraint_det_WZW}
\ee
and the derivative $U^\dagger \diff U$ is also replaced by the familiar form, 
\be
U^\dagger D_L U=U^\dagger (\diff + \im A_L)U. 
\ee
The constraint and the covariant derivative are both invariant under the $U(1)$ zero-form gauge transformation, 
\be
U\mapsto \mathrm{e}^{-\im \psi}U,\; \varphi\mapsto \varphi+N\psi,\; A_L\mapsto A_L+\diff \psi,  
\ee
where $\psi$ is the gauge parameter and $2\pi$-periodic scalar. 

Now, let us gauge $(SU(N)/\mathbb{Z}_N)_V\times (\mathbb{Z}_N)_L$. For this procedure, we perform the two gauging procedures explained above simultaneously. We introduce the left and right gauge fields by 
\be
L:=\im(\widetilde{A}_V+A_L),\; R:=\im \widetilde{A}_V, \label{eq:left_right_gauge}
\ee
respectively, and the covariant derivative is defined as 
\be
U^\dagger D U=U^\dagger (\diff U + L U - U R). 
\ee

As we have mentioned, the gauged action obtained by this procedure is manifestly gauge-invariant, but the Wess-Zumino term with the covariant derivative is no longer the $2$D action mod $2\pi$: 
\begin{widetext}
\begin{multline}
{\im\over 12\pi}\int_{M_3} \tr[(U^\dagger D U)^3]
={\im\over 12\pi}\int_{M_3} \tr[(U^\dagger \diff U)^3]\\+{\im\over 4\pi}\int_{M_3}\tr\Bigl[-\diff (U\diff U^\dagger) L-\diff (\diff U^\dagger U) R
-U\diff U^\dagger L^2+U^\dagger \diff U R^2\\+L U R\diff U^\dagger-L (\diff U)R U^\dagger-U^\dagger L^2 U R+U^\dagger L U R^2+{1\over 3}(L^3-R^3)\Bigr]. 
\end{multline}
In order to eliminate the $3$-dimensional mixed term of $U$ and gauge fields without breaking the gauge invariance, we add the following $3$-dimensional gauge-invariant term. 
\be
{\im\over 4\pi}\int_{M_3} \tr[(U D U^\dagger)F_L-(U^\dagger D U)F_R], 
\ee
where $F_{L}=\diff L+L^2$ and $F_R=\diff R+R^2$. 
Note that this term is invariant under the $U(1)$ 1-form gauge symmetry, because one can show that $\tr[U^\dagger DU-U DU^{\dagger}]=2\tr[L-R+U^\dagger \diff U]=0$.  
As a consequence, we obtain
\begin{multline}
\Gamma_{\mathrm{WZW}}[U,(\widetilde{A}_V,B),A_L]={\im\over 12\pi}\int_{M_3}\tr[(U^\dagger \diff U)^3]+{\im\over 4\pi}\int_{M_3}\diff\left\{\tr[L U\diff U^\dagger-R U^\dagger \diff U-R U^\dagger L U]\right\}\\
\hspace{-1em}
+{\im\over 4\pi}\int_{M_3}\tr\left[\left(R\diff R+{2\over 3}R^3\right)-\left(L\diff L+{2\over 3}L^3\right)\right]\;. 
\label{eq:gauged_WZW}
\end{multline}
Let us substitute (\ref{eq:left_right_gauge}) into this expression to find the anomaly.  We obtain   
\bea
\Gamma_{\mathrm{WZW}}[U,(\widetilde{A}_V,B),A_L]
&=&{\im\over 12\pi}\int_{M_3}\tr[(U^\dagger \diff U)^3]
+{\im\over 4\pi}\int_{M_3}\diff\left\{\tr[(\im \widetilde{A}_V)(U\diff U^\dagger-U^\dagger \diff U-\im U^\dagger \widetilde{A}_V U)]\right\}\nonumber\\
&&
+{\im\over 4\pi}\int_{M_3}\tr\left[A_L(\diff \widetilde{A}_V+\im \widetilde{A}_V^2)\right]. 
\label{eq:gauged_WZW_2}
\eea
\end{widetext}
The first two terms on the right hand side of (\ref{eq:gauged_WZW_2}) give the well-defined action on $M_2$ mod $2\pi$, but the last one cannot be written as a local term in two-dimensions. Indeed, it is equal to the three-dimensional topological action, 
\be
S_{\mathrm{SPT}}={\im N\over 2\pi}\int_{M_3} A_L\wedge B, \label{eq:3D_SPT}
\ee
which describes the $(2+1)$D SPT phase protected by the $\mathbb{Z}_N$ zero-form and $\mathbb{Z}_N$ one-form symmetries. 
We have shown that the gauged WZW partition function, $Z_{\mathrm{WZW},k}[(A_V,B),A_L]$, is not gauge invariant as a two-dimensional field theory. Adding the three-dimensional SPT phase (\ref{eq:3D_SPT}), then the combined system,
\be
Z_{\mathrm{WZW},k}[(A_V,B),A_L]\exp\left(k S_{\mathrm{SPT}}[A_L,B]\right),
\ee
is gauge-invariant. As a coefficient of the topological term, there is the identification $k\sim k+N$.  
As a consequence, the level-$k$ $SU(N)$ WZW model has a mixed 't~Hooft anomaly between $(SU(N)/\mathbb{Z}_N)_V$ and $(\mathbb{Z}_N)_L$ for $k\not=0$ mod $N$. 

\subsection{Possible renormalization-group flows between $SU(N)$ Wess-Zumino-Witten models}

We can discuss the possible renormalization-group (RG) flow of the $SU(N)_k$ WZW model under the perturbations preserving $PSU(N)\times (\mathbb{Z}_N)_L$. Since the 't~Hooft anomaly is RG invariant, the low-energy effective theory must also have the anomaly given by $k S_{\mathrm{SPT}}[A_L,B]$. 
If $k$ is a multiple of $N$, then $k S_{\mathrm{SPT}}=0 \bmod 2\pi$ and thus the system can be gapped with the unique ground state. 
In the following, we assume that $1\le\mathrm{gcd}(N,k)<N$, where $\mathrm{gcd}(N,k)$ is the greatest common divisor of $N$ and $k$. 

To be specific, let us ask if $SU(N)_k$ WZW model can flow into $SU(N)_{k'}$ WZW model in view of anomaly matching. 
For that discussion, we specify how the symmetry $PSU(N)\times \mathbb{Z}_N$ is realized in UV and IR theories. In both limits, $PSU(N)$ is assumed to be realized as the vector-like symmetry. 
For $SU(N)_k$ WZW theory, $\mathbb{Z}_N$ is generated by
\be
U_{\mathrm{UV}}\mapsto \mathrm{e}^{2\pi \im /N}U_{\mathrm{UV}}. 
\ee
In this setting, as we have discussed, the anomaly is given by 
\be
k S_{\mathrm{SPT}}[A_L,B]={k N\over 2\pi}\int A_L\wedge B. 
\ee
At low energies, we assume that $\mathbb{Z}_N$ is generated by 
\be
U_{\mathrm{IR}}\mapsto \mathrm{e}^{2\pi \im q/N}U_{\mathrm{IR}}, 
\ee
for some integer $q$ with $\mathrm{gcd}(N,q)=1$. In this case, the anomaly is given by 
\be
k' S_{\mathrm{SPT}}[q A_L, B]={k' q N\over 2\pi}\int A_L\wedge B. 
\ee
The anomaly matching claims that we have to have
\be
k=k' q \bmod N. 
\ee
It is important to notice that the above constraints are satisfied especially for 
\be
k'=\gcd(N,k), 
\ee
by choosing $q=k/\gcd(N,k)\bmod N$. 
The $c$-theorem~\cite{Zamolodchikov:1986gt} tells us that the modulus of level-$k$ must decrease along the renormalization group flow since~\cite{Knizhnik:1984nr} $c=k(N^2-1)/(k+N)$. 
If $k'< \gcd(N,k)$, then there is no integer $q$ satisfying both $\gcd(N,q)=1$ and $k=k' q\bmod N$. Therefore, $k'=\gcd(N,k)$ is the minimal level of the $SU(N)$ WZW model satisfying the constraint of anomaly matching. 
When preserving the $SU(N)_V/\mathbb{Z}_N\times (\mathbb{Z}_N)_L$ symmetry, we thus conclude that the $SU(N)_k$ WZW model can flow into the $SU(N)_{\mathrm{gcd}(N,k)}$ WZW model.  

We point out that our conclusion is consistent with the previous conjecture~\cite{Lecheminant:2015iga} about the RG flow of WZW models: Based on the assumption of adiabatic continuity for certain spin systems, it has been conjectured that $SU(N)_k$ WZW model can be deformed into $SU(N)_1$ WZW model if $N$ and $k$ are coprime, i.e. $\gcd(N,k)=1$. 
This is consistent with the anomaly matching condition discussed above by redefining the generator of $(\mathbb{Z}_N)_L$ symmetry as $q=k/\gcd(N,k)=k$. 
We note that the anomaly matching argument gives the constraint also when $\gcd(N,k)>1$. 

Let us make several remarks about spin chains. 
The $SU(2)/U(1)$ nonlinear sigma model at $\theta=\pi$ shows the conformal behavior in the long-range limit, and that behavior is described by the $SU(2)_1$ WZW model. The nonlinear sigma model has a mixed 't~Hooft anomaly between $SU(2)/\mathbb{Z}_2$ flavor symmetry and $\mathbb{Z}_2$ charge conjugation symmetry~\cite{Komargodski:2017dmc}. The anomaly polynomial is exactly given by (\ref{eq:3D_SPT}) for $N=2$ and $k=1$, so the anomaly matching is satisfied for the Haldane conjecture~\cite{Furuya:2015coa}. 

We generalize this Haldane conjecture to the $SU(N)/U(1)^{N-1}$ nonlinear sigma models.
If no other unknown protection for the RG flow occurs, we can conjecture for the $SU(N)/U(1)^{N-1}$ nonlinear sigma model at $\theta_\ell=2\pi p\ell/N$: When $p\not=0$ mod $N$ and the low-energy behavior is conformal, it is given by the $SU(N)_{\gcd(N,p)}$ WZW model.

\subsection{Connection between $SU(N)$ WZW model and $SU(N)/U(1)^{N-1}$ model}

We here would like to argue in a complementary way the $SU(N)$ WZW model has the same 't~Hooft anomaly of $SU(N)/U(1)^{N-1}$ nonlinear sigma model at the specific theta angles. 
We will make a direct connection between those two models preserving the relevant symmetry, $SU(N)/\mathbb{Z}_N\times \mathbb{Z}_N\subset \frac{SU(N)_L\times SU(N)_R}{\mathbb Z_N}$, and therefore provide an independent and intuitive proof for the matching of anomalies. 
For this purpose, let us consider the potential term breaking $[SU(N)_L\times SU(N)_R]/\mathbb{Z}_N$ to $SU(N)_V/\mathbb{Z}_N\times (\mathbb{Z}_N)_L$, following the idea of Ref.~\cite{Affleck:1987ch}:
\be
V=g\int_{M_2}\diff^2 x\left(\tr[U]^{2N}+\tr[U^\dagger]^{2N}\right). 
\ee
Since this perturbation respect $SU(N)_V/\mathbb{Z}_N\times (\mathbb{Z}_N)_L$, the 't~Hooft anomaly for this symmetry does not change for any values of the coupling $g$. 
For $g<0$, the classical minima are given by $U=\exp(2\pi\im n/N)\bm{1}$ with $n=0,1,\ldots,N-1$, and thus $\mathbb{Z}_N$ is spontaneously broken and $N$ degenerate vacua ensue. 
For $g>0$, the classical minimum satisfies $\tr[U]=0$, and $SU(N)_V/\mathbb{Z}_N\times (\mathbb{Z}_N)_L$ is not spontaneously broken. To study this region, we take the limit $g\to+\infty$ and set the strict constraint, 
\be
\tr[U]=0,
\label{eq:traceless_condition}
\ee
on $M_2$. 

We consider the decomposition of $U\in SU(N)$ satisfying (\ref{eq:traceless_condition}) on $M_2$ as 
\be
U(x_1,x_2)=\calU(x_1,x_2)\Omega_0 \calU(x_1,x_2)^\dagger, 
\ee
where $\Omega_0=\omega^{-(N-1)/2}\mathrm{diag}[1,\omega,\ldots,\omega^{N-1}]$ with $\omega=\mathrm{e}^{2\pi\im/N}$. We take its extension to $M_3$ as\footnote{We are allowed to take this specific extension to $M_3$ since any extensions give the same value of $\Gamma_{\mathrm{WZ}}$ up to $2\pi\im$. } 
\be
U(x_1,x_2,x_3)=\calU(x_1,x_2)\Omega(x_3) \calU(x_1,x_2)^\dagger
\ee
with $M_3=M_2\times I_-$, where $I_-=(-\infty,0]$ is the half line. The boundary condition is
\be
\Omega(0)=\Omega_0, \;\Omega(-\infty)=\bm{1}.
\ee 
We have set $\Omega(-\infty)=\bm{1}$ so that $U(x_1,x_2,-\infty)=\bm{1}$, and we can regard $M_2\times \{-\infty\}$ as a point. 
This decomposition of $U$ to $\mathcal{U}$ has a redundancy given by the maximal Abelian subgroup $U(1)^{N-1}$ of $SU(N)$. 
We want to think of the matrix $\mathcal{U}$ as corresponding to the complex scalar fields $\bm\phi_\ell$ of $SU(N)/U(1)^{N-1}$ nonlinear sigma model, i.e., 
\be
\mathcal{U}=[\bm\phi_1,\bm\phi_2,\ldots,\bm\phi_N]
\ee
To justify that, let us first make the correspondence of symmetries, and compute the WZW action. 

The redundancy of the decomposition is related to the gauge invariance under $\bm\phi_\ell\mapsto \mathrm{e}^{\im\varphi_\ell} \bm\phi_\ell$, with $\sum_{\ell=1}^N \varphi_l=0$. 
The flavor symmetry corresponds to $SU(N)_V/\mathbb{Z}_N$ because $V\in SU(N)$ acts as $\mathcal{U}\mapsto V\cdot \mathcal{U}$ and the center subgroup $\mathbb{Z}_N\subset SU(N)_V$ can be compensated by the $U(1)^{N-1}$ gauge invariance. 
The $\mathbb{Z}_N$ permutation symmetry corresponds to $(\mathbb{Z}_N)_L$. Let us introduce the permutation matrix $\mathcal{P}$ by
\be
\mathcal{U}=[\bm\phi_1,\bm\phi_2,\ldots,\bm\phi_N] \mapsto [\bm\phi_2,\ldots,\bm\phi_N,\bm\phi_1]=: \calU \mathcal{P}, 
\ee
then it satisfies $\mathcal{P}\Omega_0\mathcal{P}^{-1}=\omega \Omega_0$. Applying the $\mathbb{Z}_N$ permutation to $U$, we get 
\be
U=\calU\Omega_0\calU^\dagger \mapsto \calU \mathcal{P} \Omega_0 \mathcal{P}^{-1} \calU^\dagger=\omega U, 
\ee
which is nothing but $(\mathbb{Z}_N)_L$. 

Next, let us compute the WZW action using this parametrization. 
For the $SU(N)/U(1)^{N-1}$ nonlinear sigma model, the origin of 't~Hooft anomaly is the topological theta term, so we have to reproduce the correct theta terms from the Wess-Zumino term $\Gamma_{\mathrm{WZ}}$. 
Thus, the kinetic term is unimportant, but let us write down its result just for completeness: 
\bea
&&{1\over 2}\int_{M_2}\tr[\diff U\wedge \star \diff U^\dagger]\nonumber\\
&=&\sum_{\ell}\int_{M_2}(\diff \overline{\bm\phi}_\ell\wedge \star \diff \bm\phi_\ell-(\overline{\bm\phi}_\ell\cdot \diff \bm\phi_\ell)\wedge \star (\diff \overline{\bm\phi}_\ell\cdot \bm\phi_\ell))\nonumber\\
&&-\sum_{\ell\not=m}\int_{M_2}\omega^{m-\ell}(\overline{\bm\phi}_\ell\cdot \diff \bm\phi_m)\wedge \star (\diff \overline{\bm\phi}_m\cdot \bm\phi_\ell). 
\eea
The first term on the right hand side is the usual kinetic term with the $U(1)$ gauge fields $a_\ell=\im \overline{\bm\phi}_\ell\cdot \diff \bm\phi_\ell$. The second one did not exist in our $SU(N)/U(1)^{N-1}$ sigma model, but it is gauge invariant and does not break any global symmetries. Therefore, we can add it without any problem and it does not affect the discussion of the 't~Hooft anomalies. 

In order to compute the Wess-Zumino term conveniently, let us specify our extension to the $x_3$ direction in more concrete way. We parametrize $\Omega(x_3)$ as 
\be
\Omega(x_3)=\mathrm{diag}[\mathrm{e}^{\im\theta_1(x_3)},\mathrm{e}^{\im \theta_2(x_3)},\ldots, \mathrm{e}^{\im \theta_N(x_3)}]. 
\ee
The boundary condition on $\Omega(x_3)$ can be rephrased as
\be
\theta_\ell(-\infty)=0,\; \theta_\ell(0)={2\pi \ell\over N}, 
\ee
up to an overall shift of $\theta_\ell(0)$'s, but such overall shift does not change the following argument so we can take this convention. 
Since $\mathcal{U}$ does not depend on $x_3$, we obtain that 
\begin{align}
&\tr[(U^\dagger \diff U)^3]\nonumber\\
&=3\,\tr[(\calU^\dagger \diff U)^2 \diff \Omega \Omega^\dagger+(\diff \calU^\dagger \calU)^2 \Omega^\dagger \diff \Omega]\nonumber\\
&+3\,\tr[(\calU^\dagger \diff \calU)\Omega^\dagger (\diff \calU^\dagger \calU)\diff \Omega-(\calU^\dagger \diff \calU)\Omega (\diff \calU^\dagger \calU)\diff \Omega^\dagger]. 
\end{align}
As we shall see, the first line on the right hand side gives the theta term, while the second one gives the generalization of the $\lambda$-term. 
The computation of the first term can be done as follows: 
\bea
&&{\im\over 12\pi}\int_{M_3} 3\, \tr[(\calU^\dagger \diff \calU)^2 \diff \Omega \Omega^\dagger+(\diff \calU^\dagger \calU)^2 \Omega^\dagger \diff \Omega]\nonumber\\
&=&{\im\over 2\pi}\int_{M_2} \sum_{\ell,m}(\overline{\bm\phi}_m\cdot \diff \bm\phi_\ell)(\overline{\bm\phi}_\ell\cdot \diff \bm\phi_m) \int_{I_-}\im \diff \theta_m(x_3)\nonumber\\
&=&\sum_m {\theta_m(0)\over 2\pi}\int_{M_2}\diff \overline{\bm\phi}_m\cdot \diff \bm\phi_m. 
\eea
This is exactly the theta term, and the theta angles are given by $\theta_\ell=\theta_\ell(0)=2\pi\ell/N$. This is nothing but the $\mathbb{Z}_N$ symmetric point of the $SU(N)/U(1)^{N-1}$ nonlinear sigma model. Doing the similar computation of the second term gives 
\begin{align}
&{\im\over 4\pi}\int_{M_3}\tr[(\calU^\dagger \diff \calU)\Omega^\dagger (\diff \calU^\dagger \calU)\diff \Omega
 -(\calU^\dagger \diff \calU)\Omega (\diff \calU^\dagger \calU)\diff \Omega^\dagger]\nonumber\\
&=
{1\over 4\pi}\sum_{m\not=\ell}\sin(\theta_m(0)-\theta_\ell(0))\int_{M_2}(\overline{\bm\phi}_m\cdot \diff \bm\phi_\ell)(\overline{\bm\phi}_\ell\cdot \diff \bm\phi_m) . 
\end{align}
For the case of the $\lambda$ term, $\sin(\theta_m(0)-\theta_\ell(0))$ should be replaced by $\lambda(\delta_{m,\ell+1}-\delta_{m,\ell-1})$, so this gives its generalization. The generalization does not break any symmetry, so it does not change the argument of the 't~Hooft anomaly matching.

\section{Linear sigma models}\label{sec:linear_model}

In this section, we construct the linear sigma model version of the $SU(N)/U(1)^{N-1}$ nonlinear sigma model, and show that they have the same 't Hooft anomaly and global inconsistency explicitly. 
In certain limits of linear sigma models, we can perform the analytic computation of the partition function, and we can check the conjecture on the phase diagram given in Sec.~\ref{sec:anomaly}.

\subsection{Linear realization of $SU(N)/U(1)^{N-1}$ nonlinear sigma model}

We here consider the case $N=3$ for simplicity of the presentation, and the generalization is straightforward. 
Instead of the three copies of $\mathbb{C}P^2$ obeying orthogonality conditions, let us take 3 copies of an $SU(3)$ triplet, without orthogonality constraints on them. The kinetic term of the Lagrangian is
\be
S_{\mathrm{kin}}=-{1\over 2g}\int \sum_{\ell=1,2,3}|(\diff+\im a_\ell)\bm\Phi_\ell|^2, 
\ee
where $\bm\Phi_\ell=(\Phi_{1\ell},\Phi_{2\ell},\Phi_{3\ell})$, and we put the strict constraint on the $U(1)$ gauge fields,
\be
a_3=-a_1-a_2. \label{eq:constraint_U(1)_field_linear_1}
\ee
What this means is that the triplet $\bm\Phi_3$ is charged as $(-1,-1)$ under the gauge charges $a_1$ and $a_2$. 

The analogue of the $\lambda$-term is given as follows: We first define the gauge-covariant one-form, 
\be
\omega_{i,j}=\overline{\bm\Phi}_i \cdot (\diff+\im a_j)\bm\Phi_j - \bm\Phi_j \cdot(\diff-\im a_i) \overline{\bm\Phi}_i, 
\ee
and write down the gauge-invariant term as 
\be
S_{\lambda}={\lambda\over 2\pi}\int \sum_{\ell=1}^{3}\omega_{\ell+1,\ell}\wedge \omega_{\ell,\ell+1}. 
\ee
The model above, $S_{\mathrm{kin}}+S_{\lambda}$, has the following symmetries, which are the same with those of $SU(3)/[U(1)\times U(1)]$ nonlinear sigma model:
\begin{itemize}
\item $PSU(3)=SU(3)/\mathbb{Z}_3$ flavor symmetry, acting projectively on $\bm{\Phi}_\ell$ as $\bm\Phi_\ell\mapsto U \bm\Phi_\ell$ with $U\in SU(3)$, 
\item Time reversal $\mathsf{T}$ that sends $\bm\Phi_\ell(x,t)\mapsto \overline{\bm\Phi}_\ell(x,-t)$, 
\item $\mathbb Z_3$ exchange symmetry, which cyclically permutes $\bm\Phi_{\ell}\mapsto\bm\Phi_{\ell+1}$ and $a_\ell\mapsto a_{\ell+1}$,
\item Charge conjugations $\mathsf{C}_k$ that sends $\bm\Phi_\ell\mapsto -\overline{\bm\Phi}_{-\ell-k}$ and $a_\ell\mapsto -a_{-\ell-k}$. 
\end{itemize}
In addition, there is an extra $U(1)/\mathbb{Z}_3$ symmetry, given by 
\be
\bm\Phi_\ell\mapsto \mathrm{e}^{\im \varphi}\bm\Phi_\ell
\ee
with $\varphi\sim \varphi+2\pi$, but the physical identification on gauge-invariant operators is $\varphi\sim \varphi+2\pi/3$. 

In order to match the symmetry with that of nonlinear sigma model, we shall break $U(1)/{\mathbb{Z}_3}$ symmetry explicitly by the potential term. 
The invariant tensors of $SU(3)$ are the Kronecker delta and the epsilon tensors. The gauge-invariant quadratic invariants of $SU(3)$ made of $\bm\Phi_\ell$ are $\overline{\bm\Phi}_\ell\cdot \bm\Phi_{\ell'}$. 
Using the epsilon tensor, we also have the gauge-invariant $SU(3)$ invariant, 
\be
g=\ve_{abc}\Phi_{a1}\Phi_{b2}\Phi_{c3},\; \overline{g}=\ve_{abc}\overline{\Phi}_{a1}\overline{\Phi}_{b2}\overline{\Phi}_{c3}.
\ee
This operator is invariant under $U(1)$ gauge symmetries, and has the unit charge under $U(1)/\mathbb{Z}_3$ global symmetry, and 
\be
\mathsf{T}:g(x,t)\mapsto \overline{g}(x,-t),\; \mathsf{C}_k:g\mapsto \overline{g}. 
\ee
We therefore add the following potential term, 
\bea
&&S_{\mathrm{pot}}=\int \diff^2 x\left\{\sum_\ell V_1\left(|\bm\Phi_\ell|^2\right)+\sum_{\ell>\ell'} V_2\left(|\overline{\bm\Phi}_\ell\cdot \bm\Phi_{\ell'}|^2\right)\right\}\nonumber\\
&&\quad +\int \diff^2 x\,  V_3\left(\ve_{abc}(\Phi_{a1}\Phi_{b2}\Phi_{c3}+\overline{\Phi}_{a1}\overline{\Phi}_{b2}\overline{\Phi}_{c3})\right). 
\eea
By taking a certain limit of $V_1$ and $V_2$, we can reproduce the orthonormality constraint (\ref{eq:orthonormality}) of the nonlinear sigma model, and the matrix $[\bm\Phi_1,\bm\Phi_2,\bm\Phi_3]\in U(3)$. The potential $V_3$ gives the condition on its determinant as in (\ref{eq:constraint_det}), and $[\bm\Phi_1,\bm\Phi_2,\bm\Phi_3]\in SU(3)$.  Since $U(1)\times U(1)$ gauge invariance says that this target space is redundant by $U(1)\times U(1)$, we can obtain the nonlinear $SU(3)/[U(1)\times U(1)]$ sigma model as a low-energy effective theory of $S_{\mathrm{kin}+\mathrm{pot}+\lambda}$ in this limit. 
As we shall see, the anomaly discussed in Sec.~\ref{sec:anomaly} exists for the generic potential $V_1$, $V_2$, and $V_3$.

As in the case of nonlinear sigma model, we introduce the topological term that breaks $\mathbb{Z}_3$ and $\mathsf{C}_k$ for general values: 
\be
S_{\mathrm{top}}=\sum_{\ell=1}^{3}{\im \theta_\ell\over 2\pi}\int \diff a_\ell. 
\ee
Because of the constraint on the gauge charge, we can choose one of the $\theta$ angles to be zero, and we set $\theta_3=0$ in this section. 

The model above has all the same symmetry, and, save for the more parameter freedom, largely the same structure as the $SU(3)/U(1)^2$ nonlinear sigma model. It therefore has the same anomalies that we have been discussing so far. 

Such models, which can be supplemented with arbitrary local terms in the Lagrangian, are of interest as they better capture all the possible phases of relevant spin chains. The $SU(3)/U(1)^2$ nonlinear sigma model on the other hand is supposed to describe only a Heisenberg spin chain, and even that one can be reliably related to it only via the limit of large dimension of the $SU(3)$ representations (i.e.~large spin limit). The statement that the $SU(3)/U(1)^2$ nonlinear sigma model is the effective model of the spin chains is therefore imprecise. Rather the more precise statement is that the effective theory of general spin chains is described by a linear sigma model, with a priori unknown couplings.  Still anomalies and inconsistencies give constraint on possible vacuum realizations of such models so the phase diagram is guaranteed to be interesting. They also allow for more semi-classical regimes, because they have more tunable parameters. In particular, we can add a mass to the $\bm \Phi_\ell $ fields, preserving all the symmetries and, therefore, all the anomalies. Upon taking this mass to large values, a free photon ensues. We discuss this next.

\subsection{Free photon limit}\label{sec:large_mass}

Using the linear sigma model description, let us take the limit to compute the free energy analytically. This will provide a check and deepen the understanding of how the anomaly and global inconsistency matching discussed in Sec.~\ref{sec:phase_diagram_SU(3)} is realized.
 
The easiest thing we can do is to send the mass of the $\Phi_{\ell,f}$ fields to be large, then the matter fields can be integrated out, and we obtain the local field theory of photons.

The effective theory is a free $U(1)^{N-1}$ gauge theory, with the (real-time) Lagrangian given by
\be
\mathcal{L}=\sum_{\ell=1}^N \frac{1}{2e^2}F_\ell^2+\sum_{\ell=1}^N \frac{\theta_\ell}{2\pi} F_\ell, 
\label{eq:free_photon}
\ee
where $F_{\ell}=\p_t a_{x\ell}-\p_x a_{t\ell}$ and $e$ is the effective coupling constant. The gauge fields satisfy the constraint $a_1+\cdots+a_N=0$, and we can set $\theta_N=0$.

\begin{figure*}[htbp] 
   \centering
   \includegraphics[width=3in]{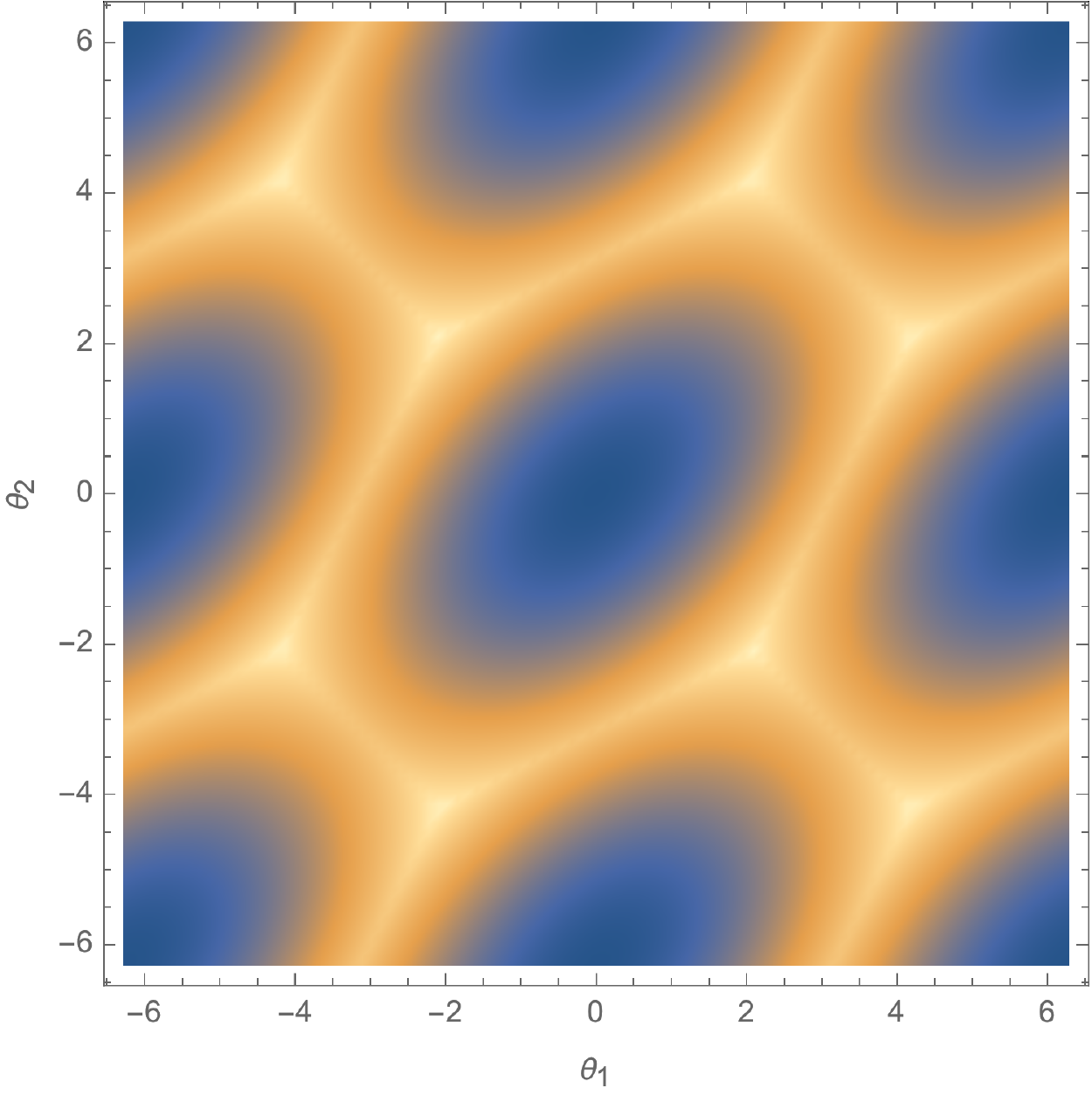} \includegraphics[width=3in]{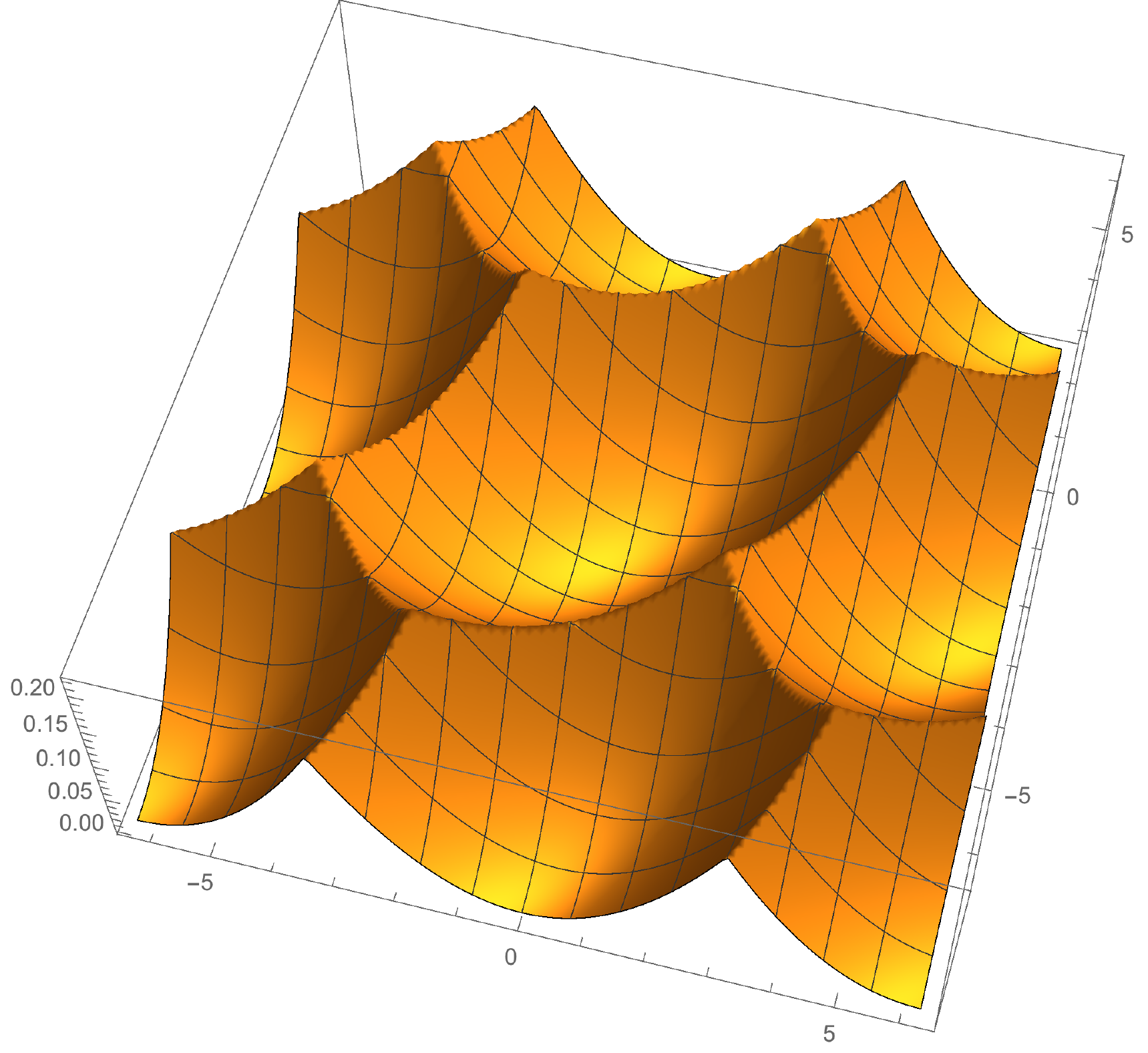} 
   \caption{The energy density of the ground state as a function of the two $\theta$-parameters. Notice that the same picture emerges as discussed in Sec.~\ref{sec:anomaly}. On the 3D plot on the right, it is clear that level crossings occur at the $\mathsf C$-symmetric lines, which meet at $\mathbb Z_3$-cyclic permutation symmetric points which carry a 't Hooft anomaly. }
   \label{fig:free_photons_phase_diag}
\end{figure*}

Now, we canonically quantize the system in order to find energy eigenstates, and we take the temporal gauge $a_{t\ell}=0$ for this purpose. The coordinates $a_{x \ell}$ ($\ell=1,\ldots,N-1$) have the canonical momentum 
\be
\pi_x^\ell=\frac{\partial \mathcal{L}}{\partial (\partial_t a_{x \ell})}=\frac{1}{e^2}F_\ell+\frac{1}{e^2}\sum_{\tilde \ell=1}^{N-1}F_{\tilde \ell}+\frac{\theta_\ell}{2\pi}
\ee
Solving it for $F_\ell$, we have
\be
F_\ell=e^2 \Pi^\ell_x-\frac{e^2}{N}\sum_{\tilde \ell=1}^{N-1}\Pi_x^{\tilde \ell},
\ee
where
\be
\Pi^\ell_x=\pi_x^\ell-\frac{\theta_\ell}{2\pi}, 
\ee
for $\ell=1,\dots,N-1$. 
The Hamiltonian density is given by
\begin{align}
\mathcal{H}&=\sum_{\ell=1}^{N-1} \pi^\ell_x (\p_t a_{x\ell})-\mathcal{L}\nonumber\\
&=\sum_{\ell=1}^{N-1} \Pi_x^\ell F_\ell-\frac{1}{2e^2}\sum_{\ell=1}^{N-1} F_\ell^2-{1\over 2e^2}\left(\sum_{\ell=1}^{N-1} F_\ell\right)^2+\pi_x^\ell \partial_x a_{t\ell}\nonumber\\
&=\frac{e^2}{2}\left[\sum_{\ell=1}^{N-1}(\Pi_x^\ell)^2-\frac{1}{N}\left(\sum_{\ell=1}^{N-1} \Pi_x^\ell\right)^2\right]+\pi_x^\ell \partial_x a_{t\ell}.
\end{align}
When demanding that $[H,\pi_0^\ell]$, the last term causes the secondary constraint $\partial_x \pi_x^\ell=0$ --- the Gauss law. Further
the spectrum is simply given by the eigenvalues of $\pi_x^\ell$, which are integers $m_\ell$, i.e.
\bea\label{eq:en_density}
&&\mathcal{E}_{\{m_k\}}(\theta_\ell)\nonumber\\
&=&
\frac{e^2}{2}\left[\sum_{\ell=1}^{N-1}\left(m_\ell-\frac{\theta_\ell}{2\pi}\right)^2-\frac{1}{N}\left(\sum_{\ell=1}^{N-1} \left(m_\ell-\frac{\theta_\ell}{2\pi}\right)\right)^2\right]. \nonumber\\
\label{eq:energy_spectrum_free}
\eea
The ground state energy is given by the minimum among those sectors:
\be
\mathcal{E}(\theta_\ell)=\min_{\{m_k\}\in\mathbb{Z}^{N-1}} \mathcal{E}_{\{m_k\}}(\theta_\ell). 
\label{eq:ground_state_energy_free}
\ee
Using this expression for $N=3$, we can confirm the phase diagram of Fig.~\ref{fig:inconsistency} (or the left one of Fig.~\ref{fig:phase_diagrams} in Sec.~\ref{sec:phase_diagram_SU(3)}). 
We plot the $N=3$ energy density given by \eqref{eq:en_density} for the ground state (Fig.~\ref{fig:free_photons_phase_diag}), we can clearly see the pattern which emerged from our general discussion.

Now consider the $\mathbb{Z}_N$ permutation symmetry sends $F_\ell\mapsto F_{\ell+1}$, where $F_N=-F_1-F_2\cdots-F_{N-1}$. This symmetry acts on the canonical momentum as
\begin{align}
&\pi_x^\ell \mapsto \pi_x^{\ell+1}-\pi_x^1+\frac{-\theta_{\ell+1}+\theta_\ell+\theta_1}{2\pi}\;, \ell=1,2\dots,N-2\\
&\pi_x^{N-1}\mapsto -\pi_x^1+\frac{\theta_{N-1}+\theta_1}{2\pi}. 
\end{align}
If we replace $\theta_\ell=2\pi p \ell/N$, we find the action of the $\mathbb{Z}_N$ permutation on the eigenvalues $\{m_\ell\}_{\ell=1,\ldots,N-1}$:
\begin{align}
&m_\ell\mapsto  m_{\ell+1}-m_1\;, \ell=1,2\dots,N-2\\
&m_{N-1}\mapsto -m_1+p\;.
\end{align}
We now look for a fixed point of the transformation, i.e. that
\begin{align}
&m_\ell= m_{\ell+1}-m_1\;, \ell=1,2\dots,N-2\\
&m_{N-1}= -m_1+p\;.
\end{align}
The first equation implies that 
\be
m_\ell=\ell m_1\;, \ell=1,\dots, N-1
\ee
and, in particular, that
\be
m_{N-1}=(N-1)m_1, 
\ee
while the second one implies
\be
m_{N-1}=-m_1+p\;.
\ee
Consistency of the two equations demands that
\be
N m_1=p\;,
\ee
which is only possible if $p=0$ mod $N$, so that $m_1=p/N\in \mathbb{Z}$. This is precisely the case where there is no anomaly in the full theory, so we get consistency. When $p\ne0$ mod $N$, there is no fixed point of the $\mathbb Z_N$ transformation acting on integers $m_\ell$, so all states (and in particular the ground state) are degenerate, and the anomaly is saturated by breaking the $\mathbb Z_N$ global symmetry.

Let us discuss this a bit more from the point of view of anomalies. Originally, the theory has the $PSU(N)$ flavor symmetry, but it is gone in the low-energy effective theory since matter fields are very massive. Instead, the theory acquires the emergent $\mathbb{Z}_N$ one-form symmetry, which is further enhanced to $U(1)^{N-1}$ one-form symmetry in the free-photon Lagrangian (\ref{eq:free_photon}). 
We can understand the above energy spectrum (\ref{eq:energy_spectrum_free}) by gauging this $U(1)^{N-1}$ one-form symmetry. To see it, we introduce the $U(1)$ two-form gauge fields, $B_\ell$, for $\ell=1,\ldots,N-1$ and impose the invariance under the $U(1)$ one-form gauge transformations, \be
B_\ell\mapsto B_\ell+\diff \lambda_\ell,\; a_\ell\mapsto a_\ell+\lambda_\ell. 
\ee
We have to replace the field strength $\diff a_\ell$ by $\diff a_\ell-B_\ell$ in order to satisfy this invariance. 
We can further add the local gauge-invariant terms of $B_\ell$ in the gauging procedure, so we add 
\be
\im \sum_\ell m_\ell \int B_\ell
\ee
for integers $\{m_\ell\}\in\mathbb{Z}^{N-1}$. We can easily find that  
\bea
&&\exp(-V \mathcal{E}_{\{m_k\}})\nonumber\\
&=&\int \Diff B_{\ell}\int\Diff a_{\ell} \exp\sum_{\ell}\left[-{1\over 4e^2}\int_{M_2} |\diff a_\ell-B_\ell|^2\right.\nonumber\\
&&\left.+{\im \theta_\ell\over 2\pi}\int_{M_2} (\diff a_\ell-B_\ell)+\im m_\ell \int_{M_2} B_\ell\right], 
\eea
where the path integral is done with the constraint $a_1+\cdots+a_N=0$ and $B_1+\cdots +B_N=0$, and $V$ is the volume of $M_2$. 
The labels $\{m_\ell\}$ of the energy eigenstate is now understood as the coefficient of the counter term for gauging $U(1)^{N-1}$ one-form symmetry~\cite{Tanizaki:2017bam, Kikuchi:2017pcp}. 
Since the original $PSU(N)$ symmetry corresponds to the diagonal subgroup $\mathbb{Z}_N\subset U(1)^{N-1}$, we would like to set 
\be
B\equiv B_1=B_2=\cdots=B_{N-1},\; N B=\diff C,
\ee
with some $U(1)$ gauge field $C$. The corresponding local counter term becomes  
\be
\im (m_1+\cdots+m_{N-1})\int B, 
\ee
and the coefficient $m_{\mathrm{tot}}=(m_1+\cdots+m_N)$ is meaningful only in $\mathbb{Z}_N$. 
This means that the interaction term coming out of the matter fields $\bm\Phi_{\ell}$ can mix the states with the same $m_{\mathrm{tot}}$ mod $N$, but the different ones cannot be mixed by such interactions. In this sense, we can regard $m_{\mathrm{tot}}$ as the $\mathbb{Z}_N$ charge of the $\mathbb{Z}_N$ one-form symmetry, or $PSU(N)$ symmetry of the original theory.

Now notice that the $\mathbb{Z}_N$ permutation acts on the $\mathbb{Z}_N$ charge of the one-form symmetry, $m_{\mathrm{tot}}=(m_1+\cdots+m_{N-1})$, as 
\be
m_{\mathrm{tot}}\mapsto m_{\mathrm{tot}}+p\;\bmod N. 
\label{eq:realization_anomaly}
\ee
This means that we cannot find the simultaneous eigenstate of the $\mathbb{Z}_N$ one-form symmetry and the $\mathbb{Z}_N$ permutation symmetry for $p\not=0$ mod $N$. Furthermore, if $p$ and $N$ are relatively prime, then all of the states must be $N$-fold degenerate. 
Since the $\mathbb{Z}_N$ one-form symmetry emerges from $PSU(N)$ symmetry, (\ref{eq:realization_anomaly}) should be regarded as the consequence of $PSU(N)$-$\mathbb{Z}_N$ 't~Hooft anomaly. 
We can repeat the similar discussion for $PSU(N)$-$\mathsf{C}$ global inconsistency\footnote{Note however that on the $\mathsf C$-invariant lines of pure $U(1)^{N-1}$ gauge theory, there is a genuine anomaly between the $\mathsf C$-symmetry and the $U(1)^{N-1}$ center symmetry. 
It therefore guarantees that the phase transition lines are as in Fig.~\ref{fig:free_photons_phase_diag}. 
Only if the center symmetry is reduced down to $\mathbb Z_N$ with odd $N$, does the anomaly turn to a global inconsistency. }.

\section{Circle compactification with persistent 't~Hooft anomaly}\label{sec:compactification}

In the previous section~\ref{sec:large_mass}, we give the linear sigma model description of $SU(3)/[U(1)\times U(1)]$ model, and it is computable in the limit where the matter fields $\bm\Phi$ are very massive. The original interest of the model is the case where the matter fields are would-be Nambu-Goldstone bosons, and thus it is very appealing if we can consider a setup to study that regime analytically. 
In this section, we provide a setup for reliable semiclassical computations of $SU(3)/[U(1)\times U(1)]$ nonlinear sigma model. 

The nonlinear sigma models in two dimension shows the asymptotic freedom in general when the target space has positive curvature~\cite{Polyakov:1975rr, Polyakov:1987ez}, which means that they become strongly coupled in the infrared regime. It is therefore quite difficult to extract the low-energy behavior of the theory analytically. For example, the semiclassical analysis using instantons suffers from the severe IR divergences, and gives the wrong results even qualitatively~\cite{Witten:1978bc}. 
One possible way to evade IR divergences is to put the theory on a small circle $\mathbb{R}\times S^1$. The size of the circle $L$ provides an energy scale $1/L\gg \Lambda$ which can be arbitrarily high, and hence has a potential to render the asymptotically free theories weakly coupled. However, if compactification is done naively, then typically the phase transition along the circle size $L$ takes place, and the wanted low-energy behavior cannot be found in a semiclassical way~\cite{Affleck:1979gy}. 

The idea of semiclassical analysis with $S^1$ compactification can be revived by compactifying the theory with twisted boundary conditions or, equivalently with a nontrivial holonomy background~\cite{Dunne:2016nmc, Dunne:2012ae,Unsal:2007jx,Unsal:2008ch} (see also 
Refs.~\cite{ Unsal:2007vu, Kovtun:2007py,  Shifman:2008ja, Shifman:2009tp, Cossu:2009sq, Cossu:2013ora, Argyres:2012ka, Argyres:2012vv, Dunne:2012zk, Poppitz:2012sw, Anber:2013doa, Basar:2013sza, Cherman:2013yfa, Cherman:2014ofa, Misumi:2014raa, Misumi:2014jua, Misumi:2014bsa, Dunne:2015ywa,Misumi:2016fno, Cherman:2016hcd, Fujimori:2016ljw, Sulejmanpasic:2016llc, Yamazaki:2017ulc, Buividovich:2017jea, Aitken:2017ayq}). Such systems often exhibit a weakly coupled regimes for small $L$ and thus can be treated semiclassically without IR divergences.  Their properties look remarkably similar to the low-energy behavior expected for uncompactified theories. 
It is therefore conjectured that the large and small circles are adiabatically connected thanks to the nontrivial holonomy, but the role of the nontrivial holonomy was not so clear when it was proposed. One of the author (T.~S.) has shown that such holonomies can lead to a vast cancellations in the spectrum preventing a would-be thermal phase transition~\cite{Sulejmanpasic:2016llc}. In Ref.~\cite{Sulejmanpasic:2016llc} it was explicitly shown how such cancellations lead to a large $N$ volume independence for $\mathbb{C}P^{N-1}$ and $O(N)$ sigma models, and the conjecture acquired solid ground for certain models. 
The problem was revisited by the another author (Y.~T.) from the viewpoint of 't~Hooft anomaly matching, and it is shown that the nontrivial holonomy is essential for persistence of 't~Hooft anomaly under $S^1$ compactification~\cite{Tanizaki:2017qhf, Tanizaki:2017mtm, Dunne:2018hog}.

In this section, we discuss the $S^1$ compactification, under which the $SU(3)/\mathbb{Z}_3$-$\mathbb{Z}_3$ mixed anomaly and $SU(3)/\mathbb{Z}_3$-$\mathsf{C}$ global inconsistency survives following Ref.~\cite{Tanizaki:2017qhf}. This provides an opportunity for future works to study the low-energy behavior of the $SU(3)/[U(1)\times U(1)]$ nonlinear sigma model by an analytic semiclassical computations. 

We take $M_2=M_1\times S^1$, and the circumference of $S^1$ is $L$. Using the clock matrix $C=\mathrm{diag}[1,\omega,\omega^2]$ with $\omega=\mathrm{e}^{2\pi\im/3}$, we define the boundary condition, 
\be
\bm\phi_\ell(x,t+L)=C\cdot \bm\phi_\ell(x,t). \label{eq:S1_compactification}
\ee
We take the periodic boundary condition for the gauge field, $a_\ell(x,t+L)=a_\ell(x,t)$. This defines our $S^1$-compactified theory. 

The above boundary condition is equivalent to introducing the background $SU(3)$ holonomy along the compactified direction. To see this, let us define $\widetilde{\bm\phi}_\ell(x,t)$ obeying the periodic boundary condition by 
\be
\phi_{f,\ell}(x,t)=\mathrm{e}^{2\pi \im f t/3L}\widetilde{\phi}_{f,\ell}(x,t)
\ee
for $f=1,2,3$. Then, the covariant time derivative is given as 
\be
|D_t \phi_{f,\ell}|^2=\left|\left(\p_0+a_{\ell,0}+{2\pi \im f\over 3L}\right)\widetilde{\phi}_{f,\ell}\right|^2, 
\ee
and we can see that the $SU(3)$-flavor background gauge field is introduced in addition to the $U(1)$ gauge field. 

Because of the flavor-dependent boundary condition, the $SU(3)/\mathbb{Z}_3$ flavor symmetry is explicitly broken down to its maximal Abelian subgroup $[U(1)\times U(1)]/\mathbb{Z}_3$. 
In addition, the system has the symmetry involving the shift matrix and one-form transformation~\cite{Tanizaki:2017qhf} (see also Ref.~\cite{Cherman:2017tey}). 
Since $SC=\mathrm{e}^{2\pi\im /3}CS$, the shift matrix itself does not generate the symmetry of the $S^1$-compactified theory. Indeed, the kinetic term is changed as 
\bea
&&\sum_f \left|\left(\p_0+a_{\ell,0}+{2\pi \im f\over 3L}\right)\widetilde{\phi}_{f,\ell}\right|^2\nonumber\\
&\mapsto& \sum_f \left|\left(\p_0+a_{\ell,0}+{2\pi \im (f+1)\over 3L}\right)\widetilde{\phi}_{f,\ell}\right|^2. 
\eea
In order to compensate the difference, we have to perform 
\be
a_{\ell,0}\mapsto a_{\ell,0}-{2\pi\im\over 3L},
\ee 
which is nothing but the $\mathbb{Z}_3$ one-form transformation on $U(1)$ Polyakov loops. 
Let us call this $\mathbb{Z}_3$ symmetry as the intertwined shift symmetry, $(\mathbb{Z}_3)_{\mathrm{shift}}$.

Now, we want to gauge the $\mathbb{Z}_3$ intertwined shift symmetry, and we denote the $\mathbb{Z}_3$ one-form gauge field $B^{(1)}$. Since it acts on the one-form gauge field $a_\ell$, it should be related to the $\mathbb{Z}_3$ two-form gauge field $B$ for $SU(3)/\mathbb{Z}_3$ flavor symmetry. Indeed, $B$ and $B^{(1)}$ is related by 
\be
B=B^{(1)}\wedge L^{-1}\diff t. 
\ee
We thus denote the partition function with $B^{(1)}$ as $Z_{M_1\times S^1}[B^{(1)}]$. 
Using this result, we can obtain the mixed 't Hooft anomaly and global inconsistency of $S^1$-compactified theory just by substituting this correspondence into the 't~Hooft anomaly and global inconsistency in two dimensions~\cite{Tanizaki:2017qhf}: 
Under $\mathbb{Z}_3$ permutation, $\bm\phi_\ell\mapsto \bm\phi_{\ell+1}$, the $SU(3)/\mathbb{Z}_3$-$\mathbb{Z}_3$ anomaly at $(\theta_1,\theta_3)=(2\pi/3,-2\pi/3)$ implies 
\be
Z_{M_1\times S^1}[B^{(1)}]\mapsto Z_{M_1\times S^1}[B^1]\exp\left(-\im \int_{M_1}B^{(1)}\right),  
\ee
and the $(\mathbb{Z}_3)_{\mathrm{shift}}$-$(\mathbb{Z}_3)_{\mathrm{permutation}}$ anomaly is found for the $S^1$-compactified theory. Similarly, the $(\mathbb{Z}_3)_{\mathrm{shift}}$-$\mathsf{C}$ global inconsistency can be found from the $SU(3)/\mathbb{Z}_3$-$\mathsf{C}$ global inconsistency. 

We have shown that the phase diagram of the $S^1$-compactified theory with the boundary condition (\ref{eq:S1_compactification}) is constrained by the same 't~Hooft anomaly and global inconsistency with that of $2$-dimensional $SU(3)/[U(1)\times U(1)]$ sigma model. 
Since the phase diagrams in Fig.~\ref{fig:inconsistency} and Fig.~\ref{fig:phase_diagrams} are found by the $SU(3)/\mathbb{Z}_3$-$\mathbb{Z}_3$ anomaly and the $SU(3)/\mathbb{Z}_3$-$\mathsf{C}$ global inconsistency matching arguments, we claim that the circle-compactified model will have the same structure of the phase diagram. 
It is thus an interesting future study to consider the analytic semiclassical computation of this model in order to get more physical insights on the $SU(3)/[U(1)\times U(1)]$ sigma model. 

\section{The 2+1D systems and domain walls}\label{sec:2+1D}
We here briefly discuss the anomalies of the QFT system when it is lifted to 2+1D. Our discussion will be cursory, leaving a more detailed discussion for the future. We will restrict ourselves to the case of $N=3$. 

In 2+1D we can no longer have $\theta$ terms. Instead the two $U(1)$ gauge fields now have a $[U(1)\times U(1)]_T$ topological symmetry, generated by the charges 
\be
Q_{1,2}=\frac{1}{2\pi}\int \diff a_{1,2}\;.
\ee
The relevant $U(1)$ Noether currents are just
\be
j_{1,2}=\frac{1}{2\pi}\star \diff a_{1,2}\;.
\ee

If we now couple the currents to a background gauge fields via the minimal coupling, we have to add a term
\be
\mathcal S=\frac{\im}{2\pi} \int_{M_3} A_1\wedge \diff a_1+\frac{\im}{2\pi} \int_{M_3} A_2\wedge \diff a_2\;.
\ee
The above action is invariant under the $U(1)\times U(1)$ gauge transformation sending $A_{1,2}\rightarrow A_{1,2}+\diff \varphi_{1,2}$ because of the quantization of the fluxes $\int \diff a_1$ and $\int \diff a_2\in 2\pi\mathbb{Z}$. However if we now gauge the $PSU(3)$ symmetry, the fluxes will fail to be quantized in multiples of $2\pi$ by the amount $\int B\in {2\pi\over 3}\mathbb{Z}$, where $B\in H^2(M_3,\pi_1(PSU(3)))$ is the $\mathbb{Z}_3$ 2-form gauge field, indicating an anomaly. However if we gauge transform with the choice $\varphi_1=-\varphi_2=\varphi$, the action is still invariant. This indicates that while there is an anomaly between the diagonal part $U(1)_V\subset [U(1)\times U(1)]_T$, which we will refer to as the ``vector'' part of the global topological symmetry, there is no anomaly involving only the $U(1)_A$  ($A$ is for ``\emph{axial}'') symmetry which is generated by the conserved charge $Q_1-Q_2$, and the $PSU(3)$ spin symmetry only. So there is a mixed 't Hooft anomaly between the $U(1)_V$ and the $PSU(3)$ global symmetries. 

Let us therefore gauge the $U(1)_V$ symmetry, by setting the vector-like gauge field $A_1=A_2=V$: $\mathcal S={\im\over 2\pi}\int V\wedge \{(\diff a_1+B)+(\diff a_2+B)\}$. The gauge transformation $V\rightarrow V+\diff \varphi$ causes a change in the action
\be\label{eq:DS_topV}
\Delta \mathcal S=\frac{\im}{2\pi}\int \diff \varphi\wedge (2B)\;\bmod 2\pi\im.
\ee
To fix this, we may consider adding a term $\frac{i}{2\pi}\int V\wedge B$ which would make the action invariant under the $V\rightarrow  V+\diff \varphi$, but term is not gauge invariant under the transformation $B\rightarrow B+\diff \xi$, where $\xi$ is a $U(1)$ gauge field. 
{ In order to achieve the invariance under both gauge transformations, we must put the $(3+1)$D SPT action, $S_{4\mathrm{D}}={\im \over 2\pi}\int V\wedge \diff B$. }

The background fields $A_1,A_2$ generally break explicitly the $\mathbb Z_3$ exchange symmetry, which takes $a_1\rightarrow a_2$ and $a_2\rightarrow -a_1-a_2$.
{ Note that this breaking of $\mathbb{Z}_3$ permutation is not subject to 't~Hooft anomaly matching, although it is the breaking of symmetry due to the background gauge field. As an example, let us again gauge the vector part, $A_1=A_2=V$, then the $\mathbb{Z}_3$ permutation changes the action as 
\bea
\mathcal{S}\mapsto \mathcal{S}-{\im\over 2\pi}\int V\wedge \{2(\diff a_1+B)+(\diff a_2+B)\}. 
\eea
Since the breaking term of the symmetry contains the dynamical gauge fields, we cannot prepare the $(3+1)$D SPT phase canceling this anomaly, and thus this is not a 't~Hooft anomaly\footnote{In Ref.~\cite{Kapustin:2014zva}, this is called the 't~Hooft anomaly \textit{not} of Dijkgraaf-Witten type. The usual 't~Hooft anomaly corresponds to the 't~Hooft anomaly of the Dijkgraaf-Witten type. }.  
}
But if we define $\mathbb Z_3: A_1\rightarrow A_2-A_1, A_2\rightarrow -A_1$, the action is $\mathbb Z_3$ invariant. Still a generic fixed background of the $A_1,A_2$ fields will break the $\mathbb Z_3$ cyclic permutation symmetry.

Consider a $\mathbb Z_3$ preserving background, given by the axial-vector-like gauge field $A_1=-A_2=A$, where $A$ is a $\mathbb Z_3$ gauge field now, i.e. it can be written as $3A=\diff\alpha$, where $\alpha\in [0,2\pi)$ is an angle-valued field.  We then have the action,
\be
\mathcal S={\im\over 2\pi}\int A\wedge (\diff a_1-\diff a_2). 
\ee
The above action corresponds to gauging a $\mathbb Z_3^A\subset U(1)_A$. Indeed, it can be checked that because of the $2\pi$ quantization of the fluxes $F_{1,2}$, such a background preserves the $\mathbb Z_3$ cyclic permutation symmetry. However if we now gauge the $PSU(3)$ symmetry, the cyclic permutation symmetry will induce a change in the action
\be
\Delta\mathcal S= \frac{\im}{2\pi}\int 3A\wedge (\diff a_2+B)=\frac{3\im }{2\pi}\int A\wedge B\bmod 2\pi \im\;,
\ee
which indicates an anomaly among three symmetries; the $\mathbb Z_3^A$, the $\mathbb Z_3$ cyclic permutation symmetry and $PSU(3)$. 

So we have found two anomalies, which both must be saturated. Both of them include the $PSU(3)$ spin-symmetry, and so both can be saturated by breaking the $PSU(3)$. This is the N\'eel phase which, unsurprisingly, saturates both anomalies. If $PSU(3)$ symmetry is restored, barring topologically ordered phase, the vector topological symmetry must be broken. At the same time, either the $\mathbb{Z}_3$ axial symmetry or the $\mathbb Z_3$ cyclic permutation symmetry must be spontaneously broken. 

In a realistic spin system, the $U(1)\times U(1)$ topological symmetry will be explicitly broken to some discrete subgroup. Let us assume that the only $\mathbb Z_n^V\subset U(1)_V$ survives. Then we have that the $A_1=A_2=V$, where now $V$ is a $\mathbb Z_n$ gauge field (i.e. $nV=\diff \alpha$). However now we are allowed a local counter-term of the form
\be
\mathcal S_{\mathrm{counter}}={\im n p\over 2\pi}\int V\wedge B\;, p\in \mathbb Z\;, 
\ee
which is invariant under the $PSU(3)$ gauge transformation, $B\mapsto B+\diff \xi$. 
If we can satisfy the condition
\be
pn=2\bmod 3\;,
\ee
the anomaly (\ref{eq:DS_topV}) can be canceled. 
When $n=3k$ for $k\in\mathbb Z$, the above condition can never be satisfied. So we conclude that as long as the symmetry $\mathbb Z_{3k}\subset U(1)_V$ is preserved, the anomaly between $U(1)_V$ and $PSU(3)$ persists. If $n=3k+l$, where $l=1,2$ we can choose $p=2,1$ respectively and so there is no anomaly. On the other hand, we have already seen that there is an anomaly between the $\mathbb Z_3^A\subset U(1)_A$, the $\mathbb Z_3$ cyclic permutation and the $PSU(3)$ symmetry. 

We expect that this $(2+1)$D model corresponds to an effective theory of some $SU(3)$ quantum magnet. The anomaly in such systems is saturated either by breaking $PSU(3)$ symmetry (the N\'eel order) or by breaking the topological or $\mathbb Z_3$ cyclic permutation symmetry, which is related to the breaking of lattice symmetries and onset of the valence-bond-solid (VBS) order. It is interesting to explore its phase diagram and relation to the microscopic theory, as well as whether the N\'eel to VBS transition supports quantum criticality which was proposed for the $SU(2)$ spin systems \cite{Senthil1490}. 
Furthermore, when discrete global symmetries are spontaneously broken, there exist domain walls connecting different vacua. Under the setup with the domain walls, we can perform the anomaly inflow argument to uncover the property of domain walls.  We leave these interesting subjects for future works. 

\section{Conclusion}\label{sec:conclusion}
In this work we discussed a number of particular quantum field theories in 1+1D, which are related to the antiferromagnetic $SU(3)$ chains in the $p$-box symmetric representations. In the large $p$ limit, the effective model is an $SU(3)/[U(1)\times U(1)]$ nonlinear sigma model~\cite{Lajko:2017wif}. To consider a general spin chain, our discussion covers a linearized version, given by a particular $U(1)^{2}$ Abelian-Higgs model, with a $PSU(3)$ global (spin) symmetry.

These models have two theta angles, and we studied the phase diagram of those theta angles by using not only the 't~Hooft anomaly matching but also the global inconsistency matching. 
These findings are generalized to $SU(N)/U(1)^{N-1}$ nonlinear sigma models and their linearized cousins. 

We first found the $SU(3)/\mathbb{Z}_3$-$\mathbb{Z}_3$ mixed 't~Hooft anomaly for special theta angles, which provides the field-theoretic description of the LSM theorem for $SU(N)$ spin chains~\cite{Affleck:1986pq, Lajko:2017wif}. 
The anomaly matching tells us that the ground states must be three-fold degenerate or there must exist gappless excitations, so the symmetric gapped vacuum is ruled out from possible low-energy behaviors. 
We also found that distinct regions of the phase diagram are globally inconsistent $SU(3)/\mathbb{Z}_3$-$\mathsf{C}$, indicating the presence of the phase transition lines in the phase diagram. We discussed possible scenarios which satisfy the global inconsistency and the anomaly matching. A minimal scenario is consistent with the proposal of Ref.~\cite{Lajko:2017wif} as well as the calculation of the pure-gauge limit of the linear sigma model, 
with phase-transition lying along the charge-conjugation-invariant lines; the charge-conjugation symmetry is spontaneously broken on the phase transition lines. However, the global inconsistency matching also leaves open another possibility; along the charge-conjugation-invariant lines the ground state could be a nontrivial SPT phase protected by $SU(3)/\mathbb{Z}_3$, which means that it must be separated from the origin of the phase-diagram by a phase transition line.  
These two conditions, combined with the $2\pi$ periodicity of theta angles, restrict the possible phase diagrams strongly to these two scenarios.

At the nontrivial $\mathbb{Z}_N$ symmetric point, the $SU(N)/U(1)^{N-1}$ nonlinear sigma model is believed to show conformal behavior. 
We therefore study the $SU(N)$ WZW model, and have shown that the level $k$ $SU(N)$ WZW model and the $SU(N)/U(1)^{N-1}$ sigma model at $\theta_\ell=2\pi p \ell/N$ have the same 't~Hooft anomaly if $k q=p$ mod $N$ for some $q$ with $\gcd(N,q)=1$. 
Combining the constraint from the $c$-theorem, we conjecture that if $SU(N)/U(1)^{N-1}$ sigma model is conformal then it is generically described by $SU(N)_{\gcd(N,p)}$ WZW model. 

We constructed the linear sigma model corresponding to the $SU(N)/U(1)^{N-1}$ nonlinear sigma model, and showed that they have the same 't Hooft anomaly and global inconsistency explicitly. 
In certain limits of linear sigma models, we can perform the analytic computation of the partition function. It therefore provides an intuitive and concrete understandings on how the anomaly and global inconsistency matching is realized, and we have checked the conjecture on the phase diagram. 
Study of nonlinear sigma models is usually tough because of the asymptotic freedom, so we also consider the adiabatic circle compactification of the model. We have shown that the 't~Hooft anomaly and global inconsistency persist under this circle compactification, and thus it is an interesting future study to analyze this circle-compactified model using reliable semiclassical analysis. 

{
We also briefly discussed the $(2+1)$-dimensional version of the model. It is expected to describe the $SU(3)$ quantum spin magnet in two spatial dimension. We show that it has various 't~Hooft anomalies involving topological symmetries, generated by the conserved abelian fluxes. Such setups leave open the possibilities of nontrivial domain walls, like the ones discussed in Refs.~\cite{Anber:2015kea,Sulejmanpasic:2016uwq,Komargodski:2017smk}.
}

\vphantom{a}

\noindent{\bf Note added:} Finalizing the draft, the authors noticed that Ref.~\cite{Yao:2018kel} appears on arXiv, which partially overlaps with Sec.~\ref{sec:WZWmodel}.

\begin{acknowledgments}
The authors thank Ian Affleck, Gerald Dunne, Yuta Kikuchi, Zohar Komargodski, Mikl\'os Lajk\'o  and Fr\'ed\'eric Mila for their comments.
 The authors also thank Philippe Lecheminant for telling us the conjecture about the $SU(N)$ Wess-Zumino-Witten model in Ref.~\cite{Lecheminant:2015iga}.
The authors would like to thank the hospitality at the KITP institute and the organizers of the ``Resurgent Asymptotics in Physics and Mathematics'' workshop where a part of this was started. T.~S. would like to thank Hartmut Wittig and the Institute for Nuclear Physics at Mainz University for their hospitality during the time that this work was completed. The work at KITP is supported by the National Science Foundation under Grant No. NSF PHY17-48958. 
The work of Y.~T. is supported by RIKEN Special Postdoctoral Program.

\end{acknowledgments}

\appendix

\bibliographystyle{utphys}
\bibliography{./QFT,./refs}

\providecommand{\href}[2]{#2}\begingroup\raggedright\begin{thebibliography}{100}

\bibitem{Haldane:1982rj}
F.~D.~M. Haldane, ``{Continuum dynamics of the 1-D Heisenberg antiferromagnetic
  identification with the O(3) nonlinear sigma model},''
\href{http://dx.doi.org/10.1016/0375-9601(83)90631-X}{{\em Phys. Lett.}
  {\bfseries A93} (1983) 464--468}.

\bibitem{Haldane:1983ru}
F.~D.~M. Haldane, ``{Nonlinear field theory of large spin Heisenberg
  antiferromagnets. Semiclassically quantized solitons of the one-dimensional
  easy Axis Neel state},''
\href{http://dx.doi.org/10.1103/PhysRevLett.50.1153}{{\em Phys. Rev. Lett.}
  {\bfseries 50} (1983) 1153--1156}.

\bibitem{Lieb:1961fr}
E.~H. Lieb, T.~Schultz, and D.~Mattis, ``{Two soluble models of an
  antiferromagnetic chain},''
\href{http://dx.doi.org/10.1016/0003-4916(61)90115-4}{{\em Annals Phys.}
  {\bfseries 16} (1961) 407--466}.

\bibitem{Affleck:1986pq}
I.~Affleck and E.~H. Lieb, ``{A Proof of Part of Haldane's Conjecture on Spin
  Chains},''
\href{http://dx.doi.org/10.1007/BF00400304}{{\em Lett. Math. Phys.} {\bfseries
  12} (1986) 57}.

\bibitem{PhysRevLett.84.1535}
M.~Oshikawa, ``Commensurability, excitation gap, and topology in quantum
  many-particle systems on a periodic lattice,''
  \href{http://dx.doi.org/10.1103/PhysRevLett.84.1535}{{\em Phys. Rev. Lett.}
  {\bfseries 84} (Feb, 2000) 1535--1538},
  \href{http://arxiv.org/abs/cond-mat/9911137}{{\ttfamily
  arXiv:cond-mat/9911137 [cond-mat.str-el]}}.

\bibitem{Hastings:2003zx}
M.~B. Hastings, ``{Lieb-Schultz-Mattis in higher dimensions},''
  \href{http://dx.doi.org/10.1103/PhysRevB.69.104431}{{\em Phys. Rev.}
  {\bfseries B69} (2004) 104431},
\href{http://arxiv.org/abs/cond-mat/0305505}{{\ttfamily arXiv:cond-mat/0305505
  [cond-mat]}}.

\bibitem{Bethe:1931hc}
H.~Bethe, ``{On the theory of metals. 1. Eigenvalues and eigenfunctions for the
  linear atomic chain},''
\href{http://dx.doi.org/10.1007/BF01341708}{{\em Z. Phys.} {\bfseries 71}
  (1931) 205--226}.

\bibitem{Affleck:1987vf}
I.~Affleck, T.~Kennedy, E.~H. Lieb, and H.~Tasaki, ``{Rigorous Results on
  Valence Bond Ground States in Antiferromagnets},''
\href{http://dx.doi.org/10.1103/PhysRevLett.59.799}{{\em Phys. Rev. Lett.}
  {\bfseries 59} (1987) 799}.

\bibitem{Affleck:1985wb}
I.~Affleck, ``{Exact Critical Exponents for Quantum Spin Chains, Nonlinear
  Sigma Models at Theta = pi and the Quantum Hall Effect},''
\href{http://dx.doi.org/10.1016/0550-3213(86)90167-7}{{\em Nucl. Phys.}
  {\bfseries B265} (1986) 409--447}.

\bibitem{Affleck:1988wz}
I.~Affleck, ``{Critical Behavior of SU($n$) Quantum Chains and Topological
  Nonlinear $\sigma$ Models},''
\href{http://dx.doi.org/10.1016/0550-3213(88)90117-4}{{\em Nucl. Phys.}
  {\bfseries B305} (1988) 582--596}.

\bibitem{PhysRevLett.91.186402}
C.~Wu, J.-p. Hu, and S.-c. Zhang, ``{Exact SO(5) Symmetry in the Spin-$3/2$
  Fermionic System},''
  \href{http://dx.doi.org/10.1103/PhysRevLett.91.186402}{{\em Phys. Rev. Lett.}
  {\bfseries 91} (Oct, 2003) 186402}.

\bibitem{PhysRevLett.92.170403}
C.~Honerkamp and W.~Hofstetter, ``{Ultracold Fermions and the $\mathrm{SU}(N)$
  Hubbard Model},'' \href{http://dx.doi.org/10.1103/PhysRevLett.92.170403}{{\em
  Phys. Rev. Lett.} {\bfseries 92} (Apr, 2004) 170403}.

\bibitem{1367-2630-11-10-103033}
M.~A. Cazalilla, A.~F. Ho, and M.~Ueda, ``Ultracold gases of ytterbium:
  ferromagnetism and mott states in an su(6) fermi system,''
  \href{http://dx.doi.org/10.1088/1367-2630/11/10/103033}{{\em New Journal of
  Physics} {\bfseries 11} no.~10, (2009) 103033}.

\bibitem{gorshkov2010two}
A.~V. Gorshkov, M.~Hermele, V.~Gurarie, C.~Xu, P.~S. Julienne, J.~Ye,
  P.~Zoller, E.~Demler, M.~D. Lukin, and A.~Rey, ``Two-orbital su (n) magnetism
  with ultracold alkaline-earth atoms,'' {\em Nature Physics} {\bfseries 6}
  no.~4, (2010) 289.

\bibitem{PhysRevB.86.224409}
S.~Bieri, M.~Serbyn, T.~Senthil, and P.~A. Lee, ``Paired chiral spin liquid
  with a fermi surface in $s=1$ model on the triangular lattice,''
  \href{http://dx.doi.org/10.1103/PhysRevB.86.224409}{{\em Phys. Rev. B}
  {\bfseries 86} (Dec, 2012) 224409}.

\bibitem{scazza2014observation}
F.~Scazza, C.~Hofrichter, M.~H{\"o}fer, P.~De~Groot, I.~Bloch, and
  S.~F{\"o}lling, ``{Observation of two-orbital spin-exchange interactions with
  ultracold SU (N)-symmetric fermions},'' {\em Nature Physics} {\bfseries 10}
  no.~10, (2014) 779.

\bibitem{taie20126}
S.~Taie, R.~Yamazaki, S.~Sugawa, and Y.~Takahashi, ``{An SU (6) Mott insulator
  of an atomic Fermi gas realized by large-spin Pomeranchuk cooling},'' {\em
  Nature Physics} {\bfseries 8} no.~11, (2012) 825.

\bibitem{pagano2014one}
G.~Pagano, M.~Mancini, G.~Cappellini, P.~Lombardi, F.~Sch{\"a}fer, H.~Hu, X.-J.
  Liu, J.~Catani, C.~Sias, M.~Inguscio, {\em et~al.}, ``{A one-dimensional
  liquid of fermions with tunable spin},'' {\em Nature Physics} {\bfseries 10}
  no.~3, (2014) 198.

\bibitem{Zhang1467}
X.~Zhang, M.~Bishof, S.~L. Bromley, C.~V. Kraus, M.~S. Safronova, P.~Zoller,
  A.~M. Rey, and J.~Ye, ``{Spectroscopic observation of SU(N)-symmetric
  interactions in Sr orbital magnetism},''
  \href{http://dx.doi.org/10.1126/science.1254978}{{\em Science} {\bfseries
  345} no.~6203, (2014) 1467--1473}.

\bibitem{0034-4885-77-12-124401}
M.~Cazalilla and A.~M. Rey, ``{Ultracold Fermi gases with emergent SU(N)
  symmetry},'' {\em Reports on Progress in Physics} {\bfseries 77} no.~12,
  (2014) 124401.

\bibitem{CAPPONI201650}
S.~Capponi, P.~Lecheminant, and K.~Totsuka, ``Phases of one-dimensional su(n)
  cold atomic fermi gases—from molecular luttinger liquids to topological
  phases,'' \href{http://dx.doi.org/10.1016/j.aop.2016.01.011}{{\em Annals of
  Physics} {\bfseries 367} (2016) 50 -- 95}.

\bibitem{Bykov:2011ai}
D.~Bykov, ``{Haldane limits via Lagrangian embeddings},''
  \href{http://dx.doi.org/10.1016/j.nuclphysb.2011.10.005}{{\em Nucl. Phys.}
  {\bfseries B855} (2012) 100--127},
\href{http://arxiv.org/abs/1104.1419}{{\ttfamily arXiv:1104.1419 [hep-th]}}.

\bibitem{Bykov:2012am}
D.~Bykov, ``{The geometry of antiferromagnetic spin chains},''
  \href{http://dx.doi.org/10.1007/s00220-013-1702-5}{{\em Commun. Math. Phys.}
  {\bfseries 322} (2013) 807--834},
\href{http://arxiv.org/abs/1206.2777}{{\ttfamily arXiv:1206.2777 [hep-th]}}.

\bibitem{Lajko:2017wif}
M.~Lajk\'o, K.~Wamer, F.~Mila, and I.~Affleck, ``{Generalization of the Haldane
  conjecture to SU(3) chains},''
  \href{http://dx.doi.org/10.1016/j.nuclphysb.2017.09.015}{{\em Nucl. Phys.}
  {\bfseries B924} (2017) 508--577},
\href{http://arxiv.org/abs/1706.06598}{{\ttfamily arXiv:1706.06598
  [cond-mat.str-el]}}.

\bibitem{PhysRevD.77.056008}
B.~All\'es and A.~Papa, ``Mass gap in the 2d o(3) nonlinear sigma model with a
  $\ensuremath{\theta}=\ensuremath{\pi}$ term,''
  \href{http://dx.doi.org/10.1103/PhysRevD.77.056008}{{\em Phys. Rev. D}
  {\bfseries 77} (2008) 056008}.

\bibitem{tHooft:1979rat}
G.~'t~Hooft,
  \href{http://dx.doi.org/10.1007/978-1-4684-7571-5_9}{``{Naturalness, chiral
  symmetry, and spontaneous chiral symmetry breaking},''} in {\em {Recent
  Developments in Gauge Theories. Proceedings, Nato Advanced Study Institute,
  Cargese, France, August 26 - September 8, 1979}}, vol.~59, pp.~135--157.
\newblock
1980.
\newblock

\bibitem{Frishman:1980dq}
Y.~Frishman, A.~Schwimmer, T.~Banks, and S.~Yankielowicz, ``{The Axial Anomaly
  and the Bound State Spectrum in Confining Theories},''
\href{http://dx.doi.org/10.1016/0550-3213(81)90268-6}{{\em Nucl. Phys.}
  {\bfseries B177} (1981) 157--171}.

\bibitem{Coleman:1982yg}
S.~R. Coleman and B.~Grossman, ``{'t Hooft's Consistency Condition as a
  Consequence of Analyticity and Unitarity},''
\href{http://dx.doi.org/10.1016/0550-3213(82)90028-1}{{\em Nucl. Phys.}
  {\bfseries B203} (1982) 205--220}.

\bibitem{Vishwanath:2012tq}
A.~Vishwanath and T.~Senthil, ``{Physics of three dimensional bosonic
  topological insulators: Surface Deconfined Criticality and Quantized
  Magnetoelectric Effect},''
  \href{http://dx.doi.org/10.1103/PhysRevX.3.011016}{{\em Phys. Rev.}
  {\bfseries X3} no.~1, (2013) 011016},
\href{http://arxiv.org/abs/1209.3058}{{\ttfamily arXiv:1209.3058
  [cond-mat.str-el]}}.

\bibitem{Wen:2013oza}
X.-G. Wen, ``{Classifying gauge anomalies through symmetry-protected trivial
  orders and classifying gravitational anomalies through topological orders},''
  \href{http://dx.doi.org/10.1103/PhysRevD.88.045013}{{\em Phys. Rev.}
  {\bfseries D88} no.~4, (2013) 045013},
\href{http://arxiv.org/abs/1303.1803}{{\ttfamily arXiv:1303.1803 [hep-th]}}.

\bibitem{Kapustin:2014lwa}
A.~Kapustin and R.~Thorngren, ``{Anomalies of discrete symmetries in three
  dimensions and group cohomology},''
  \href{http://dx.doi.org/10.1103/PhysRevLett.112.231602}{{\em Phys. Rev.
  Lett.} {\bfseries 112} no.~23, (2014) 231602},
\href{http://arxiv.org/abs/1403.0617}{{\ttfamily arXiv:1403.0617 [hep-th]}}.

\bibitem{Kapustin:2014zva}
A.~Kapustin and R.~Thorngren, ``{Anomalies of discrete symmetries in various
  dimensions and group cohomology},''
\href{http://arxiv.org/abs/1404.3230}{{\ttfamily arXiv:1404.3230 [hep-th]}}.

\bibitem{Cho:2014jfa}
G.~Y. Cho, J.~C.~Y. Teo, and S.~Ryu, ``{Conflicting Symmetries in Topologically
  Ordered Surface States of Three-dimensional Bosonic Symmetry Protected
  Topological Phases},''
  \href{http://dx.doi.org/10.1103/PhysRevB.89.235103}{{\em Phys. Rev.}
  {\bfseries B89} no.~23, (2014) 235103},
\href{http://arxiv.org/abs/1403.2018}{{\ttfamily arXiv:1403.2018
  [cond-mat.str-el]}}.

\bibitem{Wang:2014pma}
J.~C. Wang, Z.-C. Gu, and X.-G. Wen, ``{Field theory representation of
  gauge-gravity symmetry-protected topological invariants, group cohomology and
  beyond},'' \href{http://dx.doi.org/10.1103/PhysRevLett.114.031601}{{\em Phys.
  Rev. Lett.} {\bfseries 114} no.~3, (2015) 031601},
\href{http://arxiv.org/abs/1405.7689}{{\ttfamily arXiv:1405.7689
  [cond-mat.str-el]}}.

\bibitem{Witten:2015aba}
E.~Witten, ``{Fermion Path Integrals And Topological Phases},''
  \href{http://dx.doi.org/10.1103/RevModPhys.88.035001}{{\em Rev. Mod. Phys.}
  {\bfseries 88} no.~3, (2016) 035001},
\href{http://arxiv.org/abs/1508.04715}{{\ttfamily arXiv:1508.04715
  [cond-mat.mes-hall]}}.

\bibitem{Seiberg:2016rsg}
N.~Seiberg and E.~Witten, ``{Gapped Boundary Phases of Topological Insulators
  via Weak Coupling},'' \href{http://dx.doi.org/10.1093/ptep/ptw083}{{\em Prog.
  Theor. Exp. Phys.} {\bfseries 2016} no.~12, (2016) 12C101},
\href{http://arxiv.org/abs/1602.04251}{{\ttfamily arXiv:1602.04251
  [cond-mat.str-el]}}.

\bibitem{Witten:2016cio}
E.~Witten, ``{The "Parity" Anomaly On An Unorientable Manifold},''
  \href{http://dx.doi.org/10.1103/PhysRevB.94.195150}{{\em Phys. Rev.}
  {\bfseries B94} no.~19, (2016) 195150},
\href{http://arxiv.org/abs/1605.02391}{{\ttfamily arXiv:1605.02391 [hep-th]}}.

\bibitem{Tachikawa:2016cha}
Y.~Tachikawa and K.~Yonekura, ``{On time-reversal anomaly of 2+1d topological
  phases},'' \href{http://dx.doi.org/10.1093/ptep/ptx010}{{\em PTEP} {\bfseries
  2017} no.~3, (2017) 033B04},
\href{http://arxiv.org/abs/1610.07010}{{\ttfamily arXiv:1610.07010 [hep-th]}}.

\bibitem{Tachikawa:2016nmo}
Y.~Tachikawa and K.~Yonekura, ``{More on time-reversal anomaly of 2+1d
  topological phases},''
  \href{http://dx.doi.org/10.1103/PhysRevLett.119.111603}{{\em Phys. Rev.
  Lett.} {\bfseries 119} no.~11, (2017) 111603},
\href{http://arxiv.org/abs/1611.01601}{{\ttfamily arXiv:1611.01601 [hep-th]}}.

\bibitem{Gaiotto:2017yup}
D.~Gaiotto, A.~Kapustin, Z.~Komargodski, and N.~Seiberg, ``{Theta, Time
  Reversal, and Temperature},''
  \href{http://dx.doi.org/10.1007/JHEP05(2017)091}{{\em JHEP} {\bfseries 05}
  (2017) 091},
\href{http://arxiv.org/abs/1703.00501}{{\ttfamily arXiv:1703.00501 [hep-th]}}.

\bibitem{Wang:2017txt}
C.~Wang, A.~Nahum, M.~A. Metlitski, C.~Xu, and T.~Senthil, ``{Deconfined
  quantum critical points: symmetries and dualities},''
  \href{http://dx.doi.org/10.1103/PhysRevX.7.031051}{{\em Phys. Rev.}
  {\bfseries X7} no.~3, (2017) 031051},
\href{http://arxiv.org/abs/1703.02426}{{\ttfamily arXiv:1703.02426
  [cond-mat.str-el]}}.

\bibitem{Tanizaki:2017bam}
Y.~Tanizaki and Y.~Kikuchi, ``{Vacuum structure of bifundamental gauge theories
  at finite topological angles},''
  \href{http://dx.doi.org/10.1007/JHEP06(2017)102}{{\em JHEP} {\bfseries 06}
  (2017) 102},
\href{http://arxiv.org/abs/1705.01949}{{\ttfamily arXiv:1705.01949 [hep-th]}}.

\bibitem{Komargodski:2017dmc}
Z.~Komargodski, A.~Sharon, R.~Thorngren, and X.~Zhou, ``{Comments on Abelian
  Higgs Models and Persistent Order},''
\href{http://arxiv.org/abs/1705.04786}{{\ttfamily arXiv:1705.04786 [hep-th]}}.

\bibitem{Komargodski:2017smk}
Z.~Komargodski, T.~Sulejmanpasic, and M.~Unsal, ``{Walls, anomalies, and
  deconfinement in quantum antiferromagnets},''
  \href{http://dx.doi.org/10.1103/PhysRevB.97.054418}{{\em Phys. Rev.}
  {\bfseries B97} no.~5, (2018) 054418},
\href{http://arxiv.org/abs/1706.05731}{{\ttfamily arXiv:1706.05731
  [cond-mat.str-el]}}.

\bibitem{Cho:2017fgz}
G.~Y. Cho and S.~Ryu, ``{Anomaly Manifestation of Lieb-Schultz-Mattis Theorem
  and Topological Phases},''
  \href{http://dx.doi.org/10.1103/PhysRevB.96.195105}{{\em Phys. Rev.}
  {\bfseries B96} no.~19, (2017) 195105},
\href{http://arxiv.org/abs/1705.03892}{{\ttfamily arXiv:1705.03892
  [cond-mat.str-el]}}.

\bibitem{Shimizu:2017asf}
H.~Shimizu and K.~Yonekura, ``{Anomaly constraints on deconfinement and chiral
  phase transition},'' \href{http://dx.doi.org/10.1103/PhysRevD.97.105011}{{\em
  Phys. Rev.} {\bfseries D97} no.~10, (2018) 105011},
\href{http://arxiv.org/abs/1706.06104}{{\ttfamily arXiv:1706.06104 [hep-th]}}.

\bibitem{Wang:2017loc}
J.~Wang, X.-G. Wen, and E.~Witten, ``{Symmetric Gapped Interfaces of SPT and
  SET States: Systematic Constructions},''
\href{http://arxiv.org/abs/1705.06728}{{\ttfamily arXiv:1705.06728
  [cond-mat.str-el]}}.

\bibitem{Metlitski:2017fmd}
M.~A. Metlitski and R.~Thorngren, ``{Intrinsic and emergent anomalies at
  deconfined critical points},''
\href{http://arxiv.org/abs/1707.07686}{{\ttfamily arXiv:1707.07686
  [cond-mat.str-el]}}.

\bibitem{Kikuchi:2017pcp}
Y.~Kikuchi and Y.~Tanizaki, ``{Global inconsistency, 't~Hooft anomaly, and
  level crossing in quantum mechanics},''
  \href{http://dx.doi.org/10.1093/ptep/ptx148}{{\em Prog. Theor. Exp. Phys.}
  {\bfseries 2017} (2017) 113B05},
\href{http://arxiv.org/abs/1708.01962}{{\ttfamily arXiv:1708.01962 [hep-th]}}.

\bibitem{Gaiotto:2017tne}
D.~Gaiotto, Z.~Komargodski, and N.~Seiberg, ``{Time-reversal breaking in
  QCD$_{4}$, walls, and dualities in 2 + 1 dimensions},''
  \href{http://dx.doi.org/10.1007/JHEP01(2018)110}{{\em JHEP} {\bfseries 01}
  (2018) 110},
\href{http://arxiv.org/abs/1708.06806}{{\ttfamily arXiv:1708.06806 [hep-th]}}.

\bibitem{Tanizaki:2017qhf}
Y.~Tanizaki, T.~Misumi, and N.~Sakai, ``{Circle compactification and 't Hooft
  anomaly},'' \href{http://dx.doi.org/10.1007/JHEP12(2017)056}{{\em JHEP}
  {\bfseries 12} (2017) 056},
\href{http://arxiv.org/abs/1710.08923}{{\ttfamily arXiv:1710.08923 [hep-th]}}.

\bibitem{Tanizaki:2017mtm}
Y.~Tanizaki, Y.~Kikuchi, T.~Misumi, and N.~Sakai, ``{Anomaly matching for phase
  diagram of massless $\mathbb{Z}_N$-QCD},''
  \href{http://dx.doi.org/10.1103/PhysRevD.97.054012}{{\em Phys. Rev.}
  {\bfseries D97} (2018) 054012},
\href{http://arxiv.org/abs/1711.10487}{{\ttfamily arXiv:1711.10487 [hep-th]}}.

\bibitem{Guo:2017xex}
M.~Guo, P.~Putrov, and J.~Wang, ``{Time Reversal, SU(N) Yang-Mills and
  Cobordisms: Interacting Topological Superconductors/Insulators and Quantum
  Spin Liquids in 3+1D},''
\href{http://arxiv.org/abs/1711.11587}{{\ttfamily arXiv:1711.11587
  [cond-mat.str-el]}}.

\bibitem{Sulejmanpasic:2018upi}
T.~Sulejmanpasic and Y.~Tanizaki, ``{C-P-T anomaly matching in bosonic quantum
  field theory and spin chains},''
  \href{http://dx.doi.org/10.1103/PhysRevB.97.144201}{{\em Phys. Rev.}
  {\bfseries B97} (2018) 144201},
\href{http://arxiv.org/abs/1802.02153}{{\ttfamily arXiv:1802.02153 [hep-th]}}.

\bibitem{Aitken:2018kky}
K.~Aitken, A.~Cherman, and M.~Unsal, ``{Dihedral symmetry in $SU(N)$ Yang-Mills
  theory},''
\href{http://arxiv.org/abs/1804.05845}{{\ttfamily arXiv:1804.05845 [hep-th]}}.

\bibitem{Kobayashi:2018yuk}
R.~Kobayashi, K.~Shiozaki, Y.~Kikuchi, and S.~Ryu, ``{Lieb-Schultz-Mattis type
  theorem with higher-form symmetry and the quantum dimer models},''
\href{http://arxiv.org/abs/1805.05367}{{\ttfamily arXiv:1805.05367
  [cond-mat.stat-mech]}}.

\bibitem{PhysRevA.93.021606}
H.~T. Ueda, Y.~Akagi, and N.~Shannon, ``Quantum solitons with emergent
  interactions in a model of cold atoms on the triangular lattice,''
  \href{http://dx.doi.org/10.1103/PhysRevA.93.021606}{{\em Phys. Rev. A}
  {\bfseries 93} (2016) 021606}.

\bibitem{Witten:2000nv}
E.~Witten, ``{Supersymmetric index in four-dimensional gauge theories},''
  \href{http://dx.doi.org/10.4310/ATMP.2001.v5.n5.a1}{{\em Adv. Theor. Math.
  Phys.} {\bfseries 5} (2002) 841--907},
\href{http://arxiv.org/abs/hep-th/0006010}{{\ttfamily arXiv:hep-th/0006010
  [hep-th]}}.

\bibitem{Kapustin:2014gua}
A.~Kapustin and N.~Seiberg, ``{Coupling a QFT to a TQFT and Duality},''
  \href{http://dx.doi.org/10.1007/JHEP04(2014)001}{{\em JHEP} {\bfseries 04}
  (2014) 001},
\href{http://arxiv.org/abs/1401.0740}{{\ttfamily arXiv:1401.0740 [hep-th]}}.

\bibitem{Gaiotto:2014kfa}
D.~Gaiotto, A.~Kapustin, N.~Seiberg, and B.~Willett, ``{Generalized Global
  Symmetries},'' \href{http://dx.doi.org/10.1007/JHEP02(2015)172}{{\em JHEP}
  {\bfseries 02} (2015) 172},
\href{http://arxiv.org/abs/1412.5148}{{\ttfamily arXiv:1412.5148 [hep-th]}}.

\bibitem{Aharony:2013hda}
O.~Aharony, N.~Seiberg, and Y.~Tachikawa, ``{Reading between the lines of
  four-dimensional gauge theories},''
  \href{http://dx.doi.org/10.1007/JHEP08(2013)115}{{\em JHEP} {\bfseries 08}
  (2013) 115},
\href{http://arxiv.org/abs/1305.0318}{{\ttfamily arXiv:1305.0318 [hep-th]}}.

\bibitem{PhysRevB.83.035107}
X.~Chen, Z.-C. Gu, and X.-G. Wen, ``Classification of gapped symmetric phases
  in one-dimensional spin systems,''
  \href{http://dx.doi.org/10.1103/PhysRevB.83.035107}{{\em Phys. Rev. B}
  {\bfseries 83} (2011) 035107},
  \href{http://arxiv.org/abs/1008.3745}{{\ttfamily arXiv:1008.3745
  [cond-mat.str-el]}}.

\bibitem{Coleman:1973ci}
S.~R. Coleman, ``{There are no Goldstone bosons in two-dimensions},''
\href{http://dx.doi.org/10.1007/BF01646487}{{\em Commun. Math. Phys.}
  {\bfseries 31} (1973) 259--264}.

\bibitem{mermin1966absence}
N.~D. Mermin and H.~Wagner, ``{Absence of ferromagnetism or antiferromagnetism
  in one-or two-dimensional isotropic Heisenberg models},''
  \href{http://dx.doi.org/10.1103/PhysRevLett.17.1133}{{\em Phys.~Rev.~Lett.}
  {\bfseries 17} (1966) 1133}.

\bibitem{Gattringer:2015baa}
C.~Gattringer, T.~Kloiber, and M.~M\"uller-Preussker, ``{Dual simulation of the
  two-dimensional lattice U(1) gauge-Higgs model with a topological term},''
  \href{http://dx.doi.org/10.1103/PhysRevD.92.114508}{{\em Phys. Rev.}
  {\bfseries D92} no.~11, (2015) 114508},
\href{http://arxiv.org/abs/1508.00681}{{\ttfamily arXiv:1508.00681 [hep-lat]}}.

\bibitem{Bruckmann:2015sua}
F.~Bruckmann, C.~Gattringer, T.~Kloiber, and T.~Sulejmanpasic, ``{Dual lattice
  representations for O(N) and CP(N−1) models with a chemical potential},''
  \href{http://dx.doi.org/10.1016/j.physletb.2015.08.015,
  10.1016/j.physletb.2015.10.033}{{\em Phys. Lett.} {\bfseries B749} (2015)
  495--501}, \href{http://arxiv.org/abs/1507.04253}{{\ttfamily arXiv:1507.04253
  [hep-lat]}}.
[Erratum: Phys. Lett.B751,595(2015)].

\bibitem{Bruckmann:2015hua}
F.~Bruckmann, C.~Gattringer, T.~Kloiber, and T.~Sulejmanpasic, ``{Grand
  Canonical Ensembles, Multiparticle Wave Functions, Scattering Data, and
  Lattice Field Theories},''
  \href{http://dx.doi.org/10.1103/PhysRevLett.115.231601}{{\em Phys. Rev.
  Lett.} {\bfseries 115} no.~23, (2015) 231601},
\href{http://arxiv.org/abs/1509.05189}{{\ttfamily arXiv:1509.05189 [hep-lat]}}.

\bibitem{Sulejmanpasic2017}
T.~Sulejmanpasic, ``Lattice 2017 talk: Theta-angle, global symmetries and sign
  problem in compact 1+1d u(1) gauge theories,'' 2017.

\bibitem{PhysRevB.93.155134}
P.~Nataf and F.~Mila, ``{Exact diagonalization of Heisenberg $\mathrm{SU}(N)$
  chains in the fully symmetric and antisymmetric representations},''
  \href{http://dx.doi.org/10.1103/PhysRevB.93.155134}{{\em Phys. Rev. B}
  {\bfseries 93} (2016) 155134}.

\bibitem{PhysRevB.75.060401}
M.~Greiter, S.~Rachel, and D.~Schuricht, ``{Exact results for SU(3) spin
  chains: Trimer states, valence bond solids, and their parent Hamiltonians},''
  \href{http://dx.doi.org/10.1103/PhysRevB.75.060401}{{\em Phys. Rev. B}
  {\bfseries 75} (2007) 060401}.

\bibitem{PhysRevB.75.184441}
M.~Greiter and S.~Rachel, ``{Valence bond solids for $\mathrm{SU}(n)$ spin
  chains: Exact models, spinon confinement, and the Haldane gap},''
  \href{http://dx.doi.org/10.1103/PhysRevB.75.184441}{{\em Phys. Rev. B}
  {\bfseries 75} (2007) 184441}.

\bibitem{PhysRevB.80.180420}
S.~Rachel, R.~Thomale, M.~F\"uhringer, P.~Schmitteckert, and M.~Greiter,
  ``{Spinon confinement and the Haldane gap in $\text{SU}(n)$ spin chains},''
  \href{http://dx.doi.org/10.1103/PhysRevB.80.180420}{{\em Phys. Rev. B}
  {\bfseries 80} (2009) 180420}.

\bibitem{Anber:2015kea}
M.~M. Anber, E.~Poppitz, and T.~Sulejmanpasic, ``{Strings from domain walls in
  supersymmetric Yang-Mills theory and adjoint QCD},''
  \href{http://dx.doi.org/10.1103/PhysRevD.92.021701}{{\em Phys. Rev.}
  {\bfseries D92} no.~2, (2015) 021701},
\href{http://arxiv.org/abs/1501.06773}{{\ttfamily arXiv:1501.06773 [hep-th]}}.

\bibitem{Sulejmanpasic:2016uwq}
T.~Sulejmanpasic, H.~Shao, A.~Sandvik, and M.~Unsal, ``{Confinement in the
  bulk, deconfinement on the wall: infrared equivalence between compactified
  QCD and quantum magnets},''
  \href{http://dx.doi.org/10.1103/PhysRevLett.119.091601}{{\em Phys. Rev.
  Lett.} {\bfseries 119} no.~9, (2017) 091601},
\href{http://arxiv.org/abs/1608.09011}{{\ttfamily arXiv:1608.09011 [hep-th]}}.

\bibitem{chen2011}
X.~Chen, Z.-C. Gu, and X.-G. Wen, ``Classification of gapped symmetric phases
  in one-dimensional spin systems,'' {\em Physical review b} {\bfseries 83}
  no.~3, (2011) 035107.

\bibitem{PhysRevB.93.104425}
Y.~Fuji, ``Effective field theory for one-dimensional valence-bond-solid phases
  and their symmetry protection,''
  \href{http://dx.doi.org/10.1103/PhysRevB.93.104425}{{\em Phys. Rev. B}
  {\bfseries 93} (2016) 104425}.

\bibitem{Wess:1971yu}
J.~Wess and B.~Zumino, ``{Consequences of anomalous Ward identities},''
\href{http://dx.doi.org/10.1016/0370-2693(71)90582-X}{{\em Phys. Lett.}
  {\bfseries 37B} (1971) 95--97}.

\bibitem{Witten:1983tw}
E.~Witten, ``{Global Aspects of Current Algebra},''
\href{http://dx.doi.org/10.1016/0550-3213(83)90063-9}{{\em Nucl. Phys.}
  {\bfseries B223} (1983) 422--432}.

\bibitem{Witten:1983ar}
E.~Witten, ``{Nonabelian Bosonization in Two-Dimensions},''
  \href{http://dx.doi.org/10.1007/BF01215276}{{\em Commun. Math. Phys.}
  {\bfseries 92} (1984) 455--472}.
[,201(1983)].

\bibitem{Gepner:1986wi}
D.~Gepner and E.~Witten, ``{String Theory on Group Manifolds},''
\href{http://dx.doi.org/10.1016/0550-3213(86)90051-9}{{\em Nucl. Phys.}
  {\bfseries B278} (1986) 493--549}.

\bibitem{Furuya:2015coa}
S.~C. Furuya and M.~Oshikawa, ``{Symmetry Protection of Critical Phases and a
  Global Anomaly in $1+1$ Dimensions},''
  \href{http://dx.doi.org/10.1103/PhysRevLett.118.021601}{{\em Phys. Rev.
  Lett.} {\bfseries 118} no.~2, (2017) 021601},
\href{http://arxiv.org/abs/1503.07292}{{\ttfamily arXiv:1503.07292
  [cond-mat.stat-mech]}}.

\bibitem{Numasawa:2017crf}
T.~Numasawa and S.~Yamaguchi, ``{Mixed global anomalies and boundary conformal
  field theories},''
\href{http://arxiv.org/abs/1712.09361}{{\ttfamily arXiv:1712.09361 [hep-th]}}.

\bibitem{Zamolodchikov:1986gt}
A.~B. Zamolodchikov, ``{Irreversibility of the Flux of the Renormalization
  Group in a 2D Field Theory},'' {\em JETP Lett.} {\bfseries 43} (1986)
  730--732.
[Pisma Zh. Eksp. Teor. Fiz.43,565(1986)].

\bibitem{Knizhnik:1984nr}
V.~G. Knizhnik and A.~B. Zamolodchikov, ``{Current Algebra and Wess-Zumino
  Model in Two-Dimensions},''
  \href{http://dx.doi.org/10.1016/0550-3213(84)90374-2}{{\em Nucl. Phys.}
  {\bfseries B247} (1984) 83--103}.
[,690(1984)].

\bibitem{Lecheminant:2015iga}
P.~Lecheminant, ``{Massless renormalization group flow in SU(N)$_k$ perturbed
  conformal field theory},''
  \href{http://dx.doi.org/10.1016/j.nuclphysb.2015.11.004}{{\em Nucl. Phys.}
  {\bfseries B901} (2015) 510--525},
\href{http://arxiv.org/abs/1509.01680}{{\ttfamily arXiv:1509.01680
  [cond-mat.str-el]}}.

\bibitem{Affleck:1987ch}
I.~Affleck and F.~D.~M. Haldane, ``{Critical Theory of Quantum Spin Chains},''
\href{http://dx.doi.org/10.1103/PhysRevB.36.5291}{{\em Phys. Rev.} {\bfseries
  B36} (1987) 5291--5300}.

\bibitem{Polyakov:1975rr}
A.~M. Polyakov, ``{Interaction of Goldstone Particles in Two-Dimensions.
  Applications to Ferromagnets and Massive Yang-Mills Fields},''
\href{http://dx.doi.org/10.1016/0370-2693(75)90161-6}{{\em Phys. Lett.}
  {\bfseries 59B} (1975) 79--81}.

\bibitem{Polyakov:1987ez}
A.~Polyakov, {\em Gauge Fields and Strings (Mathematical Reports,)}.
\newblock CRC Press, 1~ed., 9, 1987.

\bibitem{Witten:1978bc}
E.~Witten, ``{Instantons, the Quark Model, and the 1/n Expansion},''
\href{http://dx.doi.org/10.1016/0550-3213(79)90243-8}{{\em Nucl.Phys.}
  {\bfseries B149} (1979) 285}.

\bibitem{Affleck:1979gy}
I.~Affleck, ``{The Role of Instantons in Scale Invariant Gauge Theories},''
\href{http://dx.doi.org/10.1016/0550-3213(80)90350-8}{{\em Nucl. Phys.}
  {\bfseries B162} (1980) 461--477}.

\bibitem{Dunne:2016nmc}
G.~V. Dunne and M.~Unsal, ``{New Nonperturbative Methods in Quantum Field
  Theory: From Large-N Orbifold Equivalence to Bions and Resurgence},''
  \href{http://dx.doi.org/10.1146/annurev-nucl-102115-044755}{{\em Ann. Rev.
  Nucl. Part. Sci.} {\bfseries 66} (2016) 245--272},
\href{http://arxiv.org/abs/1601.03414}{{\ttfamily arXiv:1601.03414 [hep-th]}}.

\bibitem{Dunne:2012ae}
G.~V. Dunne and M.~Unsal, ``{Resurgence and Trans-series in Quantum Field
  Theory: The CP(N-1) Model},''
  \href{http://dx.doi.org/10.1007/JHEP11(2012)170}{{\em JHEP} {\bfseries 11}
  (2012) 170},
\href{http://arxiv.org/abs/1210.2423}{{\ttfamily arXiv:1210.2423 [hep-th]}}.

\bibitem{Unsal:2007jx}
M.~Unsal, ``{Magnetic bion condensation: A New mechanism of confinement and
  mass gap in four dimensions},''
  \href{http://dx.doi.org/10.1103/PhysRevD.80.065001}{{\em Phys. Rev.}
  {\bfseries D80} (2009) 065001},
\href{http://arxiv.org/abs/0709.3269}{{\ttfamily arXiv:0709.3269 [hep-th]}}.

\bibitem{Unsal:2008ch}
M.~Unsal and L.~G. Yaffe, ``{Center-stabilized Yang-Mills theory: Confinement
  and large N volume independence},''
  \href{http://dx.doi.org/10.1103/PhysRevD.78.065035}{{\em Phys. Rev.}
  {\bfseries D78} (2008) 065035},
\href{http://arxiv.org/abs/0803.0344}{{\ttfamily arXiv:0803.0344 [hep-th]}}.

\bibitem{Unsal:2007vu}
M.~Unsal, ``{Abelian duality, confinement, and chiral symmetry breaking in
  QCD(adj)},'' \href{http://dx.doi.org/10.1103/PhysRevLett.100.032005}{{\em
  Phys. Rev. Lett.} {\bfseries 100} (2008) 032005},
\href{http://arxiv.org/abs/0708.1772}{{\ttfamily arXiv:0708.1772 [hep-th]}}.

\bibitem{Kovtun:2007py}
P.~Kovtun, M.~Unsal, and L.~G. Yaffe, ``{Volume independence in large N(c)
  QCD-like gauge theories},''
  \href{http://dx.doi.org/10.1088/1126-6708/2007/06/019}{{\em JHEP} {\bfseries
  06} (2007) 019},
\href{http://arxiv.org/abs/hep-th/0702021}{{\ttfamily arXiv:hep-th/0702021
  [HEP-TH]}}.

\bibitem{Shifman:2008ja}
M.~Shifman and M.~Unsal, ``{QCD-like Theories on R(3) x S(1): A Smooth Journey
  from Small to Large r(S(1)) with Double-Trace Deformations},''
  \href{http://dx.doi.org/10.1103/PhysRevD.78.065004}{{\em Phys. Rev.}
  {\bfseries D78} (2008) 065004},
\href{http://arxiv.org/abs/0802.1232}{{\ttfamily arXiv:0802.1232 [hep-th]}}.

\bibitem{Shifman:2009tp}
M.~Shifman and M.~Unsal, ``{Multiflavor QCD* on R(3) x S(1): Studying
  Transition From Abelian to Non-Abelian Confinement},''
  \href{http://dx.doi.org/10.1016/j.physletb.2009.10.060}{{\em Phys. Lett.}
  {\bfseries B681} (2009) 491--494},
\href{http://arxiv.org/abs/0901.3743}{{\ttfamily arXiv:0901.3743 [hep-th]}}.

\bibitem{Cossu:2009sq}
G.~Cossu and M.~D'Elia, ``{Finite size phase transitions in QCD with adjoint
  fermions},'' \href{http://dx.doi.org/10.1088/1126-6708/2009/07/048}{{\em
  JHEP} {\bfseries 07} (2009) 048},
\href{http://arxiv.org/abs/0904.1353}{{\ttfamily arXiv:0904.1353 [hep-lat]}}.

\bibitem{Cossu:2013ora}
G.~Cossu, H.~Hatanaka, Y.~Hosotani, and J.-I. Noaki, ``{Polyakov loops and the
  Hosotani mechanism on the lattice},''
  \href{http://dx.doi.org/10.1103/PhysRevD.89.094509}{{\em Phys. Rev.}
  {\bfseries D89} no.~9, (2014) 094509},
\href{http://arxiv.org/abs/1309.4198}{{\ttfamily arXiv:1309.4198 [hep-lat]}}.

\bibitem{Argyres:2012ka}
P.~C. Argyres and M.~Unsal, ``{The semi-classical expansion and resurgence in
  gauge theories: new perturbative, instanton, bion, and renormalon effects},''
  \href{http://dx.doi.org/10.1007/JHEP08(2012)063}{{\em JHEP} {\bfseries 08}
  (2012) 063},
\href{http://arxiv.org/abs/1206.1890}{{\ttfamily arXiv:1206.1890 [hep-th]}}.

\bibitem{Argyres:2012vv}
P.~Argyres and M.~Unsal, ``{A semiclassical realization of infrared
  renormalons},'' \href{http://dx.doi.org/10.1103/PhysRevLett.109.121601}{{\em
  Phys. Rev. Lett.} {\bfseries 109} (2012) 121601},
\href{http://arxiv.org/abs/1204.1661}{{\ttfamily arXiv:1204.1661 [hep-th]}}.

\bibitem{Dunne:2012zk}
G.~V. Dunne and M.~Unsal, ``{Continuity and Resurgence: towards a continuum
  definition of the $\mathbb{CP}$(N-1) model},''
  \href{http://dx.doi.org/10.1103/PhysRevD.87.025015}{{\em Phys. Rev.}
  {\bfseries D87} (2013) 025015},
\href{http://arxiv.org/abs/1210.3646}{{\ttfamily arXiv:1210.3646 [hep-th]}}.

\bibitem{Poppitz:2012sw}
E.~Poppitz, T.~Schafer, and M.~Unsal, ``{Continuity, Deconfinement, and (Super)
  Yang-Mills Theory},'' \href{http://dx.doi.org/10.1007/JHEP10(2012)115}{{\em
  JHEP} {\bfseries 10} (2012) 115},
\href{http://arxiv.org/abs/1205.0290}{{\ttfamily arXiv:1205.0290 [hep-th]}}.

\bibitem{Anber:2013doa}
M.~M. Anber, S.~Collier, E.~Poppitz, S.~Strimas-Mackey, and B.~Teeple,
  ``{Deconfinement in $\mathcal{N}=1$ super Yang-Mills theory on $\mathbb{R}^3
  \times \mathbb{S}^1$ via dual-Coulomb gas and "affine" XY-model},''
  \href{http://dx.doi.org/10.1007/JHEP11(2013)142}{{\em JHEP} {\bfseries 11}
  (2013) 142},
\href{http://arxiv.org/abs/1310.3522}{{\ttfamily arXiv:1310.3522 [hep-th]}}.

\bibitem{Basar:2013sza}
G.~Basar, A.~Cherman, D.~Dorigoni, and M.~\"{U}nsal, ``{Volume Independence in
  the Large $N$ Limit and an Emergent Fermionic Symmetry},''
  \href{http://dx.doi.org/10.1103/PhysRevLett.111.121601}{{\em Phys. Rev.
  Lett.} {\bfseries 111} no.~12, (2013) 121601},
\href{http://arxiv.org/abs/1306.2960}{{\ttfamily arXiv:1306.2960 [hep-th]}}.

\bibitem{Cherman:2013yfa}
A.~Cherman, D.~Dorigoni, G.~V. Dunne, and M.~Unsal, ``{Resurgence in Quantum
  Field Theory: Nonperturbative Effects in the Principal Chiral Model},''
  \href{http://dx.doi.org/10.1103/PhysRevLett.112.021601}{{\em Phys. Rev.
  Lett.} {\bfseries 112} (2014) 021601},
\href{http://arxiv.org/abs/1308.0127}{{\ttfamily arXiv:1308.0127 [hep-th]}}.

\bibitem{Cherman:2014ofa}
A.~Cherman, D.~Dorigoni, and M.~Unsal, ``{Decoding perturbation theory using
  resurgence: Stokes phenomena, new saddle points and Lefschetz thimbles},''
  \href{http://dx.doi.org/10.1007/JHEP10(2015)056}{{\em JHEP} {\bfseries 10}
  (2015) 056},
\href{http://arxiv.org/abs/1403.1277}{{\ttfamily arXiv:1403.1277 [hep-th]}}.

\bibitem{Misumi:2014raa}
T.~Misumi and T.~Kanazawa, ``{Adjoint QCD on $\mathbb{R}^3\times S^1$ with
  twisted fermionic boundary conditions},''
  \href{http://dx.doi.org/10.1007/JHEP06(2014)181}{{\em JHEP} {\bfseries 06}
  (2014) 181},
\href{http://arxiv.org/abs/1405.3113}{{\ttfamily arXiv:1405.3113 [hep-ph]}}.

\bibitem{Misumi:2014jua}
T.~Misumi, M.~Nitta, and N.~Sakai, ``{Neutral bions in the ${\mathbb C}P^{N-1}$
  model},'' \href{http://dx.doi.org/10.1007/JHEP06(2014)164}{{\em JHEP}
  {\bfseries 06} (2014) 164},
\href{http://arxiv.org/abs/1404.7225}{{\ttfamily arXiv:1404.7225 [hep-th]}}.

\bibitem{Misumi:2014bsa}
T.~Misumi, M.~Nitta, and N.~Sakai, ``{Classifying bions in Grassmann sigma
  models and non-Abelian gauge theories by D-branes},''
  \href{http://dx.doi.org/10.1093/ptep/ptv009}{{\em PTEP} {\bfseries 2015}
  (2015) 033B02},
\href{http://arxiv.org/abs/1409.3444}{{\ttfamily arXiv:1409.3444 [hep-th]}}.

\bibitem{Dunne:2015ywa}
G.~V. Dunne and M.~Unsal, ``{Resurgence and Dynamics of O(N) and Grassmannian
  Sigma Models},'' \href{http://dx.doi.org/10.1007/JHEP09(2015)199}{{\em JHEP}
  {\bfseries 09} (2015) 199},
\href{http://arxiv.org/abs/1505.07803}{{\ttfamily arXiv:1505.07803 [hep-th]}}.

\bibitem{Misumi:2016fno}
T.~Misumi, M.~Nitta, and N.~Sakai, ``{Non-BPS exact solutions and their
  relation to bions in ${\mathbb C}P^{N-1}$ models},''
  \href{http://dx.doi.org/10.1007/JHEP05(2016)057}{{\em JHEP} {\bfseries 05}
  (2016) 057},
\href{http://arxiv.org/abs/1604.00839}{{\ttfamily arXiv:1604.00839 [hep-th]}}.

\bibitem{Cherman:2016hcd}
A.~Cherman, T.~Schafer, and M.~Unsal, ``{Chiral Lagrangian from Duality and
  Monopole Operators in Compactified QCD},''
  \href{http://dx.doi.org/10.1103/PhysRevLett.117.081601}{{\em Phys. Rev.
  Lett.} {\bfseries 117} no.~8, (2016) 081601},
\href{http://arxiv.org/abs/1604.06108}{{\ttfamily arXiv:1604.06108 [hep-th]}}.

\bibitem{Fujimori:2016ljw}
T.~Fujimori, S.~Kamata, T.~Misumi, M.~Nitta, and N.~Sakai, ``{Nonperturbative
  contributions from complexified solutions in $\mathbb{C}P^{N-1}$models},''
  \href{http://dx.doi.org/10.1103/PhysRevD.94.105002}{{\em Phys. Rev.}
  {\bfseries D94} no.~10, (2016) 105002},
\href{http://arxiv.org/abs/1607.04205}{{\ttfamily arXiv:1607.04205 [hep-th]}}.

\bibitem{Sulejmanpasic:2016llc}
T.~Sulejmanpasic, ``{Global Symmetries, Volume Independence, and Continuity in
  Quantum Field Theories},''
  \href{http://dx.doi.org/10.1103/PhysRevLett.118.011601}{{\em Phys. Rev.
  Lett.} {\bfseries 118} no.~1, (2017) 011601},
\href{http://arxiv.org/abs/1610.04009}{{\ttfamily arXiv:1610.04009 [hep-th]}}.

\bibitem{Yamazaki:2017ulc}
M.~Yamazaki and K.~Yonekura, ``{From 4d Yang-Mills to 2d $\mathbb{CP}^{N-1}$
  model: IR problem and confinement at weak coupling},''
  \href{http://dx.doi.org/10.1007/JHEP07(2017)088}{{\em JHEP} {\bfseries 07}
  (2017) 088},
\href{http://arxiv.org/abs/1704.05852}{{\ttfamily arXiv:1704.05852 [hep-th]}}.

\bibitem{Buividovich:2017jea}
P.~V. Buividovich and S.~N. Valgushev, ``{Lattice study of continuity and
  finite-temperature transition in two-dimensional SU(N) x SU(N) Principal
  Chiral Model},''
\href{http://arxiv.org/abs/1706.08954}{{\ttfamily arXiv:1706.08954 [hep-lat]}}.

\bibitem{Aitken:2017ayq}
K.~Aitken, A.~Cherman, E.~Poppitz, and L.~G. Yaffe, ``{QCD on a small
  circle},'' \href{http://dx.doi.org/10.1103/PhysRevD.96.096022}{{\em Phys.
  Rev.} {\bfseries D96} no.~9, (2017) 096022},
\href{http://arxiv.org/abs/1707.08971}{{\ttfamily arXiv:1707.08971 [hep-th]}}.

\bibitem{Dunne:2018hog}
G.~V. Dunne, Y.~Tanizaki, and M.~Unsal, ``{Quantum Distillation of Hilbert
  Spaces, Semi-classics and Anomaly Matching},''
\href{http://arxiv.org/abs/1803.02430}{{\ttfamily arXiv:1803.02430 [hep-th]}}.

\bibitem{Cherman:2017tey}
A.~Cherman, S.~Sen, M.~Unsal, M.~L. Wagman, and L.~G. Yaffe, ``{Order
  parameters and color-flavor center symmetry in QCD},''
  \href{http://dx.doi.org/10.1103/PhysRevLett.119.222001}{{\em Phys. Rev.
  Lett.} {\bfseries 119} no.~22, (2017) 222001},
\href{http://arxiv.org/abs/1706.05385}{{\ttfamily arXiv:1706.05385 [hep-th]}}.

\bibitem{Senthil1490}
T.~Senthil, A.~Vishwanath, L.~Balents, S.~Sachdev, and M.~P.~A. Fisher,
  ``Deconfined quantum critical points,''
  \href{http://dx.doi.org/10.1126/science.1091806}{{\em Science} {\bfseries
  303} no.~5663, (2004) 1490--1494}.

\bibitem{Yao:2018kel}
Y.~Yao, C.-T. Hsieh, and M.~Oshikawa, ``{Anomaly matching and
  symmetry-protected critical phases in $SU(N)$ spin systems in 1+1
  dimensions},''
\href{http://arxiv.org/abs/1805.06885}{{\ttfamily arXiv:1805.06885
  [cond-mat.str-el]}}.

\end{thebibliography}\endgroup

\end{document}